\documentclass[usenatbib]{mnras}
\usepackage{setspace}
\usepackage{epsf}
\usepackage{graphicx}
\usepackage{color}
\usepackage{txfonts}
\usepackage{tabularx}
\usepackage{here}
\usepackage{float}

\usepackage{lscape}
\usepackage{pdflscape}

\newcommand{\oi}{[O\,{\sc i}]}
\newcommand{\oii}{[O\,{\sc ii}]}
\newcommand{\oiii}{[O\,{\sc iii}]}
\newcommand{\oiv}{[O\,{\sc iv}]}

\newcommand{\Ni}{[N\,{\sc i}]}
\newcommand{\nii}{[N\,{\sc ii}]}

\newcommand{\sii}{[S\,{\sc ii}]}
\newcommand{\siii}{[S\,{\sc iii}]}
\newcommand{\siv}{[S\,{\sc iv}]}

\newcommand{\hei}{He\,{\sc i}}
\newcommand{\heii}{He\,{\sc ii}}

\newcommand{\neii}{[Ne\,{\sc ii}]}
\newcommand{\neiii}{[Ne\,{\sc iii}]}
\newcommand{\neiv}{[Ne\,{\sc iv}]}

\newcommand{\ariii}{[Ar\,{\sc iii}]}
\newcommand{\ariv}{[Ar\,{\sc iv}]}

\newcommand{\cliii}{[Cl\,{\sc iii}]}
\newcommand{\cliv}{[Cl\,{\sc iv}]}

\newcommand{\feiii}{[Fe\,{\sc iii}]}

\newcommand{\cii}{C\,{\sc ii}}
\newcommand{\ciii}{C\,{\sc iii}}
\newcommand{\civ}{C\,{\sc iv}}

\newcommand{\ha}{H$\alpha$}
\newcommand{\hb}{H$\beta$}

\usepackage{color} 
\newcommand{\kms}{km s$^{-1}$}

\newcommand{\te}{$T_{\epsilon}$}
\newcommand{\Ne}{$n_{\epsilon}$}

\newcolumntype{C}{>{\centering\arraybackslash}X}
\newcolumntype{L}{>{\raggedright\arraybackslash}X}
\newcolumntype{R}{>{\raggedleft\arraybackslash}X}

\title[Chemical Abundances and Dust in the PN Wray16-423]
{Chemical Abundances in the PN Wray16-423 in the Sagittarius Dwarf
Spheroidal Galaxy: Constraining the Dust Composition\thanks{
Based on data collected by the Subaru Telescope,
operated by the National Astronomical Observatory of Japan
(NAOJ) under program IDs: S12A-126S and S14A-174 (PI: M.~Otsuka).}
\thanks{
Based on observations made with the MPG/ESO 2.2-m Telescope at the La
Silla Observatory under program ID 91.D-0055A (PI: M.~Otsuka).}
}

\author[M.~Otsuka]{Masaaki
Otsuka$^{1,2}$\thanks{E-mail:otsuka@asiaa.sinica.edu.tw}
\\
\\
\begin{minipage}{1.0\linewidth}
$^{1}$Institute of Astronomy and Astrophysics, Academia Sinica, 
11F of Astronomy-Mathematics Building, AS/NTU. No.1, Sec. 4, Roosevelt
Rd, Taipei 10617, Taiwan, R.O.C.\\
$^{2}$Subaru Telescope, 650 North A'ohoku Place, Hilo, HI 96720, USA
\end{minipage}
}

\begin{document}

\date{}

\pagerange{\pageref{firstpage}--\pageref{lastpage}} \pubyear{2015}

\maketitle

\label{firstpage}

\begin{abstract}
We performed a detailed analysis of elemental abundances, dust features, 
and polycyclic aromatic hydrocarbons (PAHs) in the C-rich planetary
 nebula (PN) Wray16-423 in the Sagittarius dwarf spheroidal galaxy,
 based on a unique dataset taken from the Subaru/HDS, MPG/ESO FEROS,
 \emph{HST}/WFPC2, and \emph{Spitzer}/IRS. We performed 
the first measurements of Kr, Fe, and recombination O abundance 
in this PN. The extremely small [Fe/H] implies that most Fe atoms are in
 the solid phase, considering into account the abundance of [Ar/H]. The
 \emph{Spitzer}/IRS spectrum displays broad 16-24\,$\mu$m and 30\,$\mu$m
 features, as well as PAH bands at 6-9\,$\mu$m and 10-14\,$\mu$m. The
 unidentified broad 16-24\,$\mu$m feature may not be related to iron
 sulfide (FeS), amorphous silicate, or PAHs. Using the spectral energy
 distribution model, we derived the luminosity and effective temperature of the
 central star, and the gas and dust masses. The observed elemental
 abundances and derived gas mass are in good agreement with asymptotic
 giant branch nucleosynthesis models for an initial mass of
 1.90\,M$_{\odot}$ and a metallicity of $Z$=0.004. We infer that
 respectively about 80\,\%, 50\,\%, and 90\,\% of the Mg, S, and Fe
 atoms are in the solid phase. We also assessed the maximum possible
 magnesium sulfide (MgS) and iron-rich sulfide (Fe50S) masses and tested
 whether these species can produce the band flux of the observed 30\,$\mu$m
 feature. Depending on what fraction of the sulfur is in sulfide
 molecules such as CS, we conclude that MgS and Fe50S could be possible
 carriers of the 30\,$\mu$m feature in this PN.

  \end{abstract}

\section{Introduction}

Planetary nebulae (PNe) represent the final stage in the evolution of 
initially $\sim$1-8\,M$_{\odot}$ stars. During their evolution, these
mass stars eject a large amount of their mass into circumstellar shells
via strong stellar winds. Eventually, the ejecta blend with 
interstellar media (ISM). The evolution history
of the progenitors is imprinted in the central stars of PNe (CSPNe), in
the atomic gas, dust, and molecule-rich circumstellar shells.
Using the emission lines of gas and thermal radiation from dust grains, one
can simultaneously obtain abundances of elements and dust, as well as the ejected
gas/dust masses from the PN progenitors. An
investigation of these materials provides insight into stellar
evolution and the ISM material-recycling mechanism of host galaxies.
Thus, PNe provide a unique laboratory for the study of stellar and galaxy evolution.

Elemental abundances, the luminosities of CSPNe,
and gas and dust masses are essential to understanding stellar and
galaxy evolution; the latter three parameters are more accurately obtained through
observations of extragalactic PNe, rather than Galactic PNe, because
the distance is well determined. Currently, there are over 4,200 known 
extragalactic PNe \citep[from Table~1 of][]{Reid:2012aa}.
In the galaxy that is nearest to us
\citep[24.8$\pm$0.8~kpc;][]{Kunder:2009aa}, 
the Sagittarius dwarf spheroidal galaxy (Sgr dSph), four PNe have been confirmed
to date; BoBn1, StWr2-21, Hen2-436,
and Wray16-423 \citep{Walsh:1997aa,Dudziak:2000aa,2006MNRAS.369..875Z,Kniazev:2008aa,
2008ApJ...682L.105O,2010ApJ...723..658O,Otsuka:2011aa}.

Sgr dSph PNe offer obvious advantages over other extragalactic PNe 
in terms of the study of stellar evolution, such as
those found in Magellanic Clouds (MCs); Sgr dSph PNe are closer to
us and therefore appear brighter than other extragalactic PNe, 
allowing the detection of rare elements such as neutron ($n$)-capture
elements. 
The final yields of neon and $n$-capture elements are
largely concerned with the $^{13}$C pocket region that forms in the 
He-rich intershell during the thermal-pulse asymptotic giant branch
(AGB) phase. The extent of the $^{13}$C pocket is an important parameter in 
AGB nucleosynthesis
models; however, it remains unconstrained. An assessment of $n$-capture and
neon abundances, and the progenitor mass derived from Sgr dSph PNe could be used to 
test the $^{13}$C pocket
mass adopted in AGB models.

The C/O ratio is an important parameter with regard to CSPN and
nebula characterisation and initial mass constraints. The C and O
abundances using recombination lines (RLs) are sometimes larger than
those using collisionally excited lines 
\citep[CELs; e.g., see][for details]{Liu:2006aa}. Moreover, 
the discrepancy factor between the RL and 
CEL abundances is generally different for O and for C 
\citep[e.g., see][]{Wang:2007aa}. As a consequence, the C/O 
ratio determined from dividing the RL C abundance by the CEL O abundance might not be
indicative of the actual C/O ratio. Therefore, we should at the very
least determine the C/O ratios using the same type of emission
lines. According to \citet{Delgado-Inglada:2014aa}, the
uncertainty of the RL C/O ratio ($\pm$0.06 dex) is considerably smaller
than that of the CEL C/O ratio ($\pm$0.2 dex). An important factor of
the CEL C/O uncertainty is the normalization between UV and optical
fluxes since CEL C abundances are determined using the UV 
$[$C\,{\sc iii}\,$\lambda\lambda$1906/09\,{\AA} lines whereas 
we measure the CEL O abundances using the
optical {\oiii}\,$\lambda\lambda$4959/5007\,{\AA} and
{\oii}\,$\lambda\lambda$3726/29\,{\AA}. By using C and O
RLs in the optical, we do not have to worry about this difficult
flux normalization and one can simultaneously determine C and O
abundances and then the C/O ratio, as performed in Sgr dSph PNe 
BoBn1 \citep{2010ApJ...723..658O} and Hen2-436 \citep{Otsuka:2011aa}.

The Hubble Space Telescope (\emph{HST}) can resolve the
central stars of Sgr dSph PNe, as reported in \citet{2006MNRAS.369..875Z}.
Therefore, the luminosities of the CSPNe can be measured directly,
without the need for empirical methods.
The mass of the progenitor star can be estimated by plotting the measured luminosity and 
temperature of the CSPN on theoretical post-AGB evolution
tracks. Thus, using this approach, the observed elemental abundances and gas
mass can be compared to predicted values from AGB nucleosynthesis
models.

Previously, we carried out
multiwavelength observational studies of the Sgr dSph PNe BoBn1 
\citep{2010ApJ...723..658O} and Hen2-436 \citep{Otsuka:2011aa}. While all Sgr
dSph PNe are metal-poor \citep{Walsh:1997aa,Dudziak:2000aa,2006MNRAS.369..875Z}, 
BoBn1 is extremely metal-poor when comparing
to Hen2-436 or Wray16-243
\citep{Otsuka:2015aa,2006MNRAS.369..875Z,Kniazev:2008aa}. 
Our study on chemical abundances and dust in
these Sgr dSph PNe has provided us with much information on PNe and
their progenitor's evolution. \citet{2010ApJ...723..658O} measured the rare
$n$-capture elements Xe and Ba abundances for first time in
extragalactic PNe. Based on the abundance pattern of BoBn1, they
proposed that the progenitor might be a binary. \citet{Otsuka:2011aa}
measured the $n$-capture Kr abundance for the first time in Hen2-436 and
found that the abundance pattern is very similar to that of
Wray16-423. Both studies furthermore also derived dust masses and
discussed the dust production in metal-poor environments. In this
paper, we focus on the Sgr dSph PN Wray 16-423.

Wray16-423 was discovered near the core of the Sgr dSph
by \citet{Zijlstra:1996aa}, based on direction and radial
velocities. \citet{Ibata:1995aa} reported an average heliocentric 
radial velocity for the Sgr dSph of 140$\pm$2\,{\kms}, with an intrinsic velocity dispersion
of 11.4$\pm$0.7\,{\kms} \citep{Ibata:1997aa}. The radial velocity of
Wray16-423 was determined to be +133.1$\pm$2\,{\kms} by
\citet{Zijlstra:1996aa}. 
The elemental abundances of the nebula and the luminosity and effective temperature of the CSPN were calculated by
\citet{Walsh:1997aa}; these values were refined by the photoionisation
models of \citet{Dudziak:2000aa}, \citet{Gesicki:2003aa}, and
\citet[][see Table~\ref{abund} and Section~\ref{S-element} for a
comparison of these values]{2006MNRAS.369..875Z}. \citet{Dudziak:2000aa} 
calculated the respective C and O abundances from the C RLs and
the O CELs. The C/O ratio using the same
type of emission lines is still unknown. \citet{Dudziak:2000aa} expected
an [Fe/H] abundance of $\sim$--0.5, although they did not report the
detection of Fe lines. \citet{2006MNRAS.369..875Z} 
determined the magnitude of the CSPN to be
$m_{V}$=18.85$\pm$0.20, based on \emph{HST}/F547M measurements. However, this
estimate seems to be too bright. Indeed, assuming a spectral energy
distribution defined by an $m_{V}$=18.85 CSPN and adopting an effective
temperature of 107\,000 K \citep{2006MNRAS.369..875Z}, the 
luminosity $L_{\ast}$ should be $\sim$12\,600\,L$_{\odot}$. 
However, this does not match their estimated value of 
$L_{\ast}$=4350\,L$_{\odot}$ obtained using a photoionisation
model. This discrepancy and a revision of the CSPN magnitude is
discussed in Section~\ref{hsts}.

\citet{Stanghellini:2012aa} reported that the \emph{Spitzer}
mid-infrared spectrum of Wray16-423 displays carbonous 
dust features. As we report later, the broad 30\,$\mu$m feature is 
seen in this PN. The broad 30\,$\lambda$m feature 
has been found in many C-rich objects. However, 
the carrier of the broad 30\,$\mu$m feature is unidentified;
\citet{Hony:2002ab} proposed that the feature originates
from magnesium sulfide (MgS) whose shapes can be described by a
continuous distribution of ellipsoids (CDE). Since then, CDE MgS
grains have been the major candidate for the 30\,$\mu$m
feature \citep[e.g.,][]{Zijlstra_06_LMC,Sloan:2014aa}. 
However, the recent study of \citet{Zhang_09_30mic} casts 
doubt on this identification for the 30\,$\mu$m feature. 
They demonstrated that the MgS mass formed from the available
S atoms, could not account for the observed strength of the 30\,$\mu$m
feature in the Galactic proto-PN HD56126, even if all the dust grains
existed as MgS. CDE iron-rich sulphide, such as Mg$_{0.5}$Fe$_{0.5}$S (Fe50S),
also show a broad feature around 30\,$\mu$m, similar to the CDE MgS 
feature \citep{Messenger:2013aa}. Using the data of Wray16-423, we can 
test whether the strength of the 30\,$\mu$m feature can be explained 
by thermal emission from the available MgS or Fe50S grains. Since 
the distance to Wray16-423 is well determined, one can directly determine
the luminosity of the CSPN as the gas and dust heating source and 
set the size of the dusty nebula. Therefore, the solid phase Mg, S, 
and Fe masses and the MgS and Fe50S grain masses 
would be more credible by using the accurately measured Mg, S and Fe abundances.

For the above reasons, we performed a detailed spectroscopic analysis of
Wray16-423 based on multiwavelength data taken from Subaru/HDS, MPG
ESO/FEROS, \emph{Spitzer}/IRS, and \emph{HST}/WFPC2: (1) to determine and refine nebular elemental
abundances, in particular, the C and O recombination abundances
and Fe and slow $n$-capture process ($s$-process) elements; (2) to investigate dust features and 
PAH bands and compare these among LMC PNe; (3)
to derive the physical conditions of the CSPN, gas, and dust by
constructing the spectral energy distribution (SED) model using the
radiative transfer code {\sc Cloudy} \citep{Ferland:1998aa}; 
(4) to estimate the progenitor mass by comparing the observed elemental
abundances and estimated gas mass with AGB model predictions; and (5)
test how much masses of the MgS and Fe50S grains are necessary to reproduce the broad
30\,$\mu$m feature.

\section{Observations and Data Reductions}

\subsection{Subaru 8.2-m Telescope HDS spectroscopy}

\begin{table}
\centering
\caption{Summary of optical spectroscopic observations.\label{obs_jour}}  
\begin{tabularx}{\linewidth}{@{}c@{\hspace{4pt}}l@{\hspace{4pt}}c@{\hspace{4pt}}l@{}}
\hline
 Date &Telescope/Instrument&Wavelength&Exposure time\\
\hline
2012/07/04&Subaru 8.2-m/HDS&4300-7100\,{\AA}&2$\times$1800\,s, 360\,s\\
2013/06/18&MPG/ESO 2.2-m/FEROS&3500-9200\,{\AA}&3$\times$3000\,s, 2$\times$300\,s\\
2014/07/09&Subaru 8.2-m/HDS&3680-5400\,{\AA}&2$\times$1800\,s, 180\,s, 30\,s\\
 \hline
\end{tabularx}
\end{table}

Optical high-dispersion spectra were taken using the High-Dispersion 
Spectrograph \citep[HDS;][]{2002PASJ...54..855N} attached to the 
Nasmyth focus of the 8.2-m Subaru Telescope on 2012 July 4 (Prop.ID:
S12A-126S, PI: M.~Otsuka) and 2014 July 9 (Prop.ID: S14A-174, the same
PI). HDS observations are summarised in Table~\ref{obs_jour}.

The weather conditions during the exposure for both runs were 
stable, and the visibility was $\sim$0.5{\arcsec} (2012) and
$\sim$0.8{\arcsec} (2014), measured using a guider CCD. An
atmospheric dispersion corrector (ADC) was used to minimise the differential 
atmospheric dispersion through the broad wavelength region. 
The slit width and length were set to 1.2{\arcsec} and 6{\arcsec}, 
respectively, and the 2$\times$2 on-chip binning mode was selected. 
The resolving power ($\lambda$/$\Delta\lambda$) reached 33\,500, which
was measured from
the average full width at half maximum (FWHM) of over 600 narrow Th-Ar
comparison lines obtained for wavelength calibrations.
For the flux calibration, blaze function correction, 
and airmass correction, we observed the standard star HD184597.

We reduced the data with the echelle spectra reduction package {\sc ECHELLE} and the
two-dimensional spectra reduction package {\sc TWODSPEC} in {\sc IRAF}\footnote{IRAF
is distributed by the National Optical Astronomy Observatories, 
operated by the Association of Universities for Research in
Astronomy (AURA), Inc., under a cooperative agreement with the National
Science Foundation.}. 
To keep a good signal-to-noise (S/N) over the
observed wavelength range, we extracted a region of 15 pixels along the slit
(corresponding to 4.14{\arcsec}) and summed all fluxes. 
For the flux calibration, we reduced the ESO archival spectra
($\sim$0.3-2.5\,$\mu$m) of HD184597, taken using XSHOOTER 
\citep[][Prop.ID: 60.A-9022C]{Vernet:2011aa}. We corrected the reduced 0.3-2.5\,$\mu$m
spectrum of HD184597 to match the $B$- and $V$-band magnitudes reported in
\citet[$B$=6.81 and $V$=6.88]{Hog:2000aa}, and then we constructed a
wavelength versus $AB$-magnitude table in small wavelength increments
(0.4\,{\AA}, for this study) for performing flux calibration of the Wray16-423 spectra.

The resulting S/N ratio in the sky-subtracted object frame of Wray16-423
is $>$5 for the continuum and $>$8 at the intensity peaks of the emission lines.

The resultant HDS spectrum is presented in Figures~\ref{hdsspec1}
and \ref{hdsspec2} of Appendix~A.

\subsection{MPG ESO 2.2-m Telescope FEROS spectroscopy \label{feros-obs}}

Optical high-dispersion spectra (3500-9200\,{\AA}) were obtained using 
the Fiber-fed Extended Range Optical Spectrograph
\citep[FEROS;][]{Kaufer:1999aa} 
attached to the MPG/ESO 2.2-m Telescope, La Silla, Chile on 2013 June 18 
(Prop.ID: 91.D-0055A, PI: M.~Otsuka), as a follow-up
up to HDS observations; auroral {\oii} lines,
nebular {\siii}, {\cliv}, and {\ariii} lines, and the Paschen
discontinuity were measured, and line identifications and flux
measurements were cross-checked for both HDS and FEROS spectra.
The FEROS observation is summarised in Table~\ref{obs_jour}.

The weather conditions were stable and clear throughout the night, and
the visibility was $\sim$1.1{\arcsec} measured from the guider CCD.
FEROS's fibers use 2.0{\arcsec} apertures, to provide simultaneous spectra of 
the object and sky frames. The ADC was used during
the observation. We selected a 2$\times$2 on-chip binning mode and low
gain mode\footnote{The gain and readout-noise in the case of 2$\times$2
binning and low gain mode were measured to be
4.03\,$e^{-}$\,ADU$^{-1}$ and 5.68\,$e^{-}$ using the IRAF task {\sc FINDGAIN}.}. 
The resolving power reached $\sim$37\,200
measured from the average FWHM of over 100 Th-Ar comparison lines. 
We also took 5$\times$3600\,sec dark frames to correct for the zero
intensity of longer exposure-time frames. 
For the same reason as described in the HDS spectroscopy, we summed over the $\pm$3
pixels centered on the fibre position. We confirmed that the extracted
spectra detected $\sim$65\,\% of the light falling into the 2{\arcsec}
aperture by comparing to the spectra made by summing up the full
aperture. For flux calibration, we observed the standard star HR7950,
before and after the Wray16-423 exposure. We referred to the HR7950 spectrum of
\citet{Valdes:2004aa} for constructing a wavelength versus $AB$-magnitude (0.4\,{\AA} step). 

We reduced the data using the same method as that adopted 
for the HDS data. The resulting S/N ratio in the sky-subtracted object
frame is $>$2 on the continuum and $>$5 at the intensity peaks of the emission lines.

\subsection{Spitzer/IRS spectroscopy \label{spit}}

\begin{figure}
\includegraphics[width=\columnwidth,bb=18 318 573 686,clip]{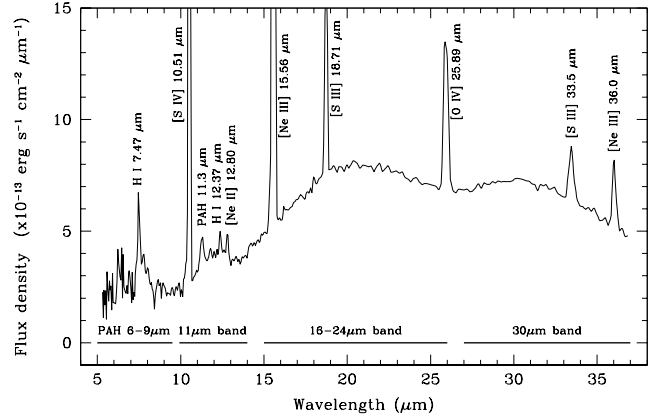}
 \caption{\emph{Spitzer}/IRS spectrum of Wray16-423.
 As-observed atomic lines, PAHs, and broadband features are indicated.
}
\label{spt_spec}
\end{figure}

We analysed the archival mid-infrared spectra taken by the 
\emph{Spitzer}/Infrared Spectrograph \citep[IRS;][]{Houck:2004aa} with the SL 
(5.2-14.5\,$\mu$m, the slit dimension: 3.6{\arcsec}$\times$57{\arcsec}) and the LL
modules (13.9-39.9\,$\mu$m, 10.6{\arcsec}$\times$168{\arcsec}).
The data were originally taken by L.~Stanghellini (AOR Key: 25833216) on
2008 June 9. We processed them using the reduction package 
{\sc SMART} v.8.2.9 \citep{Higdon:2004aa} and IRSCLEAN v.2.1.1
provided by the \emph{Spitzer} Science Center.
We performed additional correction of the IRS spectrum by scaling it up to match the flux density at the band W4
($\lambda_{\rm c}$=22.09\,$\mu$m, $\Delta\lambda$=4.10\,$\mu$m)
of the Wide-field Infrared Survey Explorer \citep[\emph{WISE};][]{Wright:2010aa}, using a constant
factor of 1.069. 

The resulting spectrum is presented in
Figure~\ref{spt_spec}. 
The spectrum clearly shows the 6-9\,$\mu$m and 
11.3\,$\mu$m PAHs, the broad 11\,$\mu$m, 16-24\,$\mu$m, and
30\,$\mu$m bands.

\subsection{HST/WFPC2 photometry and the {\ha}/{\hb} fluxes \label{hsts}}

\begin{figure}
\centering
\includegraphics[width=\columnwidth]{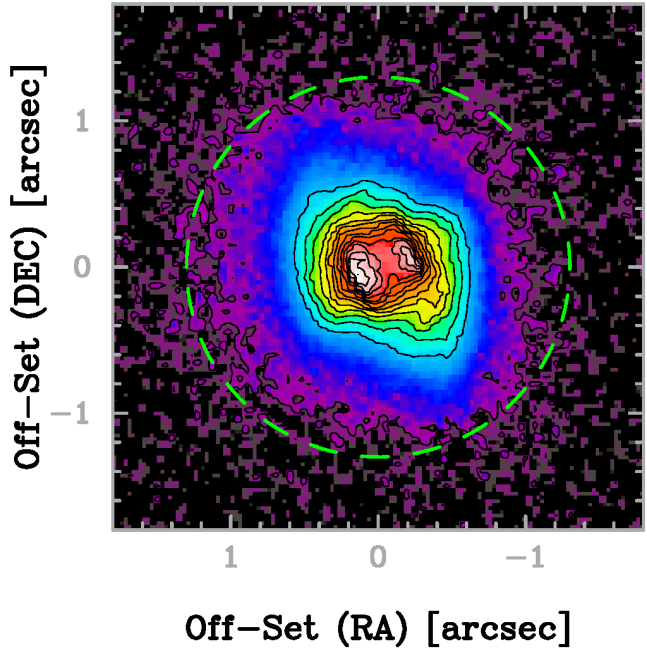}
 \caption{$HST$/WFPC2 F656N ({\ha}) image of Wray16-423. The levels of the
 contour plots are from 3-$\sigma$ above the mean background
 intensity to the peak intensity of the object. The radius of the
 green dashed circle is 1.3$''$.
\label{hst}}
\end{figure}

  \begin{table}
   \centering
   \caption{Photometry of the \emph{HST}/WFPC2 images. \label{wfpc2}}
\begin{tabularx}{\columnwidth}{@{}c@{\hspace{24pt}}c@{\hspace{24pt}}l@{\hspace{24pt}}c@{}}
\hline
Band  &$\lambda_{c}$   &Portion      &$F_{\lambda}$\\ 
      &({\AA})         &    &(erg s$^{-1}$ cm$^{-2}$\,{\AA}$^{-1}$)\\
\hline
F547M($V$)&5483.86    &CSPN only    &(3.75$\pm$0.34)$\times$10$^{-17}$\\
     &           &whole nebula  &(1.11$\pm$0.04)$\times$10$^{-15}$\\
F656N({\ha})&6563.76    &whole nebula  &(9.61$\pm$0.15)$\times$10$^{-14}$\\
\hline
\end{tabularx}
\raggedright
Note -- The $V$-band magnitude of the CSPN is 19.98$\pm$0.20, determined
   from the F547M image.  \\
\end{table}

  We reduced the \emph{HST}/Wide Field Planetary Camera 2 (WFPC2) F656N
  ({\ha}, $\lambda_{\rm c}$=6563.76\,{\AA},
  $\Delta\lambda$=53.77\,{\AA}) and F547M images ($V$-band, $\lambda_{\rm
c}$=5483.86\,{\AA}, $\Delta\lambda$=205.52\,{\AA}) using the {\sc
  MultiDrizzle} \citep{Koekemoer:2003aa} on {\sc
PYRAF}. The data were taken on 2002 July 25 (Prop.ID: 9356, PI: A.~Zijlstra).
We set the plate scales to a constant 0.025{\arcsec} pixel$^{-1}$.
The resultant F656N image is displayed in Figure~\ref{hst}.
The Wray16-423 image shows an inner cylindrical structure,
surrounded by an elliptical nebula shell that extends radially to $\sim$1.3{\arcsec}, 
as indicated by the green dashed circle. 
In the F656N and F547M images, we measured the count rates (cts) within
the aperture radius of 1.3{\arcsec}, corresponding to the sum of the CSPN's and
nebula's cts. We measured average (background) cts in an annulus with
inner radius 2.1{\arcsec} and outer radius 2.4{\arcsec}, 
centered on the CSPN. Finally, we converted the cts into flux density, using the WFPC2 
photometric zero-points in erg s$^{-1}$ cm$^{-2}$\,{\AA}$^{-1}$ cts$^{-1}$.
The results are summarised in Table~\ref{wfpc2}.

To measure the pure {\ha} flux using the F656N flux density, 
one must remove the contribution from both the local continuum and the {\nii}\,$\lambda$\,6548
{\AA}. We used the FEROS spectrum to estimate these
contributions. Taking into account the F656N filter transmission
characteristics, we compared the $F_{\lambda}$(\emph{HST},F656N) with
the counterpart FEROS spectrum, i.e.,
$F_{\lambda}$(FEROS,F656N). Finally, we determined the
scaling factor
$F_{\lambda}$(\emph{HST},F656N)/$F_{\lambda}$(FEROS,F656N) =
2.383\footnote{
We would estimate the Wray16-423's nebula to be
$\gtrsim$2.8{\arcsec} at the 1.1{\arcsec} seeing, assuming that its
actual nebula size is 2.6{\arcsec}. From our report in
Section~\ref{feros-obs}, our FEROS spectra detected $\sim$0.65$\times$(2.0{\arcsec}/2.8{\arcsec})$\times$
100$\sim$46\,$\%$ or less of the light from Wray16-423. In the actual
observation, the seeing was not a constant 1.1{\arcsec} and also the thin
clouds were passing during the exposures. This is the reason why 
we need such a correction factor.}.

Using this scaled FEROS spectrum, we measured 
a pure $F$({\ha}) of (4.70$\pm$0.03)$\times$10$^{-12}$\,erg
s$^{-1}$ cm$^{-2}$ and a pure $F$({\hb}) of 
(1.53$\pm$0.02)$\times$10$^{-12}$\,erg s$^{-1}$ cm$^{-2}$. We used the measured
pure $F$({\hb}) to normalise the flux in the flux
density scaled \emph{Spitzer}/IRS spectrum, as explained in the
previous section.

Next, we measured the $V$-band flux density of the CSPN in the F547M
image. 
In the flux density measurements of the CSPN, we need to select the
background area more carefully because the CSPN is embedded in a bright
nebula and therefore the CSPN magnitude largely depends on the
background region adopted. The background cts adopted in the flux 
density measurement of the whole nebula are not suitable 
for the flux density measurements of the CSPN. For this reason,
we measured cts in four positions near ($\sim$0.2-0.3{\arcsec}
away) the CSPN and used the average
cts as the sky background. The measured flux density of the CSPN in the
F547M band corresponds to the $V$-band magnitude $m_{V}$ of 19.98$\pm$0.20, which is
about 1 magnitude fainter than \citet[$m_{V}$=18.85$\pm$0.20]{2006MNRAS.369..875Z}.
The difference would be attributed to the adopted background region; the
$m_{V}$ is 18.73$\pm$0.10 when as the background we use the average cts measured in 
the area of the annulus centered on the PN with inner 
and outer radii of 2.1{\arcsec} and 2.4{\arcsec}, respectively.

\section{Results}

\subsection{Reddening correction, line-flux measurements, and radial velocity}

\begin{figure}
\centering
\includegraphics[width=\columnwidth]{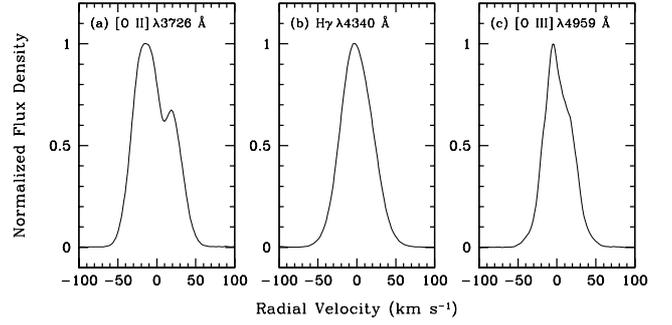}
\caption{({\it from left to right}) The emission line profiles of the
 {\oii}\,$\lambda$\,3726.03\,{\AA}, H$\gamma$\,4340.46\,{\AA}, and
 {\oiii}\,$\lambda$\,4958.91\,{\AA}, from HDS observations. 
The wavelength is converted into radial velocity with respect to the
 systemic radial velocity of +133.12\,{\kms}. The flux density is 
normalised to the flux density at the peak.}
\label{lineprof}
\end{figure}

The measured line fluxes in the obtained spectra were de-reddened using the
following formula: 

\begin{equation}
I(\lambda) = F(\lambda)\times10^{c({\rm H}\beta)(1+f(\lambda))},
\end{equation}

\noindent 
where $I$($\lambda$) is the de-reddened line-flux, $F$($\lambda$) is the observed line-flux, 
$f$($\lambda$) is the interstellar extinction function at $\lambda$
computed by the reddening law of \citet{1989ApJ...345..245C} with
$R_{V}$=3.1, and $c$(H$\beta$) is the reddening coefficient at H$\beta$. 
For the \emph{Spitzer}/IRS spectrum, we adopted the reddening
law of \citet{Fluks:1994aa}.

We measured $c$({\hb}) by comparing 
the observed Balmer line ratios of H$\gamma$ and H$\alpha$ to H$\beta$ 
with the theoretical ratios of \citet{1995MNRAS.272...41S} for an 
electron temperature {\te}=10$^{4}$\,K and electron density
{\Ne}=10$^{4}$\,cm$^{-3}$, under the Case B assumption. 
The H$\alpha$ line in the HDS short-exposure spectrum taken in the 2012 run was affected by
cosmic rays. Therefore, we used a $c$({\hb}) value of 0.132$\pm$0.035,
determined from the $F$(H$\gamma$)/$F$(H$\beta$) ratio. This 
$c$({\hb}) value is in excellent agreement with that in the 2014 HDS run
(0.133$\pm$0.063). In the FEROS spectrum, we obtained a
$c$({\hb}) value of 0.110$\pm$0.022 from the $F$(H$\alpha$)/$F$(H$\beta$) ratio.

We fitted the emission lines in the HDS and FEROS spectra using multiple Gaussian components.
Figure~\ref{lineprof} shows
the emission line profiles of the {\oii}\,$\lambda$\,3726.03\,{\AA},
H$\gamma$\,$\lambda$\,4340.46\,{\AA}, and
{\oiii}\,$\lambda$\,4958.91\,{\AA}.
We obtained a heliocentric radial velocity of +133.12$\pm$0.28\,{\kms}
(RMS of the residuals: 3.17\,{\kms}) from the central wavelengths of {\it all} the lines
detected in the HDS and FEROS spectra. Most line profiles can be fitted by a single Gaussian
component, whereas several lines show double
peaks or blue-shifted asymmetry, as seen in {\oii} and {\oiii}.

We list the observed wavelengths and de-reddened relative fluxes of
each Gaussian component, indicated by the identification (ID) number in
Table~\ref{hdstab} of Appendix~A, with respect to the
de-reddened H$\beta$ flux $I$({\hb}) of 100. For lines composed of multiple 
components, we list the de-reddened relative fluxes of each
component, as well as the sum of these components (indicated by 'T'). 
In the ions showing multiple intensity peaks such as
{\oii}\,$\lambda$3726\,{\AA}, one could derive electron temperatures
and densities in each Gaussian component. In our data, it would be difficult
to accurately derive their ionic abundances relative to the H$^{+}$ in each
component because each velocity component of the detected lines are
different. Since our aim is to investigate average ionic and elemental
abundances in the nebula, we used the integrated fluxes of
each Gaussian component. We list the results of the flux measurements for the \emph{Spitzer}/IRS 
spectrum in Table~\ref{spt_tbl} of Appendix~A, where we adopted the
$c$({\hb}) of 0.110$\pm$0.022, which is the same value applied for the
FEROS spectrum.

\subsection{Plasma diagnostics}

\begin{figure}
\centering
\includegraphics[width=\columnwidth,bb=29 202 547 587,clip]{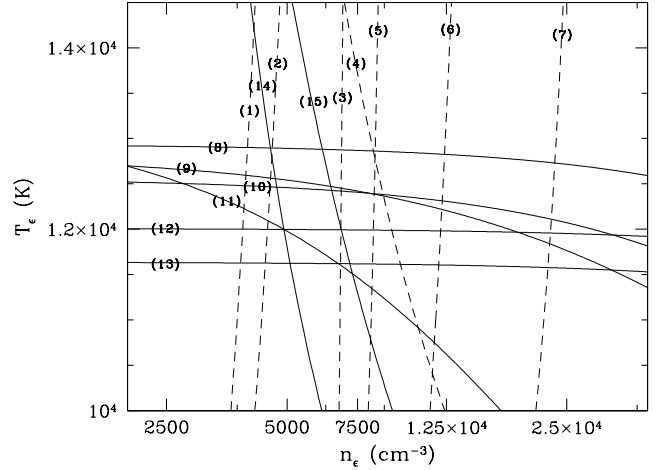}
 \caption{$n_{\epsilon}$-$T_{\epsilon}$ diagram based on diagnostic
 CEL ratios. Each curve is labelled with an ID
 number, listed in Table~\ref{diagno_table}.  The thick lines indicate the
  lines of $T_{\epsilon}$ diagnostic. The dashed lines indicate the diagnostic
 lines of $n_{\epsilon}$. The {\Ne}-{\te} curve calculated from the
 $[$N\,{\sc i}$]$ $I$($\lambda$\,5198)/$I$($\lambda$\,5200) is out of the
 plane, because {\Ne}=1010\,cm$^{-3}$. \label{diagno_figure}} 
\end{figure}

\begin{table}
\centering
\caption{Plasma diagnostics.\label{diagno_table}}
\begin{tabularx}{\columnwidth}{@{}c@{\hspace{7pt}}l@{\hspace{10pt}}l@{\hspace{10pt}}r@{}}
 \hline
ID & $n_{\epsilon}$ diagnostic&Value&Result\\
   &                          &     &(cm$^{-3}$)\\
 \hline
(1)&$[$S~{\sc ii}$]$($\lambda$\,6716)/($\lambda$\,6731) & 0.60$\pm$0.03 & 3970$\pm$690 \\ 
(2)&$[$O~{\sc ii}$]$($\lambda$\,3726)/($\lambda$\,3729) & 1.82$\pm$0.13 & 3540$\pm$790 \\ 
(3)&$[$S~{\sc iii}$]$($\lambda$\,18.7\,$\mu$m)/($\lambda$\,33.5\,$\mu$m) &2.30$\pm$0.19 &6860$\pm$1060 \\ 
(4)&$[$O~{\sc ii}$]$($\lambda$\,3726/29)/($\lambda$\,7320/30) & 7.42$\pm$0.54 & 6870$\pm$650 \\  
(5)&$[$Cl~{\sc iii}$]$($\lambda$\,5517)/($\lambda$\,5537) & 0.69$\pm$0.03 & 8280$\pm$870 \\ 
(6)&$[$Ar~{\sc iv}$]$($\lambda$\,4711)/($\lambda$\,4740) & 0.68$\pm$0.01 & 11\,320$\pm$540 \\
(7)&$[$Ne~{\sc iii}$]$($\lambda$\,15.6\,$\mu$m)/($\lambda$\,36.0\,$\mu$m) & 15.43$\pm$1.28 & 23\,250$\pm$8230 \\
   &$[$N~{\sc i}$]$($\lambda$\,5198)/($\lambda$\,5200) & 1.42$\pm$0.12 & 1010$\pm$260 \\ 
  &Paschen decrement                              &   &25\,700$\pm$6300                 \\   
\hline
ID & $T_{\epsilon}$ diagnostic&Value&Result\\
   &                          &     &(K)\\
 \hline
(8)&$[$S~{\sc iii}$]$($\lambda$\,9069)/($\lambda$\,18.7\,$\mu$m+33.5\,$\mu$m) & 0.56$\pm$0.05 & 11\,270$\pm$1060 \\
(9)&$[$Ne~{\sc iii}$]$($\lambda$\,3869+$\lambda$\,3967)/($\lambda$\,15.6\,$\mu$m) &1.87$\pm$0.12 &12\,800$\pm$300\\   
(10)&$[$O~{\sc iii}$]$($\lambda$\,4959+$\lambda$\,5007)/($\lambda$\,4363) & 106.4$\pm$1.8 & 12\,380$\pm$80 \\ 
(11)&$[$N~{\sc ii}$]$($\lambda$\,6548+$\lambda$\,6583)/($\lambda$\,5755) & 54.4$\pm$2.0 & 12\,090$\pm$230 \\ 
(12)&$[$Ar~{\sc iii}$]$($\lambda$\,7135+$\lambda$\,7751)/($\lambda$\,5191) & 107.8$\pm$5.2 & 12\,000$\pm$250 \\
(13)&$[$Cl~{\sc iv}$]$($\lambda$\,8046)/($\lambda$\,5323) & 26.8$\pm$5.7 & 11\,810$\pm$1020 \\
(14)&$[$S~{\sc ii}$]$($\lambda$\,6717+$\lambda$\,6731)/($\lambda$\,4069) & 3.81$\pm$0.16 & 12\,760$\pm$810 \\
(15)&$[$S~{\sc iii}$]$($\lambda$\,9069)/($\lambda$\,6312) & 5.62$\pm$0.48 & 12\,920$\pm$620 \\ 
    &He~{\sc i}($\lambda$\,7281)/($\lambda$\,5876) & 0.066$\pm$0.003 & 13\,600$\pm$1770 \\ 
    &He~{\sc i}($\lambda$\,7281)/($\lambda$\,6678) & 0.145$\pm$0.003 & 11\,700$\pm$550 \\ 
    &(Paschen Jump)/(P11)&0.018$\pm$0.002&11\,690$\pm$2090\\
 
 \hline
 \end{tabularx}
\raggedright Note -- Corrected recombination contribution for $[$O\,{\sc
 ii}$]$\,$\lambda\lambda$\,7320/30\,{\AA} and $[$O\,{\sc
  iii}$]$\,$\lambda$\,4363\,{\AA} lines.\\
 \end{table}

 In the following line-diagnostics and subsequent ionic element calculations,
 the adopted transition probabilities, effective collision strengths, and
recombination coefficients are the same as those listed in Tables 7 and 11 of
 \citet{2010ApJ...723..658O}.

 We calculated the electron densities {\Ne} and temperatures {\te}
 using the diagnostic line ratios of RLs and CELs. The results are
 summarised in Table~\ref{diagno_table}; the second, third, and last
 columns give the diagnostic lines, their
 line ratios, and the resulting {\Ne} and {\te}, respectively.
 The numbers in the first column
 indicate the ID number of each curve in the {\Ne}-{\te} diagram,
presented in Figure~\ref{diagno_figure}. The thick lines indicate the diagnostic
lines for {\te}, whereas the dashed lines 
correspond to {\Ne} diagnostics. Below, we explain in detail the {\Ne} and {\te} calculations using RLs and CELs.

\subsubsection{RL diagnostics}
The intensity ratio of a high-order Paschen line P$n$ (where $n$ is
the principal quantum number of the upper level) to a lower-order Paschen 
line is sensitive to the {\Ne}, as demonstrated in
\citet{Fang:2011aa}. The Paschen decrement can be an {\Ne} indicator 
for high-density regions. As such, we ran small grid models to estimate {\Ne}.

We calculated the electron temperatures derived from {\hei} lines {\te}({\hei}) using the two {\hei} line ratios
and the emissivities of these {\hei} lines from \citet{1999ApJ...514..307B} 
for the case of {\Ne}=10$^{4}$ cm$^{-3}$. For He$^{+,2+}$ abundance
calculations, we adopted the {\te}({\hei})=12\,650$\pm$1320\,K, which
is the average between the values derived from {\hei} 
$I$($\lambda$\,7281\,{\AA})/$I$($\lambda$\,5876\,{\AA}) and 
{\hei} $I$($\lambda$\,7281\,{\AA})/$I$($\lambda$\,6678\,{\AA}).

We calculated the electron temperature derived from the Paschen jump
discontinuity {\te}(PJ) by applying Equation (7) of \citet{Fang:2011aa}:

\begin{equation}
\label{pj}
T_{\epsilon}({\rm PJ}) =
8.72\times\left(1+0.52\frac{\rm He^{+}}{\rm H^{+}}+4.40\frac{\rm
    He^{2+}}{\rm H^{+}}\right)\cdot\left(
\frac{I_{\lambda}({\rm PJ})}{I({\rm P11})}\right)^{-1.77},
\end{equation}

\noindent 
where He$^{+,2+}$/H$^{+}$ is the ionic He$^{+}$ abundance of
9.63(--2)\footnote{$X(-Y)$ means $X\times10^{-Y}$ hereafter} and He$^{2+}$
abundance of 1.08(--2), respectively, $I_{\lambda}$(PJ) is the strength 
of the Paschen jump -- i.e. the difference in de-reddened flux density
between 8169\,{\AA} and 8194\,{\AA} in erg\,s$^{-1}$\,cm$^{-2}$\,{\AA}$^{-1}$, and 
$I$(P11) is the flux of the H\,{\sc i}\,$\lambda$\,8862\,{\AA} in
erg\,s$^{-1}$\,cm$^{-2}$. Because the H\,{\sc i}\,$\lambda$\,8862\,{\AA}
line is in an order gap, we estimated $I$(P11) using the theoretical ratio of $I$(P11)/$I$(P12)=1.3 in the Case B {\te}=10$^{4}$\,K and {\Ne}=10$^{4}$\,cm$^{-3}$ 
\citep{1995MNRAS.272...41S}. We adopted the {\te}(PJ) for the RL 
C$^{2+,3+,4+}$ and O$^{2+}$ abundance calculations.

\subsubsection{CEL diagnostics}

We derived the CEL {\te} and {\Ne} by solving for level 
populations using a multi-level atomic model ($\geq$5),
resulting in eight calculated {\Ne},
including {\Ne}({\Ni}), as well as eight {\te}.

For the {\oii}\,$\lambda\lambda$\,7320/30\,{\AA} and 
{\oiii}\,$\lambda$\,4363\,{\AA} lines, we
subtracted the recombination contamination from the O$^{2+}$ and 
O$^{3+}$ lines using Equations (\ref{ro2}) and 
(\ref{ro3}), originally 
given by \citet{2000MNRAS.312..585L},

\begin{equation}
\label{ro2}
\frac{I_{R}(\rm [O\,{\sc II}]\,\lambda\lambda\,7320/30)}{I(\rm H\beta)} =
9.36\left(\frac{T_{\epsilon}}{10^4}\right)^{0.44}\times\frac{\rm
O^{2+}}{\rm H^{+}},
\end{equation}
\begin{equation}
\label{ro3}
\frac{I_{R}(\rm [O\,{\sc III}]\,\lambda\,4363)}{I(\rm H\beta)} =
12.4\left(\frac{T_{\epsilon}}{10^4}\right)^{0.59}\times\frac{\rm
O^{3+}}{\rm H^{+}}.
\end{equation}

\noindent
In the $I_{R}$({\oii}\,$\lambda\lambda$\,7320/30) calculation, 
we adopted the {\te}(PJ) and recombination O$^{2+}$ abundances
(3.12$\times$10$^{-4}$, see Section~\ref{S-RLabund}). For the 
$I_{R}$({\oiii}\,$\lambda$\,4363) calculation, we used 
the O$^{3+}$ abundance derived from the fine-structure {\oiv}\,$\lambda$\,25.89\,$\mu$m line
(6.93$\times$10$^{-6}$, See section~\ref{S-CELabund}) and {\te}({\neiii}). 
$I_{R}$({\oii}\,$\lambda\lambda$\,7320/30) and 
$I_{R}$(({\oiii}\,$\lambda$\,4363) were 0.86 and 0.01, respectively.

\citet{Walsh:1997aa} reported a single {\te} of 
12\,400$\pm$400\,K and {\Ne} of 6000$\pm$1500\,cm$^{-3}$. 
In the present study, we first calculated all {\Ne}(CEL)s under their derived value
{\te}=12\,400 K, except for {\Ne}({\Ni}), where we assumed a {\te} of
10\,000\,K because the effective collisional strengths of the Mg$^{0}$
is available at {\te}=10\,000\,K (See Section~\ref{S-CELabund} for
details). 
The {\te}=10\,000\,K would be a good choice for estimates of neutral
atoms such as N$^{0}$ and O$^{0}$ even in high-excitation PNe such as
BoBn1 \citep{2010ApJ...723..658O}. We derived {\te}({\neiii}) using 
{\Ne}({\ariv}) because the ionisation potentials (IPs) of both ions are
similar. We derived {\te}({\oiii}) and {\te}({\cliv}) 
under the average {\Ne} among {\Ne}({\ariv},{\cliii},{\siii}); 
the IPs of the {\oiii} and {\cliv} are approximately median between
{\Ne}({\ciii}\&{\siii},{\ariv}) or slightly larger than these ions.  
We calculated {\te}({\ariii}) and two {\te}({\siii}) under the average {\Ne} 
between {\Ne}({\cliii},{\siii}). {\te}({\nii}) and
{\te}({\sii}) were calculated under the average {\Ne} among the two
{\Ne}({\oii}) and {\Ne}({\sii}).

\subsection{Ionic abundances}
\subsubsection{CEL ionic abundances \label{S-CELabund}}

\begin{figure}
\centering
\includegraphics[width=\columnwidth,bb=63 186 512 411,clip]{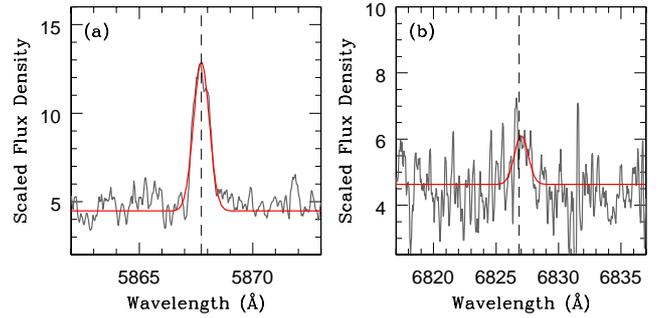}
\caption{The line profiles of the $[$Kr\,{\sc iv}$]$\,$\lambda$\,5867.74
\,{\AA} ({\it left panel}) and the $[$Kr\,{\sc
 iii}$]$\,$\lambda$\,6826.85\,{\AA} ({\it right panel}). The grey lines
 are the observed spectra at the rest wavelength, which is shifted by
 the systemic heliocentric radial velocity (+133.12\,{\kms}). The sky
 emission lines were removed out as much as possible. The red
 lines are the results of Gaussian fitting. The laboratory wavelengths
 of these lines are indicated by the dashed lines. \label{kr_line}
} 
\end{figure}

We obtained 21 ionic abundances using CELs, as listed in
Table~\ref{celabund} of Appendix~B. We carefully determined 
the adopted {\te} and {\Ne} pairs for the calculations
of each ion using the {\Ne}-{\te} diagram. The
adopted {\te} and {\Ne} values for each ion are listed in the fourth and
fifth columns of Table~\ref{celabund}, respectively. For
the Mg$^{0}$ abundance calculation, we adopted the effective collision
strengths installed
in the three-dimensional Monte Carlo photoionisation code {\sc
Moccasin} \citep{Ercolano:2003aa} and the transition probabilities
of \citet{Mendoza:1983aa}. Because the effective collision strengths of Mg$^{0}$
are only available at the case of {\te}=10$^{4}$\,K, we adopted this
temperature. {\te}=10$^{4}$\,K and $n_{\epsilon}$({\Ni}) were used for
the N$^{0}$ and Mg$^{0}$ calculations. The ionic
abundances were calculated by solving the statistical equilibrium
equations for more than five levels, given the adopted {\te} and {\Ne}, 
except for Ne$^{+}$, where we calculated its abundance using the 
two-energy level model. The last column contains the resulting
ionic abundances relative to the H$^{+}$, X$^{\rm m+}$/H$^{+}$, and
their 1-$\sigma$ errors, which include errors from line intensities, {\te}, and 
{\Ne}, but does not consider the accuracy of the atomic data.
In the O$^{+}$ and O$^{2+}$ abundance calculations, we corrected 
the line intensities of the {\oii}\,$\lambda\lambda$\,7320/30\,{\AA}
and {\oiii}\,$\lambda$\,4363\,{\AA}, respectively. 
In the last line of each ion's line series, the adopted ionic 
abundance and its error are indicated in boldface.
These values are estimated from the weighted
mean of the line intensity.

As presented in Table~\ref{celabund}, the
calculated abundances of each ion using different transition lines
(i.e., fine-structure, nebular, auroral, and trans-auroral lines) are
well consistent (within the error), indicating proper selection of the adopted 
{\te} and {\Ne} pairs for each ion and accurate measurement
of the flux.

We report the first detection of the [Kr\,{\sc
iv}]\,$\lambda$\,5867.74\,{\AA} and [Kr\,{\sc
iii}]\,$\lambda$\,6826.85\,{\AA}
in Wray16-423. The line profiles are presented in
Figure~\ref{kr_line}. The sky emission lines between $\sim$6820 and 6830\,{\AA} 
were subtracted as much as possible. Because there are
no candidate lines within 0.18-0.20\,{\AA} from the measured 
central wavelengths of the above lines (5867.74\,{\AA} and
6827.02\,{\AA} at rest), we concluded that the lines corresponded to 
krypton. Krypton lines have been observed in the Sgr dSph PNe, specifically, 
BoBn1 \citep{2010ApJ...723..658O} and Hen2-436
\citep{Otsuka:2011aa}. The Kr$^{2+}$ abundance in
Wray16-423 is comparable to that in Hen2-436, which shows the same
metallicity.

\subsubsection{RL ionic abundances \label{S-RLabund}}

We list the RL ionic abundances in Table~\ref{rlabund} of Appendix~B. To the best of our 
knowledge, RL C$^{3+,4+}$ and  O$^{2+}$ for Wray16-423 are reported 
for the first time. Because we detected C\,{\sc ii,iii,iv} and O\,{\sc ii} lines, 
we were able to calculate the elemental C/O ratio using the same type of 
emission lines.

In the abundance calculations, we adopted the Case B assumption for the lines 
from levels that have the same spin as the ground state, but 
the Case A assumption for lines of other multiplicities.  
In the last line of each ion's line series, we present the adopted final
ionic abundance and its 1-$\sigma$ error estimated from the line intensity-weighted
mean, indicated by boldface. Because the RLs
are {\Ne}-insensitive under $\lesssim$10$^{8}$\,cm$^{-3}$, 
we evaluated the recombination coefficients (provided as 
a polynomial function of {\te}) at 
a constant {\Ne}=10$^{4}$\,cm$^{-3}$ for all lines.

In the He$^{+}$ abundance calculations, using 
{\hei}\,$\lambda$\,4471\,{\AA} and {\hei}\,$\lambda$\,5876\,{\AA} lines, we
assumed Case A conditions and adopted the recombination coefficients of
\citet{1991A&A...251..680P}. We subtracted the contribution to
these lines by collisional excitation from the He$^{0}$~$2s$$^{3}$$S$
level, using the formulae given by \citet{1995ApJ...442..714K}.

We detected the different multiplet O\,{\sc ii} lines. 
Because the electron density in the 
O$^{2+}$ emitting region is $\gtrsim$10$^{4}$\,cm$^{-3}$ from our plasma diagnostics, 
the upper levels of the transitions in the V1 O\,{\sc ii} line 
would be in local thermal equilibrium (LTE). Therefore, we did not
correct the V1 O\,{\sc ii} line intensities.
To calculate the O$^{2+}$ abundances, we excluded the 
V1 O\,{\sc ii}\,$\lambda$\,4641.81/4651.33/4673.73\,{\AA} 
and V2 O\,{\sc ii}\,$\lambda$\,4325.76\,{\AA} lines. 
The first two V1 lines are likely contaminated with
N\,{\sc iii}\,$\lambda$\,4641.85\,{\AA} and C\,{\sc
iii}\,$\lambda$\,4651.47\,{\AA}, respectively. The O$^{2+}$ abundance
derived from the O\,{\sc ii}\,$\lambda$\,4673.73\,{\AA} line is larger than that 
from other V1 lines. The last 
V2 line is likely contaminated with C\,{\sc ii}\,$\lambda$\,4325.83\,{\AA}.

The RL O$^{2+}$ abundance is larger by a factor of 1.64$\pm$0.43 larger than 
the O$^{2+}$ CEL. According to \citet{Liu:2006aa}, the O$^{2+}$ discrepancy factor in Wray16-423
would be a lower limit. As proposed for the halo H4-1 \citep[O$^{2+}$
discrepancy factor=1.75$\pm$0.36;][]{Otsuka:2013aa}, we speculate that
the O$^{2+}$ discrepancy in Wray16-423 could be explained because the
estimated CEL O$^{2+}$ abundance could be increased by $\sim$0.2~dex
or less if we include temperature fluctuations, which is originally
proposed by \citet{1967ApJ...150..825P}. Since the O$^{2+}$
discrepancy factor is already low, we do not discuss further in the paper.

\subsection{Elemental abundances \label{S-element}}

   \begin{table*}
   \centering
   \caption{Elemental abundances from CELs and RLs.\label{abund}}
\begin{tabularx}{\textwidth}{@{}l@{\hspace{14pt}}c@{\hspace{14pt}}c@{\hspace{14pt}}r@{\hspace{14pt}}c@{\hspace{14pt}}r@{\hspace{14pt}}c@{\hspace{14pt}}r@{\hspace{14pt}}r@{\hspace{14pt}}r@{}}

 \hline
 (1)&
 (2)&
 (3)&
 (4)&
 (5)&
 (6)&
 (7)&
 (8)&
 (9)&
 (10)\\
 X &
Types of  & 
X/H & 
$\log_{10}$(X/H)& 
[X/H] & 
$\log_{10}$(X$_{\odot}$/H)$^{\rm a}$ & 
ICF(X)  &
$\log_{10}$(X/H)&
$\log_{10}$(X/H)&
$\log_{10}$(X/H)\\ 
  &
Emissions 
     &&+12&&+12&&+12~(Ref.1)&+12~(Ref.2)&+12~(Ref.3)\\
 \hline
He &RL &1.06(--1)$\pm$9.56(--3) &11.02$\pm$0.04 &+0.09$\pm$0.04 &10.93$\pm$0.01 &1.00&11.03$\pm$0.01&11.03$\pm$0.02&11.09$\pm$0.05\\
C &RL &8.29(--4)$\pm$1.57(--4) &8.92$\pm$0.08 &+0.53$\pm$0.09
		 &8.39$\pm$0.04 &1.00&8.86$\pm$0.06&
				 &9.08$\pm$0.35\\
N &CEL &4.94(--5)$\pm$5.04(--6) &7.69$\pm$0.04 &--0.14$\pm$0.07 &7.83$\pm$0.05&24.81$\pm$2.12&7.68$\pm$0.05&7.62$\pm$0.10&7.38$\pm$0.22\\
O &CEL &2.06(--4)$\pm$4.10(--6) &8.31$\pm$0.01 &--0.38$\pm$0.05 &8.69$\pm$0.05 &1.00&8.33$\pm$0.02&8.31$\pm$0.07&8.35$\pm$0.03\\
 &RL &3.37(--4)$\pm$8.90(--5) &8.53$\pm$0.11 &--0.16$\pm$0.13 &8.69$\pm$0.05 &1.08$\pm$0.03&$\cdots$&$\cdots$&9.56$\pm$0.11\\
Ne &CEL &4.09(--5)$\pm$2.93(--6) &7.61$\pm$0.03 &--0.26$\pm$0.10 &7.87$\pm$0.10&1.00&7.55$\pm$0.03&7.50$\pm$0.08&7.62$\pm$0.05\\
S &CEL &2.51(--6)$\pm$1.58(--7) &6.40$\pm$0.03 &--0.79$\pm$0.05 &7.19$\pm$0.04 &1.00&6.67$\pm$0.04&6.48$\pm$0.08&6.83$\pm$0.08\\
Cl &CEL &5.42(--8)$\pm$5.66(--9) &4.74$\pm$0.05 &--0.77$\pm$0.30 &5.50$\pm$0.30 &1.00&4.89$\pm$0.18&$\cdots$&4.86$\pm$0.09\\
Ar &CEL &1.00(--6)$\pm$3.85(--8) &6.00$\pm$0.02 &--0.55$\pm$0.08 &6.55$\pm$0.08 &1.00&5.95$\pm$0.07&5.88$\pm$0.08&5.95$\pm$0.32\\
K &CEL &2.02(--8)$\pm$2.30(--9) &4.31$\pm$0.05 &--0.80$\pm$0.07 &5.11$\pm$0.05 &3.06$\pm$0.23&4.65$\pm$0.22&$\cdots$&$\cdots$\\
Fe &CEL &3.62(--7)$\pm$5.62(--8) &5.56$\pm$0.07 &--1.91$\pm$0.07 &7.47$\pm$0.03 &23.81$\pm$2.12&$\cdots$&$\cdots$&5.73$\pm$0.20\\
 Kr &CEL &~~3.52(--9)$\pm$4.88(--10) &3.55$\pm$0.06 &+0.27$\pm$0.10
		 &3.28$\pm$0.08&1.00&$\cdots$&$\cdots$&3.62$\pm$0.13\\
 \hline
\end{tabularx}
   \raggedright
References (1) \citet{2006MNRAS.369..875Z}, \citet{Dudziak:2000aa}, (2)
    (2) \citet{Walsh:1997aa} for Wray16-423; (3) \citet{Otsuka:2011aa} for Hen2-436.\\
$^{a}$\citet{2009ARA&A..47..481A} for N and Cl, and \citet{2003ApJ...591.1220L} for the other elements.\\
   \end{table*}

The elemental abundances of the nebula are listed in Table~\ref{abund}. 
The types of emission lines used for the abundance estimations are 
specified in the second column.
The fifth column lists the number densities relative to the solar
value, where [X/H] corresponds to $\log_{10}$(X/H)-$\log_{10}$(X/H)$_{\odot}$.

To estimate elemental abundances using only the observed
ionic abundances, we introduced an ionisation correction factor, ICF(X),
based on the ionisation potential. 
The adopted ICF(X) for each element is listed in Table~\ref{icf} of Appendix~B, and the
value is listed in the seventh column of Table~\ref{abund}.
We calculated the adopted ICF(Kr) using Equation (5) of
\cite{Sterling:2007aa}, based on photoionisation models.

In the eighth column of Table~\ref{abund}, we list the elemental
abundances of Wray16-423 compiled by \citet{2006MNRAS.369..875Z}, 
originally calculated using the photoionisation model of
\citet{Dudziak:2000aa} based on the
analysis of optical spectra by \citet{Walsh:1997aa}. The ninth
column is the result of \citet{Walsh:1997aa}. In general, our
calculated abundances, except for S and K, are in excellent agreement with prior
results.  \citet{Walsh:1997aa} reported a C$^{2+}$/O$^{2+}$ ratio of
3.4$\pm$1.2, derived from C\,{\sc ii} and {\oiii} lines
(C$^{2+}$(RL)/O$^{2+}$(CEL)=2.48$\pm$0.74 in the present work). Assuming that
C(RL)/O(CEL) $\simeq$ C$^{2+}$/O$^{2+}$, their C abundance would be
5.99(--4)$\pm$2.17(--4), which corresponds to 8.78$\pm$0.16 dex. 
The differences of the S and K abundances between
\citet{2006MNRAS.369..875Z} and ours might be caused by adoption
of different {\te} for these ionic abundances and overestimate of the
S$^{3+}$ abundance. The {\siv} lines are not seen in optical but in
mid-infrared wavelength. Their photoionisation without 
dust grains would affect the nebula's density and temperature structure
and the fraction of each element's ion.

The RL C/O ratio of 2.46$\pm$0.80 indicates that Wray16-423 
is presumably a C-rich PN. Elemental RL and CEL O abundances seem to follow the trend; as
shown in \citet{Liu:2006aa}, \citet{Wang:2007aa}, and \citet{Delgado-Inglada:2014aa}, the RL 
C/O ratio seems to be consistent with the CEL C/O ratio within
errors (.i.e., the RL C/O $\simeq$ the CEL C/O), except for several
objects. If this is also the case for Wray16-423, the CEL C 
abundance would be 8.70$\pm$0.14, where the upper value is
consistent with the lower value of the RL C abundance.

The noble gas Ar is not easily tied up in dust grains, and is also not
synthesised in low-intermediate mass stars. Therefore, Ar can be 
used as a metallicity indicator of the progenitor star. The [Ar/H] abundance is
consistent with that in Hen2-436
\citep[--0.60$\pm$0.33,][]{Otsuka:2011aa}. Like for Hen2-436, 
[Fe/H] is significantly lower than [Ar/H]. Because [Fe\,{\sc iii}]
intensities are comparable to BoBn1 \citep{Otsuka:2011aa} and their
Fe$^{2+}$ abundances in Wray16-423 and BoBn1 are consistent,
the small Fe abundance in Wray16-423 cannot be attributed to inaccuracies
in Fe$^{2+}$ atomic data or the selection of {\te}
for their ionic abundances. Thus, a large fraction of Fe atoms would
exist as dust grains.

An important contribution of our work is the detection of the light 
$s$-process element Kr and calculation of its abundance. This required 
verification of whether or not the Kr abundance value for Wray16-423 is 
reasonable. \citet{Sterling:2008aa} examined the relationship between the [Kr/O] and 
C/O ratio, based on their $Ks$-band spectroscopic survey for Galactic PNe,
from which the following relationship was obtained: 

\begin{equation}
{\rm [Kr/O] = (0.38\pm0.10) + (0.79\pm0.29)\cdot\log_{10}(C/O)}.
\end{equation}

By substituting the observed RL $\log_{10}$(C/O) ratio, we obtained a predicted 
[Kr/O] value
of +0.69$\pm$0.42, which is in good agreement with the 
observed [Kr/O] of +0.65$\pm$0.11 from Kr and O CELs. The authors also showed 
the light $s$-enhancement ([ls/Fe], ls: Sr, Y, and Zr) versus the C/O ratio among AGB stars
and CH sub-giants:  

\begin{equation}
{\rm [ls/Fe] = (0.89\pm0.06) + (1.47\pm0.18)\cdot\log_{10}(C/O)}.
\end{equation}

As described above, the calculated Fe abundance in Wray16-423 may
not accurately reflect the correct metallicity. Given that [Fe/H] is
equal to [Ar/H], the [ls/Ar]=[Kr/Ar] is moderately enhanced, 
+0.82$\pm$0.13. This value is slightly smaller than the predicted 
[ls/Fe] of 1.46$\pm$0.44 from Equation (6), although it depends on
the adopted metallicity. There are a few results on the
$s$-process in carbon stars in the Sgr dSph. For example, \cite{de-Laverny:2006aa} 
reported $s$-process enhancements in two C-rich stars
(C/O=1.05-1.18). The respective [ls/Fe]s in IGI95-C1 
([M/H]=--0.8, [ls/M]=+0.6) and IGI95-C3 ([M/H]=--0.5,
[ls/M]=+1.0) are comparable to that in Wray16-423. In comparison with 
previous studies of PNe and AGB stars, we concluded that the Kr abundance 
value obtained for Wray16-423 is reasonable.

In the last column of Table~\ref{abund}, we list the elemental
abundance of Hen2-436 by \citet{Otsuka:2011aa}; the abundances obtained
in the present study, with the exception of N, are close to those for Hen2-436.
The O and Ar abundances indicate that both PN progenitors 
would form under a similar evolutionary stage of the Sgr dSph.
\citet{Dudziak:2000aa} pointed out that the nearly identical abundance patterns 
for these two Sgr dSph PNe provide evidence that the progenitors formed in a
single starburst event \citep[$\sim$5 Gyrs ago,][]{2006MNRAS.369..875Z} within a well mixed
ISM. The age-metallicity relationship for the Sgr dSph globular
cluster candidates by \citet{Law:2010aa} indicates that the
[Fe/H]$\sim$--0.6 corresponds to $\sim$8 Gyrs ago. 
\citet{Layden:2000aa} reported that the Sgr dSph field is
composed of several different populations; the [Fe/H]=--0.7$\pm$0.2
population is 5$\pm$1 Gyrs old and the [Fe/H]=--0.4$\pm$0.3 population is
0.5-3 Gyrs old.  Although the exact [Fe/H] abundances in Wray16-423 and
Hen2-436 are unknown,
the progenitor stars of Wray16-423 and Hen2-436 most likely
formed 0.5-6 Gyrs ago.

\subsection{[WC] type central star}

  \begin{figure}
   \includegraphics[width=\columnwidth,bb=50 185 372 449,clip]{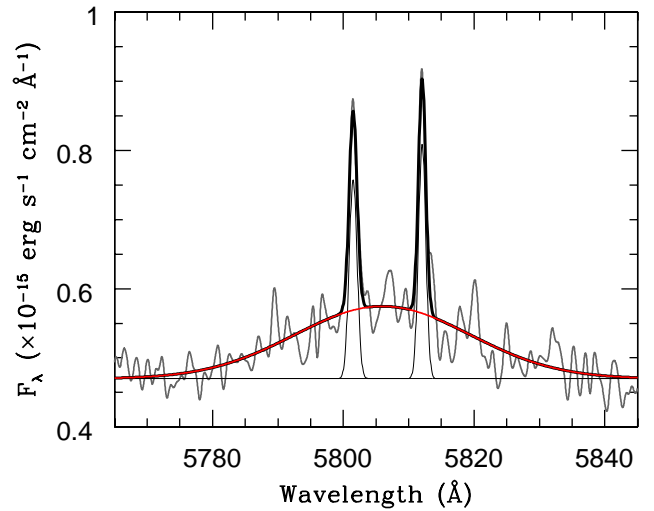}
   \caption{The observed profile of the C\,{\sc
   iv}\,$\lambda\lambda$\,5801/12\,{\AA} band indicated by the grey line.
   The observed wavelength is corrected by
   the heliocentric radial velocity of +133.12\,{\kms}. An interesting
   feature (perhaps originating from a stellar wind), is indicated
   by the red line. \label{c4}
   }
  \end{figure}

   In Figure~\ref{c4}, we present the profile of the
   C\,{\sc iv}\,$\lambda\lambda$\,5801/12\,{\AA} band composed
   of narrow nebular C\,{\sc iv}\,$\lambda\lambda$\,5801/12\,{\AA} lines
   and a weak broad wing component (perhaps originating from a stellar wind).
   We fitted the band profile by three Gaussian components: 
  in Figure~\ref{c4}, the thin black lines are the components
  of the nebular C\,{\sc iv}\,$\lambda\lambda$\,5801/12\,{\AA}, 
the red line indicates our interesting broad component, and the thick
   black line
  is the sum of these three. The FWHMs of the C\,{\sc
  iv}\,$\lambda\lambda$\,5801/5812\,{\AA} lines are 0.75$\pm$0.04 and
  0.83$\pm$0.05\,{\AA} and that of the broad component (centered at
   5805.96\,{\AA} in the rest frame) is
  32.12$\pm$2.15\,{\AA}, where its FWHM expansion velocity is
   1660$\pm$110 {\kms}.
   The C\,{\sc iv}\,$\lambda\lambda$\,5801/12\,{\AA} band profile in Wray16-423 
is similar to the Sgr dSph PN StWr2-21; \citet{Kniazev:2008aa}
   de-convolved the C\,{\sc iv}\,$\lambda\lambda$\,5801/12\,{\AA} into
   three components, which included a broad component having a FWHM of 42.3$\pm$3\,{\AA}. 
Although the nebular C\,{\sc iv}\,$\lambda\lambda$\,5801/5812\,{\AA} lines
   seem to be completely embedded in the broad component in Hen2-436,
   the FWHM of the 
C\,{\sc iv}\,$\lambda\lambda$\,5801/12\,{\AA} band is 32.8$\pm$0.2\,{\AA},
   measured from the same ESO VLT/FEROS2 spectra used in
   \citet{Otsuka:2011aa}, and consistent with the findings of
   \citet[31.1\,{\AA}]{Walsh:1997aa}. We did
   not detect any other broad components, such as O\,{\sc
   vi}\,$\lambda\lambda$\,3811/34\,{\AA}.

   A consensus has not been reached regarding the classification of the CSPN of Wray16-423.
   \citet{Walsh:1997aa} classified it as a [WC8] or a weak emission line star (WELS),  on the basis of the strengths
  of the C\,{\sc iii}\,$\lambda$\,4650\,{\AA} (FWHM=9.9\,{\AA}) and C\,{\sc
  iv}\,$\lambda\lambda$5801/12\,{\AA} (FWHM=14.7\,{\AA}) bands, using low-dispersion
   spectra ($R$ could be $\sim$1200 around 5800\,{\AA}); the broad 
component under the nebular C\,{\sc
  iv}\,$\lambda\lambda$\,5801/5812\,{\AA} lines was not detected. The
   broad component of the C\,{\sc iii} and C\,{\sc iv} complex centered 
around 4650\,{\AA}, originating from the stellar winds, is not clearly seen in our HDS and FEROS
   spectra. There is no broad component in the C\,{\sc
   ii}\,$\lambda$4267\,{\AA}, too.

Taking into account that their estimated CSPN temperature
   (85\,000 K) is hot for a [WC8] PNe, \citet{Walsh:1997aa} concluded
   that Wray16-423 could fall
   into WELS. \citet{Acker:2002aa} analysed 42 emission line nuclei
   of PNe, and reported that the [WC4] CSPNe show the highest temperatures among
   the [WC] class (54\,950-91\,200 K). The determination of CSPN temperatures
   can be difficult, as pointed out by \citet{Acker:2002aa}, which may
explain why the CSPN temperatures of the [WC] class PNe
   have a relatively wide range. Even after the CSPN temperature was
   corrected up to 107\,000 K by photoionisation models based on the low-dispersion
   spectra of \citet{Walsh:1997aa}, \citet{Gesicki:2003aa}
   and \citet{2006MNRAS.369..875Z} supported the conclusion by
   \citet{Walsh:1997aa}.

   According to the classification of \citet{Acker:2002aa}, the broad
   FWHM of the stellar C\,{\sc iv} component suggests that the CSPN
   of Wray16-423 may be a [WC4]-type central star.

\subsection{PAHs and dust features}

\begin{figure}
\centering
\includegraphics[width=\columnwidth]{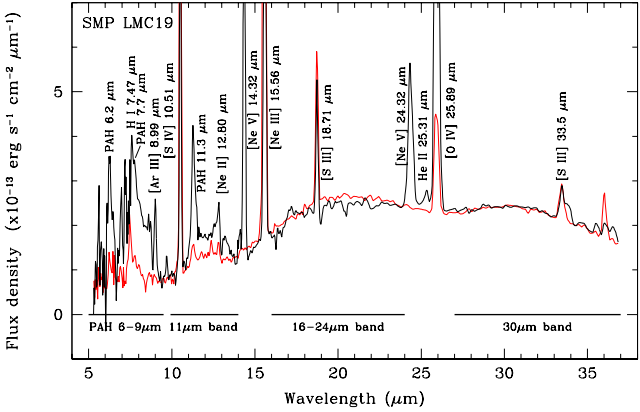}
\caption{\emph{Spitzer}/IRS spectrum of SMP LMC19 in the LMC. 
The intensity-scaled spectrum of Wray16-423 is indicated by the red
 line. The scaling factor is 2.98, for adjustment to the
 16-36\,$\mu$m flux density of LMC19. 
}
\label{spt_spec2}
\end{figure}

\begin{figure}
\centering
\includegraphics[width=\columnwidth]{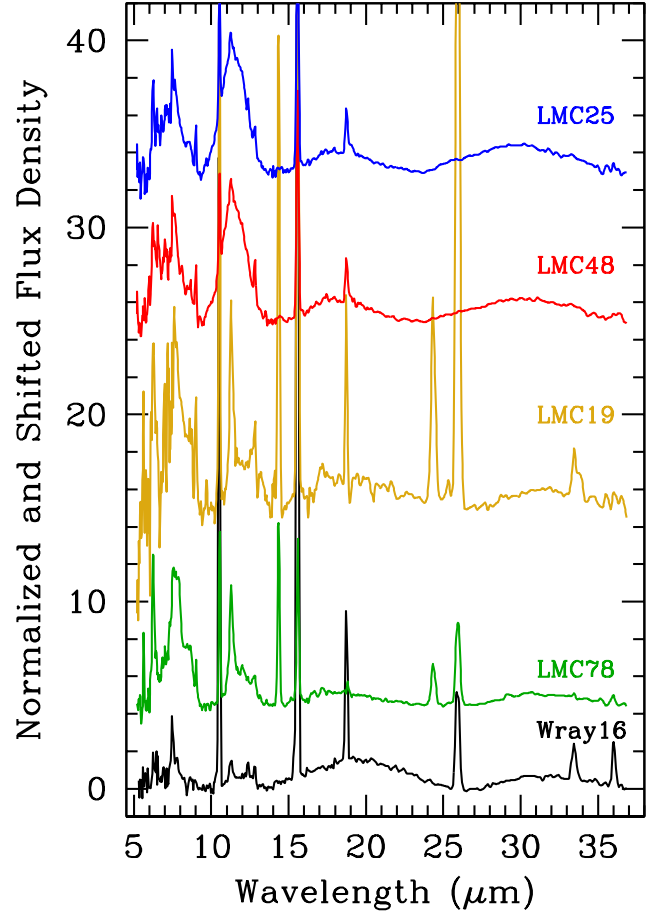}
\caption{\emph{Spitzer}/IRS spectra of SMP LMC25, 48, 19, and 78 and
 Wray16-423 (Wray16). The local continuum was subtracted by fifth-order
 spline-function fitting. Flux densities are normalised and shifted to emphasise the 16-24 and 30\,$\mu$m features.
}
\label{spt_spec1}
\end{figure}

Wray16-423 shows PAH bands and dust features, specifically, 
6-9\,$\mu$m and 10-14\,$\mu$m PAH bands and broad 11\,$\mu$m,
16-24\,$\mu$m, and 30\,$\mu$m features. 
These PAH bands and these dust features have been
found in C-rich
MC PNe
\citep{Stanghellini:2007aa,2009ApJ...699.1541B,Garcia-Hernandez:2012aa,2014MNRAS.439.1472M,Sloan:2014aa}. \citet{2009ApJ...699.1541B} 
reported broad 11\,$\mu$m, 16-24\,$\mu$m\footnote{\citet{2009ApJ...699.1541B} call this Bump 16.},
and 30\,$\mu$m features in 8 C-rich MC PNe. Because the typical LMC metallicity is
close to that of Wray16-423, LMC PNe provide a good comparison in terms of dust
production in a metal-deficient environment. Therefore, 
we looked at the LMC PN sample to compare the mid-IR 
spectrum of Wray16-423 with those of 
selected LMC PNe.

Accordingly, we found that SMP LMC19, 25, 48, and 78 exhibit 
broad 11/16-24/30\,$\mu$m features, comparable to the dust feature
strengths measured in Wray16-423. Among them, the mid-IR spectrum 
of LMC19 is most similar to that of Wray16-423
(Fig.~\ref{spt_spec2}); the strengths of the broad 16-24\,$\mu$m and 30\,$\mu$m features 
and the global $\sim$5-37\,$\mu$m SED are similar 
as well.

For a more detailed look at these PAH and broad dust 
features, we subtracted the expected local 
dust continuum, determined by fifth-order spline-function fitting. The resultant
spectra are displayed in Figure~\ref{spt_spec1}. The flux densities are
scaled and shifted to emphasise broad 16-24\,$\mu$m and 30\,$\mu$m
features. 

In the following sections, we describe 
the 6-9\,$\mu$m and 10-14\,$\mu$m 
PAH bands, the broad 11\,$\mu$m, 16-24\,$\mu$m, 
and 30\,$\mu$m\,band features in Wray16-423, and LMC19, 
25, 48, and 78. Dust and gas mass estimates for 
these LMC PNe will be discussed in a future study.

\subsubsection{The 6-9\,$\mu$m PAH band \label{S:pah6-9}}

The classification of the 6-9\,$\mu$m PAH band is based on the positions of
6.2, 7.7, and 8.6\,$\mu$m features at the intensity peaks. According
to \citet{Peeters:2002aa} and \citet{2014MNRAS.439.1472M}, 
the 6-9\,$\mu$m PAH profiles have peak positions near 
6.22-6.3\,$\mu$m, 7.6-8.22\,$\mu$m, and $\gtrsim$8.6\,
$\mu$m. Our classifications
are summarised in Table~\ref{band_measure} of Appendix~C. 
From the overview of \citet{Peeters:2002aa}, both the 6.2 and 
7.7\,$\mu$m PAHs in 
Wray16-423 are Class~$\mathcal{B}$, with corresponding peaks at
6.24$\pm$0.01\,$\mu$m, and 7.78$\pm$0.04\,$\mu$m, respectively. 
The 6.2 and 7.7\,$\mu$m PAHs in our comparison LMC PNe fall into Classes 
$\mathcal{B}$ and $\mathcal{A}$, respectively. The 8.6\,$\mu$m PAH in
Wray16-423 could not be resolved, due to a low signal. The 8.6\,$\mu$m
PAHs in LMC PNe are classified as Class~$\mathcal{B}$.

\subsubsection{The 10-14\,$\mu$m PAH band and the broad 11\,$\mu$m feature}

\begin{figure}
\centering
\includegraphics[width=\columnwidth]{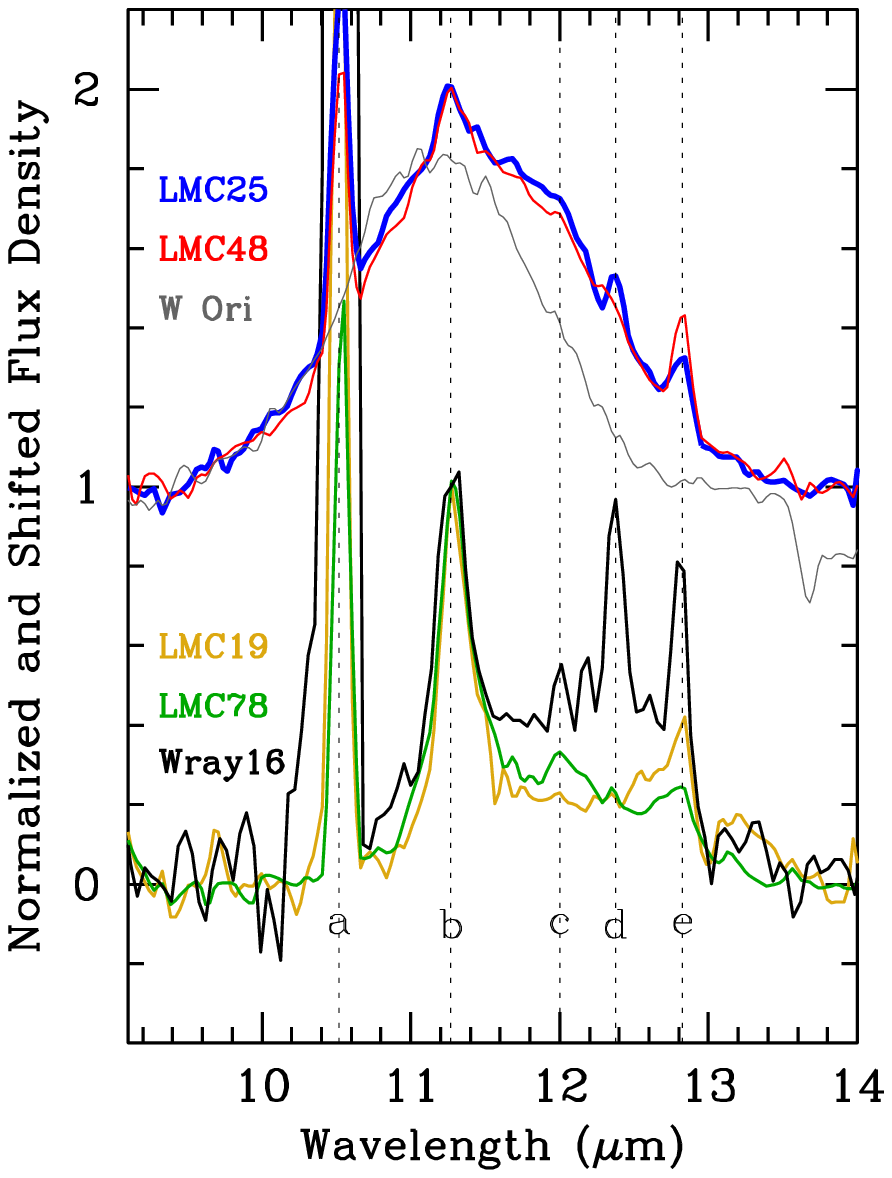}
\caption{
The 9.5-13.5\,$\mu$m emission complex
showing the 10-14\,$\mu$m PAH bands and the broad 11\,$\mu$m feature.
The vertical dashed lines mark the positions of (a)
{\siv}\,$\lambda$10.51\,$\mu$m, (b) 11.3\,$\mu$m PAH,
(c) 12\,$\mu$m PAH, (d) H\,{\sc i}\,$\lambda$\,12.38\,$\mu$m, and (e)
 the 12.7\,$\mu$m PAH + {\neii}\,$\lambda$\,12.80\,$\mu$m complex.
 The grey line shows an \emph{ISO}/SWS spectrum of the
 Galactic C-rich AGB star W Ori, with the 11\,$\mu$m SiC feature.
}
\label{fig-11um}
\end{figure}

Figure~\ref{fig-11um} shows the 10-14\,$\mu$m PAH bands complex 
and the broad 11\,$\mu$m feature. The flux densities are normalised with
respect to the 11.3-\,$\mu$m PAH emission peak. The positions of 
the PAHs and the atomic gas emission lines are shown as vertical dashed
lines with the lowercase letters a-e.

In the study on PAHs in LMC post-AGB 
stars, \citet{2014MNRAS.439.1472M} classified the 10-14\,$\mu$m PAH feature 
into four types: $\alpha$, $\beta$, $\gamma$, and $\delta$, depending on the peak wavelengths 
and the shapes of the sub-features. The Class~$\alpha$ (including 
the broad 11\,$\mu$m feature) shows a sharp feature
at 11.3\,$\mu$m, a well-isolated feature at 12.7\,$\mu$m, and a very weak
feature at 12\,$\mu$m. The 11.3\,$\mu$m PAH emission in Class~$\alpha$
is the strongest among the three features. The respective features would be 
from the solo (11.3\,$\mu$m PAH), duo (12\,$\mu$m PAH), and trio
(12.7\,$\mu$m PAH) out-of-plane C-H bending mode of PAH 
\citep{Sloan:2014aa}. We classify the 10-14\,$\mu$m features in Wray16-423, 
LMC19, and LMC78 as Class~$\alpha$. 
The flux density for these three PNe reaches its maximum 
in this range at the peak of the 11.3\,$\mu$m PAH emission 
(central wavelengths: 11.28, 11.30, and 11.29\,$\mu$m,
respectively); 12\,$\mu$m PAHs are also observed in all three.
The 12.7\,$\mu$m PAHs are observed in LMC19 and 78, 
while the presence of this PAH in Wray16-423 seems to be merged with
{\neii}\,$\lambda$\,12.80\,$\mu$m.

LMC25 and 48 are characterised as Class~$\delta$, by the broad 
feature extending from 10 to 14\,$\mu$m with the 11.3\,$\mu$m PAH on the 
top \citep{2014MNRAS.439.1472M}. Class~$\delta$ 
spectra are quite different from Class~$\alpha$ spectra, in terms of the underlying broad 11\,$\mu$m
feature
and its central position.  
We fitted Gaussians to the PAH and atomic gas emission lines, and
subtracted the fits from the spectra. We then fitted the broad 11\,$\mu$m
feature in the residual spectra was fitted using a single
Gaussian component to determine their central wavelengths $\lambda_{c}$
and FWHMs (Table~\ref{band_measure}). 
The average peak position and FWHM of the broad feature are 
11.92$\pm$0.05\,$\mu$m and 1.74$\pm$0.11\,$\mu$m, respectively, for 
Class~$\alpha$, while those in Class~$\delta$ are 11.46$\pm$0.01\,$\mu$m 
and 1.86$\pm$0.03\,$\mu$m, respectively.

\citet{2014MNRAS.439.1472M} argued that the broad 11\,$\mu$m 
feature characteristic to Class~$\delta$ should be ascribed to silicon carbide (SiC).  
They used fits for the 10-14\,$\mu$m spectra of three
post-AGB stars using the NASA/Ames PAH IR-spectral 
database \citep{Boersma:2014aa,Bauschlicher:2010aa}\footnote{http://www.astrochem.org/pahdb/}
and the \emph{ISO}/SWS spectrum of the C-rich AGB star W Ori as the 
SiC template (see also Fig.~\ref{fig-11um}). 
\citet{Sloan:2014aa} analyzed the broad 11\,$\mu$m features 
and associated PAHs in LMC and SMC post AGB stars. They call the band 
profiles similar to LMC25 and LMC48 ``Big-11''. They demonstrated that
this Big-11 is composed of SiC and PAH, and they concluded that the
Big-11 feature is due primarily to SiC (88\,$\%$ of the total flux).
\citet{Otsuka:2015aa} argued that the broad 11\,$\mu$m
feature in W Ori could be reproduced using the absorption efficiency
$Q_{\lambda}$ of a spherical $\alpha$-SiC grain.
Assuming that SiC grains greatly 
contribute to the 10-14\,$\mu$m feature of Class~$\delta$, the wider
FWHM in Class~$\delta$ compared with Class~$\alpha$ would then be due to 
the contribution from the 12 and 12.7\,$\mu$m PAHs.

\subsubsection{The broad 16-24\,$\mu$m feature \label{S:pah_size}}

\begin{figure}
\centering
\includegraphics[width=\columnwidth]{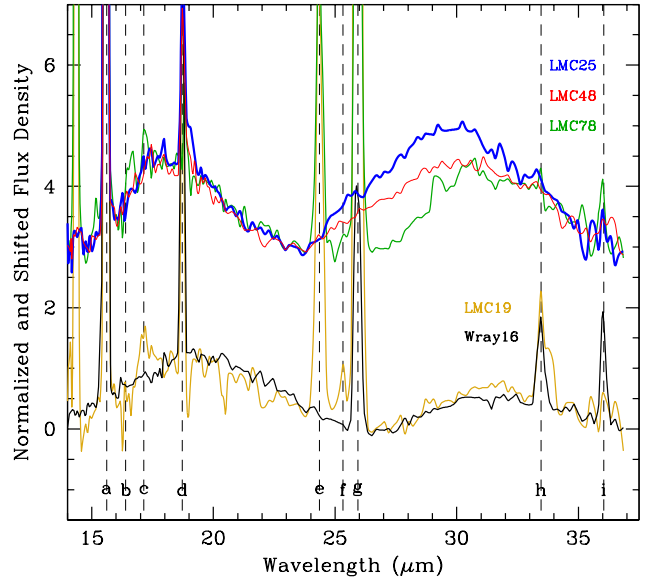}
\caption{
The 14-37.5\,$\mu$m continuum subtracted spectra, showing
 the broad 16-24\,$\mu$m and 30\,$\mu$m features. 
The vertical dashed lines mark the positions of (a)
 {\neiii}\,$\lambda$\,15.56\,$\mu$m, (b) and (c) 16.4\,$\mu$m PAH
 and 15-18\,$\mu$m plateau (possible), (d)
 {\siii}\,$\lambda$\,18.71\,$\mu$m, 
(e) {\neiv}\,$\lambda$\,24.32\,$\mu$m, 
(f) He\,{\sc ii}\,$\lambda$\,25.31\,$\mu$m, 
(g) {\oiv}\,$\lambda$\,25.89\,$\mu$m, 
(h) {\siii}\,$\lambda$\,33.50\,$\mu$m, and
(i) {\neiii}\,$\lambda$\,36.01\,$\mu$m.
}
\label{spt_spec4}
\end{figure}

\begin{figure}
\centering
\includegraphics[width=\columnwidth,bb=55 194 468 538,clip]{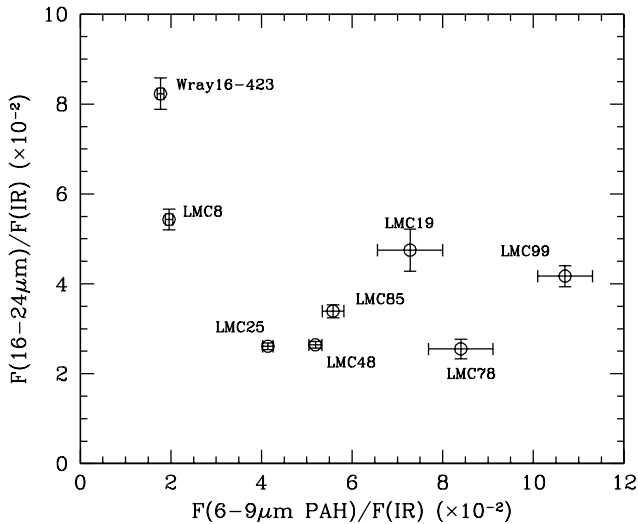}
\caption{The plot between the 6-9\,$\mu$m PAH and the broad
 16-24\,$\mu$m feature strengths.}
\label{fig-69-1624um}
\end{figure}

Broad 16-24\,$\mu$m and 30\,$\mu$m features are presented in Fig.~\ref{spt_spec4}. 
\citet{2009ApJ...699.1541B} and \citet{Garcia-Hernandez:2012aa}
reported the detection of this feature in MC PNe. \citet{2014MNRAS.439.1472M} detected this
feature in the LMC post-AGB star IRAS0588-6944. In our galaxy, a broad
16-24\,$\mu$m feature has been found in PNe M1-11
\citep{Otsuka:2013ab}, K3-17, M2-43 \citep{Hony:2002aa}, and the
proto-PN IRAS01005+7910 \citep{Zhang:2010aa}.

The 16-24\,$\mu$m feature is narrower in LMC25, 48, 78 
(See Table~\ref{band_measure}); this could be due to 
the contribution of the blue wing of the 30\,$\mu$m feature. The band profiles
in LMC25, 48, and 78 are slightly asymmetric; the
${\lambda}_{c}$s of the 16-24\,$\mu$m feature
in these three are blue-shifted, compared with those of Wray16-423 and
LMC19. The asymmetric profile could be due to a small plateau 
superposed on the broad 16-24\,$\mu$m around 15-18\,$\mu$m. The 15-18\,$\mu$m plateau
seems to be possibly in
LMC19 ($\lambda_{c}$$\sim$17.10\,$\mu$m), but its presence is unclear in Wray16-423.
In Figure~\ref{spt_spec5} of Appendix~C, we compare our sample PNe with the
Galactic PN K3-60 (AORKEY: 19903744, PI: H.~Dinnerstein, indicated by
the grey lines). We use the K3-60 spectrum to explain the
position and width of the 15-18\,$\mu$m plateau, as an example. K3-60 shows 
the 15-18\,$\mu$m plateau but no broad 16-24\,$\mu$m feature.
The 6-9 and 10-14\,$\mu$m PAH band profiles in K3-60 are well
matched to those in LMC78, 19, and Wray16-423.
The complex of the 15-18\,$\mu$m plateau and the 16.4\,$\mu$m PAH also
contribute to the broad 16-24\,$\mu$m feature; however, its contribution
seems to be negligibly small in our sample PNe. 
Even if the 15-18\,$\mu$m plateau plus the 16.4\,$\mu$m PAH are the
major contributors to the 16-24\,$\mu$m feature, this complex cannot
explain the asymmetry and difference in width between the observed
16-24\,$\mu$m features. The red and blue wings of the 16-24\,$\mu$m
feature have different slopes between LMC25/48/78 and LMC19/Wray16-423, 
and this cannot be explained by the presence of the plateau only. Note
that also the broad 30\,$\mu$m feature is clearly broader and redder in
LMC19 and Wray16-423. The different widths and positions might reflect 
the colder dust and/or larger dust grains.

The carrier of the broad 16-24\,$\mu$m feature has yet to be
identified. \citet{Van-Kerckhoven:2000aa} noted that the
6.2/7.7/8.6/11.3\,$\mu$m PAH emission is from relatively small PAH
molecules ($\sim$50 C-atoms). Additionally, they argued that
the 16-20\,$\mu$m plateau -- it should
be attributed to the 15-18\,$\mu$m plateau as described above --
is due to the destruction of large PAHs and PAH clusters, containing $\sim$2000 C-atoms.

Can the 16-24\,$\mu$m feature be also attributed to large/cluster PAHs and the 6-9\,$\mu$m
PAH form by the destruction of large PAHs and PAH clusters? If so,
then an anti-correlation may exist between the intensities of the 16-24\,$\mu$m feature and the
6-9\,$\mu$m PAH. In 7 LMC PNe listed in Table~\ref{band_measure2} of
Appendix~C and Wray16-423 showing both features,
we measured the integrated fluxes of the
6-9\,$\mu$m PAH band ($F$(6-9\,$\mu$m PAH)), the 16-24\,$\mu$m feature
($F$(16-24\,$\mu$m)) in the local continuum subtracted spectra,
and the integrated flux between 5.75-36.8\,$\mu$m ($F$(IR)). 
We define the ratios of the $F$(6-9\,$\mu$m PAH)/$F$(IR) and the
$F$(16-24\,$\mu$m PAH)/$F$(IR) as feature strengths, and we 
present between these strengths in Fig.~\ref{fig-69-1624um}.
The correlation factor is --0.50, depending on the data of
Wray16-423 because the factor is down to --0.16 without Wray16-423. 
Therefore, at this moment we cannot conclude that there is a correlation between 
the 16-24\,$\mu$m feature and the 6-9\,$\mu$m PAH band (or PAHs
themselves).

\citet{Hony:2002aa} proposed iron sulfide (FeS) to explain the
16-24\,$\mu$m feature. We note \citet{Zhukovska:2008aa, who argued that FeS is abundant
in O-rich environments because in a C-rich environment the Mg is not consumed by the even more
stable magnesium silicates as in the O-rich case and the magnesium sulfide
(MgS) is more stable than FeS.} 
FeS displays a broad feature centered
at 23\,$\mu$m and two sharp resonances at 34 and 38\,$\mu$m. 
In our sample PNe, the $\lambda_{c}$s of the 16-24\,$\mu$m feature are 
not in accordance with that of the broad component of FeS at the central
wavelength ($\sim$23\,$\mu$m), and none showed the 34\,$\mu$m resonance.
Thus, Fe depletion in Wray16-423 does not reflect the presence of
FeS as a carrier of the 16-24\,$\mu$m feature; however, it may imply
those of other Fe-based grains.

Amorphous silicates have broad emission features at two peaks
near 9.7\,$\mu$m (Si-O stretching mode) and $\sim$18\,$\mu$m (O-Si-O bending
mode). Thus, the fossil silicate dust that forms in early AGB outflow
is of interest. Dual-dust PNe have been found in the Galactic bulge 
\citep[e.g.,][]{Gorny:2010aa,Perea-Calderon:2009aa}. 
\citet{Gorny:2010aa} reported
four PNe showing amorphous silicate, PAHs, and crystalline silicate features
at 23.5, 27.5, and 33.8\,$\mu$m, respectively. Crystalline silicates are
not detected in our sample. As Figure~7 of \citet{2009ApJ...699.1541B}, the central wavelength
of the 18\,$\mu$m bump is not in accordance with
that of the 16-24\,$\mu$m feature. If amorphous
silicate is responsible for the 16-24\,$\mu$m feature, then a broad
bump should be observed in the blue shoulder of the broad 11\,$\mu$m
feature. This is not the case. Therefore, the broad 16-24\,$\mu$m features in our sample
cannot be attributed to amorphous silicate.

\subsubsection{The broad 30\,$\mu$m feature \label{S-30um}}

  \begin{figure}
   \includegraphics[width=\columnwidth,bb=41 183 461 488,clip]{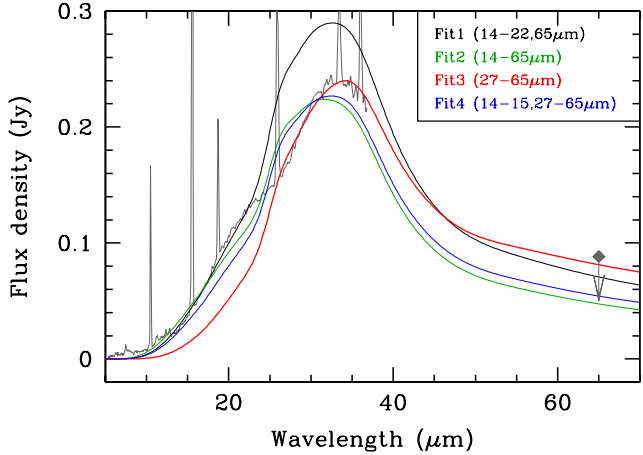}
   \caption{Reproduced SEDs by the Equation~(\ref{eq:dustem})
   (black, green, red, and blue lines) overlaid on the \emph{Spitzer}/IRS
   spectrum of Wray16-423 (grey line). The point indicated by the
   diamond shows the upper-limit flux density at \emph{AKARI}/FIS
   65\,$\mu$m. Fits 1-4 are different in terms of the fitting wavelength range. 
   The fitting results are summarised in Table~\ref{t-30um}. \label{f-30um}}
  \end{figure}

   \begin{table}
    \centering
\caption{Fitting results for the broad 30\,$\mu$m feature. The
    uncertainty of $T_{d}$(max) is within 0.6~K. \label{t-30um}}
  \begin{tabular}{@{}ccc@{}} 
\hline
Fitting &Fitting wavelength &$T_{d}$(max)\\
        &($\mu$m) &(K)\\
	\hline
Fit1    &14-22,65      &145.5\\
Fit2    &14-65        &131.3\\
Fit3    &27-65        &108.1\\
Fit4    &14-15, 27-65 &133.0\\
   \hline
  \end{tabular}
   \end{table}

Our PNe display a broad 30\,$\mu$m feature, which has been seen in
many C-rich objects. 
\citet{Hony:2002ab} performed a comprehensive analysis of
the 30\,$\mu$m feature showing objects with \emph{ISO}/SWS spectra.
Due to the wavelength coverage, our measured FWHMs were narrower than the measurements of
\citet[$\sim$10\,$\mu$m]{Hony:2002ab}.
\citet{Hony:2002ab} argued that the feature originated from a 
CDE MgS. The CDE MgS has been the major candidate for the
30\,$\mu$m feature. However, as we introduced earlier, the origin of this feature is unclear.

\citet{Messenger:2013aa} performed a detailed analysis on
the 30\,$\mu$m and the 11\,$\mu$m SiC features; the authors
concluded that the 30\,$\mu$m feature is likely due to a carbide
or sulfide component, and that the abundance of the carrier of the 30\,$\mu$m feature
is linked to SiC abundance. This conclusion seems to agree with
\citet{Sloan:2014aa}, who argued that the MgS feature strength climbs
just as the SiC strength drops, as the dust shells around the stars become
red, based on the assumption that the SiC/amorphous carbon
is coated by MgS as the dust grains move away from the central star.
Laboratory experiments by \citet{Lodders:1995aa} revealed the
condensation temperatures of different C/O ratios and pressures; for
the case of C/O = 2.0 and a total pressure of 10$^{-3}$ bar, the
condensation temperatures of graphite, SiC, and MgS grains are
1978 K, 1736 K, and 1152 K, respectively.

We measured the $F$(11\,$\mu$m SiC)/$F$(IR)
for LMC PNe LMC8, 25, 48, and 85 (Table~\ref{band_measure2}) showing SiC.
Among these PNe, an anti-correlation 
relationship exists between the SiC and 30\,$\mu$m intensities
(correlation factor=--0.88), supporting the idea of
\citet{Sloan:2014aa}. We investigated the effective temperatures of 14
LMC PNe using {\sc Cloudy}. The average effective temperature of the
four SiC-containing PNe (LMC8, 25, 48, and 85) is 50\,750 K,
while a value of 123\,000 K is determined among the non-SiC containing PNe
(LMC19, 78, and 99, and Wray16-423). The CSPNe of SiC-containing PNe
seem to be cooler and younger than those of the non-SiC-containing PNe.
Thus, the MgS grains in non-SiC-containing PNe may be far from the
central stars. In comparison, 
the average $\lambda_{c}$s of the 30\,$\mu$m feature in SiC-containing four
PNe is 30.50\,$\mu$m and that in the remaining non-SiC-containing four
PNe is 31.73\,$\mu$m.

However, \citet{Zhang_09_30mic} cast doubt on this identification for the
30\,$\mu$m feature. They demonstrated that the
MgS mass of (0.96-7.16)$\times$10$^{-5}$\,M$_{\odot}$, formed from the available
S atoms, could not account for the observed strength of the 30\,$\mu$m
feature in the proto-PN HD56126, even if all the dust grains
existed as MgS (the required MgS mass is $>$3.5$\times$10$^{-4}$\,M$_{\odot}$).

CDE iron-rich sulphide, such as Mg$_{0.5}$Fe$_{0.5}$S (Fe50S),
also show a broad feature around 30\,$\mu$m,
similar to the CDE MgS feature \citep{Messenger:2013aa}. In PNe,
\citet{Henry:2012aa} reported no correlation
between the S abundance deficiency and the occurrence of the broad
30\,$\mu$m feature, implying that iron-rich sulfide rather than
MgS may be favored as a carrier of the 30\,$\mu$m feature. 
The type of Mg-Fe-S grains that form depends on the amount of
Mg, Fe, and S atoms in the solid phase.

Another possible candidate for the 30\,$\mu$m feature is graphite, as proposed by
\citet{Jiang:2014aa}. The opacity of graphite by
\citet{Draine:1984aa} exhibits 
broad emission (FWHM$>$60\,$\mu$m) peaked at $\sim$30-40\,$\mu$m.
The theoretical opacity of graphite with a
d.c. conductivity of 100\,$\Omega$$^{-1}$ cm$^{-1}$ by \citet{Jiang:2014aa}
shows a well-consistent FWHM of $\sim$10\,$\mu$m to the observed FWHMs reported by
\citet{Hony:2002ab}. \citet{Otsuka:2014aa} found that the relative
strength of the 30\,$\mu$m feature to the underlying continuum in
fullerene C$_{60}$-containing PNe
is constant; they also succeeded in fitting both the 30\,$\mu$m feature
and the 13-160\,$\mu$m continuum using the synthesised
absorption efficiency $Q_{\mathrm{abs}}(\lambda)$, derived from
the combined data of \emph{ISO}/SWS/LWS and \emph{Spitzer}/IRS
spectra of C$_{60}$-containing PN IC418. Their derived opacity curve is
similar to that of graphite by \citet{Draine:1984aa}, with the exception
of the feature strength and FWHM. \citet{Otsuka:2014aa} concluded that
the 30\,$\mu$m feature and featureless continuum may be related to
hydrogenated amorphous carbon (HAC) or graphite. HAC 
has two resonances centered at $\sim$20\,$\mu$m and $\sim$30\,$\mu$m as demonstrated
in Fig.~9 of \citet{Otsuka:2013ab}. However, since its optical constant is strongly 
variable for different chemical and physical conditions such as
hydrogen-content and grain size, one might be able to reproduce 
the strong 30\,$\mu$m feature plus the weaker sub structures.

We utilised the method of \citet{Otsuka:2014aa} 
to verify our ability to fit the $\geq$14\,$\mu$m SED of 
Wray16-423 with their
$Q_{\mathrm{abs}}(\lambda)$ even if it will not be possible to fit the
broad 16-24\,$\mu$m feature in this way.  The model of \citet{Otsuka:2014aa} assumes that 
the dust density, as a function of the distance from the CSPN $r$, is
distributed around the CSPN with a power-law ($\propto
r^{-p}$) and that the dust temperature distribution $T_{d}$($r$) also follows a
power-law ($\propto r^{-q}$). The thermal radiation from spherical grains
can be estimated by the following: 

\begin{equation}
  F_\lambda = C \cdot Q_{\mathrm{abs}}(\lambda)
   \int^{T_{d}(\mathrm{max})}_{T_{d}(\mathrm{min})} T_{d}^{-\alpha}
   B_\lambda (T_{d})~dT_{d}, 
\label{eq:dustem}
\end{equation}   

\noindent where $C$ is a constant value, $T_{d}({\mathrm{max}})$ and
$T_{d}({\mathrm{min}})$ are the maximum and minimum temperatures of the
30\,$\mu$m feature, respectively, $\alpha$=$(3-p)/q>0$, and $B_\lambda (T_{d})$ is
the Plank function. We chose both $p$ and $q$ to be 2. 
The fit is not sensitive to 
$T_{d}({\mathrm{min}})$; thus,  we let $T_{d}({\mathrm{min}})$=20 K, a
temperature more or less characteristic of dust in the ISM.

We fit the \emph{Spitzer}/IRS spectrum for wavelengths longer than
14\,$\mu$m by the Equation~(\ref{eq:dustem}). As a constraint to set the
upper limit flux density over 36.5\,$\mu$m by the model fit, we used 
an upper limit flux density $I_{\nu}$(65\,$\mu$m)=88.3 mJy measured 
from the \emph{AKARI}/FIS Far-infrared All-Sky Survey Maps. We
performed four fits, as summarised in Table~\ref{t-30um}. The fitting
regions of each differed. The resultant SEDs
are plotted on the \emph{Spitzer}/IRS spectrum in
Figure~\ref{f-30um}. Fit1 - a test whether the 14-22\,$\mu$m dust
continuum is linked with the 30\,$\mu$m feature - predicts $\sim$1.2
times larger flux density at the peak of the 30\,$\mu$m feature. Fit2
converging at 14-65\,$\mu$m gives a better result; however, it cannot explain
$\sim$20-24\,$\mu$m of the broad 16-24\,$\mu$m feature or the
30\,$\mu$m. Fit3, focusing only the 30\,$\mu$m feature, provides
a good fitting for the 30\,$\mu$m feature, although it under-predicts the flux
density below 27\,$\mu$m. Fit4 skips the broad
16-24\,$\mu$m feature and gives results similar to those obtained in Fit2.

Our simple analysis demonstrates that the broad 16-24\,$\mu$m in
Wray16-423 is unrelated to the 30\,$\mu$m feature. If we focus on the
27\,$\mu$m or longer wavelength data, we can reproduce part of
the observed 30\,$\mu$m feature with the synthesised
$Q_{\mathrm{abs}}(\lambda)$ of \citet{Otsuka:2014aa}. More fits to
the 30\,$\mu$m feature showing PNe (but without broad 16-24\,$\mu$m
feature) are necessary to synthesize the absorption efficiency of the 30\,$\mu$m
feature for astronomical objects and to separate
the 16-24\,$\mu$m feature from the complex of 16-24\,$\mu$m
and 30\,$\mu$m features.

\subsection{SED modeling \label{Ssed}}

\begin{figure}
\includegraphics[width=\columnwidth,bb=37 339 512 645,clip]{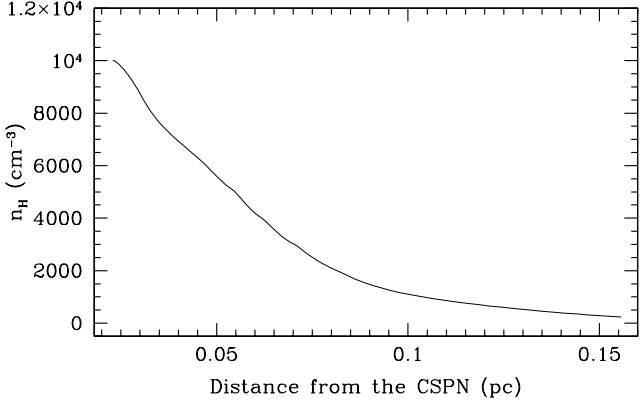}
\caption{Radial profile of the hydrogen density adopted in SED
 modeling.}
\label{density}
\end{figure}

  \begin{table}
   \centering
   \caption{Best fitting parameters and the derived properties by the {\sc Cloudy} model. \label{model}}
 \begin{tabularx}{\columnwidth}{@{}ll@{}}
\hline
{Parameters of the CSPN}      &{Values}\\
\hline
$L_{\ast}$       &4753~L$_{\odot}$\\
  $T_{\rm eff}$ &110\,110\,K        \\
  $M_{V}$       &2.691\\
$\log\,g$ &6.0~cm s$^{-2}$\\
Distance  &24.8~kpc\\
\hline
{Parameters of the Nebula}      &{Values}\\
\hline
Abundances      &He:11.02, C:8.86, N:7.69, O:8.22, Ne:7.50,\\
($\log_{10}$(X/H)+12)&S:6.40, Cl:4.71, Ar:5.92, K:4.31, Fe:5.57\\
                &[Si,Mg/H]=--2.00, Others:[X/H]=--0.65\\
Geometry        &Spherical\\
Shell size      &$R_{\rm in}$=0.20{\arcsec} (0.02~pc), 
$R_{\rm out}$=1.3{\arcsec} (0.16~pc)\\         
$n_{\rm H}$     &See Fig.\ref{density}\\
filling factor  &0.85\\
$\log I$({\hb}) &--11.705~erg s$^{-1}$ cm$^{-2}$ (de-reddened)\\
$m_{\rm g}$     &0.59\,M$_{\odot}$ \\
\hline
{Parameters of the Dust}      &{Values}\\
\hline
Composition   &3.1\,$\%$ PAHs and 96.9\,$\%$ graphite (mass frac.)\\
Grain size    &0.0004-0.011\,$\mu$m for PAH\\
              &0.005-0.25\,$\mu$m for graphite\\
$T_{d}$       &113-305\,K (PAHs), 43-194\,K (graphite)\\  
$m_{d}$      &1.32(--4)\,M$_{\odot}$ in total \\ 
$m_{d}$/$m_{\rm g}$ &2.26(--4) \\
  \hline
 \end{tabularx}
  \end{table}

We constructed an SED model to investigate the physical 
conditions of the gas and dust grains, in order to derive their 
masses using {\sc Cloudy} c10.00.

\subsubsection{Model approach}

We set the distance to Wray16-423 to 24.8 kpc
for comparison with the observed fluxes and flux densities. 
For the incident SED from the CSPN, we adopted the model grids
of non-LTE line-blanketed, plane-parallel, hydrostatic model atmospheres
of halo stars ([Z/H]=--1) by \citet{Rauch:2003aa}. Because the
{\heii} line-fluxes are sensitive to the surface gravity
$\log$\,$g$, we tested small gas-only model grids with varying
$\log$\,$g$ that matched the flux of the {\heii}\,$\lambda$\,4686\,{\AA}. 
In the small grids, we set the CSPN's effective temperature to $T_{\rm
eff}$=110\,660\,K and luminosity $L_{\ast}$ to 4683\,L$_{\odot}$ using
Equations (3.1) and (3.2) of \citet{Dopita:1991aa}, which are established
amongst Magellanic Cloud PNe, assuming optically thick nebula.
Accordingly, we set $\log$\,$g$ to 6.0~cm s$^{-2}$. We kept this surface gravity 
in the gas+dust full SED model, and adopted the above $T_{\rm eff}$ and
$L_{\ast}$ as a first guess.

For gas-phase elemental abundances, we adopted the observed values 
listed in Table~\ref{abund} as the first guess, and refined 
these to match the observed line intensities of each element.  
We considered the CEL O line fluxes to set the O abundance. 
We adopted the same transition probabilities and effective collision
strengths of CELs
used in our plasma diagnostics and abundance determinations. 
Because {\sc Cloudy} does not include any Kr lines, we did not consider
this element. The abundances for non-calculated elements, except for Mg
and Si, were held constant at [X/H]=--0.65, which is the average
between the observed [Cl/H] and [Ar/H]. Because the deficiency 
of refractory elements Mg and Si is unknown, we adopted [Mg,Si/H]=--2.

We determined the radial H-density profile of the nebula, based on the
radial intensity profile of the \emph{HST}/WFPC2 F656N image 
using an Abel transformation under the assumption of spherical
symmetry. The ionization-bound models over-predicted the {\nii} and
{\oii} lines and under-predicted the {\oiii} lines, whereas 
the matter-bound models gave better fit to the observed fluxes of these
emission lines. Therefore, we assumed the matter-bound condition. 
The outer radius $R_{\rm out}$ was set to 1.3$''$ (0.16~pc) and the inner radius 
$R_{\rm in}$ was 0.2$''$ (0.023~pc), from the \emph{HST}
image.  We adopted a constant filling
factor of 0.85. The adopted radial profile of the hydrogen
density is presented in Figure~\ref{density}. Our set hydrogen density
profile appeared to be similar to the photoionisation
models without dust grains by \citet{Dudziak:2000aa}, who adopted the constant
hydrogen densities of 9500~cm$^{-3}$ for 0.023-0.051~pc (covering factor
=0.17) and 3600~cm$^{-3}$ for 0.023-0.08~pc (0.83).

We attempted to fit the observed SED in the range from 0.1516\,$\mu$m
(\emph{GALEX}-Fuv) to 65\,$\mu$m (\emph{AKARI}/FIS), assuming that the infrared excess around 5-40\,$\mu$m
was due to the emission from PAH molecules and carbon-based grains.
There are no suitable optical constants for the broad 16-24\,$\mu$m and
30-$\mu$m features; thus, we skipped these bands. The 16-24\,$\mu$m and 27.4-36.5\,$\mu$m SEDs were
out of the fitting range in the model. Here, we adopted graphite
and PAH. We assumed a spherical shape for both the graphite and PAH. In graphite, the spheres were assumed to be randomly oriented,
and the "1/3-2/3" approximation was utilised \citep[for validity of
this approximation see,][]{Draine:1993aa}.
The optical constants were taken from \citet{2007ApJ...657..810D}
(PAH-Carbonaceous grains) for PAH 
and \citet{Draine:1984aa} for graphite. For PAH, we adopted the radius $a$ in the range from 0.0004 (30
C-atoms) to 0.0011\,$\mu$m (500 C-atoms) with the standard interstellar
dust-grain size distribution by \citet{Mathis:1977aa} (i.e.,
$a^{-3.5}$). For the graphite, we adopted the same size distribution with $a$=0.005-0.25\,$\mu$m.

To verify the model fitting accuracy, we evaluated
the chi-square value (${\chi}^{2}$) from the 44 gas emission
fluxes, 9 gas-phase abundances, 7 broad band fluxes, and the 9 
de-reddened flux densities $I_{\nu}$ from \emph{GALEX}, \emph{HST}/WFPC2,
2MASS, and \emph{WISE} bands. We used the \emph{AKARI}/FIS
$I_{\nu}$(65\,$\mu$m) to set the dust continuum above 36.5\,$\mu$m.

\subsubsection{Comments on SED model results \label{S-SEDr}}

\begin{figure*}
\includegraphics[width=\textwidth,bb=26 149 579 692]{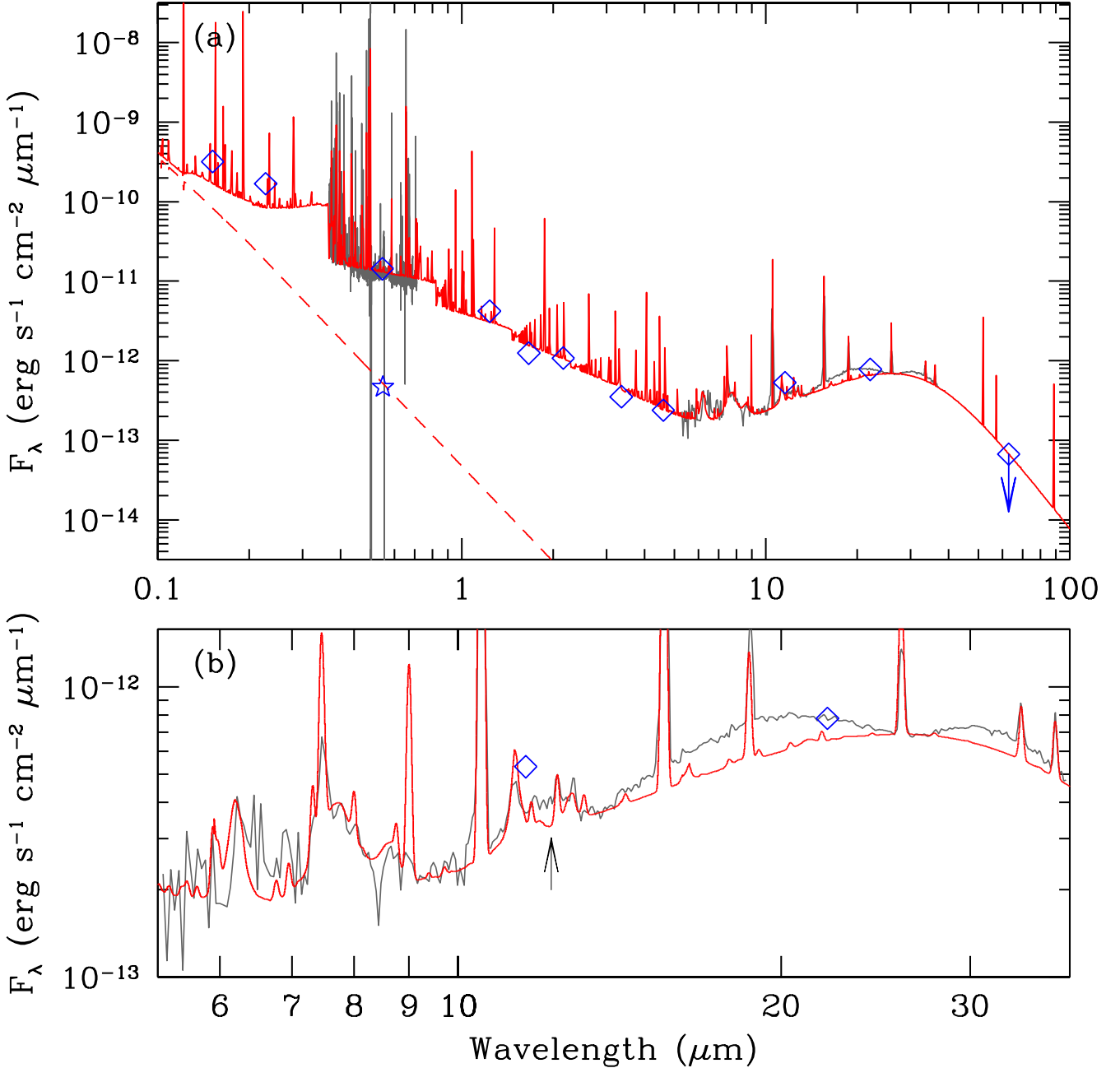}
\caption{({\it upper panel}) Predicted SED of the PN Wray16-423 by {\sc Cloudy} modeling
 (the red solid line and the dashed line). The red solid line is the
 emergent model spectrum of the CSPN plus the nebula, and the dashed line is the spectrum of the
 CSPN. The grey lines are the observed spectra taken by the
 Subaru/HDS and the \emph{Spitzer}/IRS. The blue diamonds are the
 photometry data of the CSPN plus the nebula
 from the \emph{GALEX} (Fuv and Nuv bands), \emph{HST} (F547M), 2MASS
 (near-IR $JHKs$), \emph{WISE} (11.6 and 22.1\,$\mu$m), and \emph{AKARI}/FIS
 65\,$\mu$m. The green star is the photometry of the CSPN measured by
 the \emph{HST}/(F547M) image.
 ({\it lower panel}) Closed-up plot for mid-IR
 wavelengths. The spectral resolving power ($R$) of the
 {\sc Cloudy} model spectrum was constant (100) to match that of the
 \emph{Spitzer}/IRS. The lack of emission, indicated by the arrow,
 could be filled by adding amorphous carbon grains in the model. See
 the text for details.
 }
\label{sed}
\end{figure*}

   \begin{table*}
\caption{Comparison between SED modeling and observations. \label{model2}}
\begin{tabularx}{\textwidth}{@{}LCRRLCRR@{}}
\hline
{Ion}      &{$\lambda_{\rm lab}$} &{$I$({\rm CLOUDY})}      &{$I$(Obs)}
 &{Ion}      &{$\lambda_{\rm lab}$} &{$I$({\rm CLOUDY})} &{$I$(Obs)}\\
&&($I$({\hb})=100)&($I$({\hb})=100.0)&&&($I$({\hb})=100.0)&($I$({\hb})=100.0)\\
  \hline
  {\oii}  &  3726\,{\AA}  & 19.79 & 19.52 & {\hei}  &  5876\,{\AA}  & 16.22 & 15.20 \\ 
{\oii}  &  3729\,{\AA}  & 14.17 & 10.72 & {\siii}  &  6312\,{\AA}  & 1.19 & 1.42 \\ 
{\neiii}  &  3869\,{\AA}  & 87.36 & 85.71 & {\nii}  &  6548\,{\AA}  & 5.45 & 5.50 \\ 
{\neiii}  &  3968\,{\AA}  & 26.33 & 26.15 & {\ha} & 6563\,{\AA} & 284.8 & 285.0 \\ 
{\sii}  & 4069\,{\AA} & 0.63 & 1.11 & {\nii}  &  6584\,{\AA}  & 16.08 & 16.20 \\ 
{\ciii}  &  4187\,{\AA}  & 0.17 & 0.15 & {\hei}  &  6678\,{\AA}  & 4.19 & 3.98 \\ 
{\cii}  &  4267\,{\AA}  & 0.45 & 0.47 & {\sii}  &  6716\,{\AA}  & 1.82 & 1.58 \\ 
H$\delta$ & 4341\,{\AA} & 46.69 & 46.90 & {\sii}  &  6731\,{\AA}  & 2.26 & 2.64 \\ 
{\oiii}  &  4363\,{\AA}  & 14.59 & 13.39 & {\ariii}  &  7135\,{\AA}  & 7.77 & 10.83 \\ 
{\hei}  &  4388\,{\AA}  & 0.63 & 0.57 & {\hei}  & 7281\,{\AA} & 1.06 & 1.01 \\ 
{\hei}  &  4471\,{\AA}  & 5.20 & 5.10 & {\oii}  &  7323\,{\AA}  & 1.56& 2.79 \\ 
{\feiii}  & 4701\,{\AA} & 0.013 & 0.022 & {\oii}  &  7332\,{\AA}  & 1.24 & 2.15 \\ 
{\ariv}  &  4711\,{\AA}  & 2.81 & 1.86 & {\ariii}  &  7751\,{\AA}  & 1.88 & 2.71 \\ 
{\ariv}  &  4740\,{\AA}  & 3.72 & 2.71 & {\cliv}  &  8047\,{\AA}  & 0.34 & 0.56 \\ 
{\ciii}  &  4649\,{\AA}  & 0.48 & 0.40 & {\siii}  &  9069\,{\AA}  & 4.65 & 7.99 \\ 
{\civ}  &  4659\,{\AA}  & 0.045 & 0.103 & {\siv}  &  10.51\,$\mu$m  & 55.07 & 27.80 \\ 
{\heii}  &  4686\,{\AA}  & 10.02 & 11.14 & {\neii}  &  12.80\,$\mu$m  & 0.21 & 1.07 \\ 
{\hb}  &  4863\,{\AA}  & 100.0 & 100.0 & {\neiii}  &  15.55\,$\mu$m  & 49.05 & 59.76 \\ 
{\feiii}  &  4881\,{\AA}  & 0.032 & 0.032 & {\siii}  &  18.67\,$\mu$m  & 7.65 & 9.92 \\ 
{\hei}  &  4922\,{\AA}  & 1.37 & 1.27 & {\oiv}  &  25.88\,$\mu$m  & 16.00 & 11.92 \\ 
{\oiii}  &  4931\,{\AA}  & 0.14 & 0.14 & {\siii}  &  33.47\,$\mu$m  & 3.99 & 4.31 \\ 
{\oiii}  &  4959\,{\AA}  & 336.3 & 362.4 & {\neiii}  &  36.01\,$\mu$m  & 4.03 & 3.87 \\ 
{\oiii}  &  5007\,{\AA}  & 1012.1 & 1062.1 & IRS-1  &  6.20\,$\mu$m  & 12.36 & 13.466 \\ 
{\ariii}  &  5192\,{\AA}  & 0.12 & 0.13 & IRS-2  &  6.90\,$\mu$m  &11.29 & 13.96 \\ 
{\feiii}  &  5271\,{\AA}  & 0.064 & 0.040 & IRS-3  &  7.90\,$\mu$m  & 18.95 & 18.22 \\ 
{\cliv}  &  5324\,{\AA}  & 0.012 & 0.021 & IRS-4  &  11.30\,$\mu$m  & 19.45 & 19.28 \\ 
{\cliii}  &  5518\,{\AA}  & 0.28 & 0.25 & IRS-5  &  12.30\,$\mu$m  & 19.47 & 21.30 \\ 
{\cliii}  &  5538\,{\AA}  & 0.31 & 0.37 & IRS-6  &  13.50\,$\mu$m  & 40.31 & 40.70 \\ 
{\nii}  &  5755\,{\AA}  & 0.45 & 0.40 & IRS-7  &  14.50\,$\mu$m  &
			  21.20 & 20.52 \\
  \hline
  Band &$\lambda_{\rm c}$&$I_{\nu}$({\rm CLOUDY}) &$I_{\nu}$(Obs)
	      &X&&$\log_{10}$(X/H)$_{{\rm CLOUDY}}$&$\log_{10}$(X/H)$_{\rm Obs}$\\
  & &(mJy)&(mJy) &&&+12&+12\\
  \hline
\emph{GALEX}-Fuv  &  1516\,{\AA}  & 5.61 & 2.58 & He &  & 11.02 &11.02\\ 
\emph{GALEX}-Nuv  &  2267\,{\AA}  & 3.96 & 3.07 & C &  & 8.86 &8.92\\ 
WFPC2/F547M  &  5484\,{\AA}  & 1.35 & 1.44 &     N &  & 7.69 &7.69 \\ 
2MASS-$J$  &  1.235\,$\mu$m  & 2.49 & 2.13 &    O &  & 8.22 & 8.31 \\ 
2MASS-$H$  &  1.662\,$\mu$m  & 1.60 & 1.15 &      Ne &  &7.50  & 7.61 \\ 
2MASS-$Ks$  &  2.159\,$\mu$m  & 2.47 & 1.68 &  S   &&6.40  & 6.40 \\ 
\emph{WISE}-B1  &  3.353\,$\mu$m  & 1.90 & 1.32 & Cl &&4.71  & 4.74 \\ 
\emph{WISE}-B2  &  4.603\,$\mu$m  & 2.87 & 1.69 & Ar &&5.92  & 6.00 \\ 
\emph{WISE}-B3  &  11.56\,$\mu$m  & 23.79 & 23.68 & Fe &&5.57  & 5.56 \\ 
\emph{AKARI}/FIS& 65\,$\mu$m & 88.30 & $<$88.33 &${\chi}^{2}$(reduced ${\chi}^{2}$)& & & 48.14(0.63)\\ 
\hline
\end{tabularx}
    
  \raggedright
  Note -- The values for the IRS-1, 2, 3, 4, 5, 6, and 7 bands correspond to
  integrated fluxes between the following wavelength ranges, 5.9-6.9\,$\mu$m,
  6.4-7.4\,$\mu$m, 7.4-8.4\,$\mu$m, 10.8-11.8\,$\mu$m,
  11.8-12.8\,$\mu$m, 12.5-14.5\,$\mu$m, and 14-15\,$\mu$m,
    respectively.\\
   \end{table*}

The properties of the CSPN, nebula, and dust/molecules matching the observed
quantities are listed in Table~\ref{model}. The predicted fluxes
relative to the {\hb}, the flux densities at the interest bands,
and elemental abundances compared with
observations are listed in Table~\ref{model2}.  
The reduced chi-square value indicates that our SED model can reproduce the
observed quantities.

We may have overestimated the {\neii}\,$\lambda$\,12.80\,$\mu$m in the
low-dispersion \emph{Spitzer}/IRS spectrum, because this line would be
contaminated by the 12.7\,$\mu$m PAH (see Fig.~\ref{fig-11um}).
Although we could not reproduce these line fluxes within estimated
errors, our model is globally successful in explaining
the observations. Although without the UV spectra we cannot say for
certain, the large difference in \emph{GALEX}-Fuv and Nuv 
might be due to the adopted extinction value and reddening
correction function for \emph{GALEX} wavelength.

The model predicted $M_{V}$ of the CSPN (2.685) is consistent with the
observed absolute magnitude of 2.770$\pm$0.2 in the case of 24.8 kpc
within error. The uncertainty of $T_{\rm eff}$ is 2\,000\,K.
Taking into account the distance uncertainty of 0.8~kpc
\citep{Kunder:2009aa}, our derived $L_{\ast}$ of
4753$\pm$1230\,L$_{\odot}$ is fairly consistent with
\citet{Dudziak:2000aa}, who reported 4350$\pm$150\,L$_{\odot}$ and
$T_{\rm eff}$=107\,000$\pm$10\,000\,K.
The uncertainties of the predicted elemental abundances are within 0.1~dex.

Taking the absolute \emph{Spitzer}/IRS flux calibration uncertainty
$\sim$17\,$\%$ \citep{Decin:2004aa}, the dust mass $m_{d}$ would be
(1.32$\pm$0.23)$\times$10$^{-4}$\,M$_{\odot}$. The uncertainty of
the dust mass fraction would be within $\pm$1,$\%$. The estimated dust mass
here would be the lower limit value, because we skipped the
broad 16-24\,$\mu$m and 30\,$\mu$m bands. If one of the carriers for
the 30\,$\mu$m broad band is the Fe-rich sulfide, then the dust mass and
the dust-to-gas mass ratio could increase. Indeed, the large
Fe depletion with respect to Ar is confirmed; thus, the missing Fe atoms
would exist as Fe dust grains. In the next section, we discuss how much
dust mass can be added if the broad 30\,$\mu$m feature is from
Fe-rich sulfide or MgS. The gas mass $m_{g}$ (the sum of the ionised 
and atomic gas mass within the volume occupied 
within $R_{\rm out}$) is 0.59$\pm$0.10\,M$_{\odot}$. 
We also ran the model for the case of $R_{\rm out}$=1.0{\arcsec} (0.12 pc), and we obtained the
$m_{d}$=(1.05$\pm$0.18)$\times$10$^{-4}$\,M$_{\odot}$ and $m_{g}$=0.45$\pm$0.08\,M$_{\odot}$. The estimated $m_{g}$ for 
our $R_{\rm out}$=1.0{\arcsec} and 1.5{\arcsec} models is consistent
with the result of 0.4\,M$_{\odot}$ by \citet{2006MNRAS.369..875Z}. The
ionised gas mass exceeds the prior estimate of 0.248\,M$_{\odot}$ by
\citet{Dudziak:2000aa}, which we attribute to the difference in {\hb} flux and
outer radius ($\log$~$I$({\hb})=--11.89~erg s$^{-1}$ cm$^{-2}$ and $R_{\rm out}$=0.08~pc at
25~kpc). If we adopted their $I$({\hb}), then $m_{g}$ is 
$\sim$0.29\,M$_{\odot}$ in the case of $R_{\rm out}$=0.12~pc.

The dust mass in Wray16-423 is similar to that in Hen2-436
\citep[(2.9-4.0)$\times$10$^{-4}$\,M$_{\odot}$,][]{Otsuka:2011aa}, whereas the gas mass
is 8-12 times larger. \citet{Otsuka:2011aa} and
\citet{Dudziak:2000aa} estimated 0.05-0.07\,M$_{\odot}$ (ionised +
neutral gas) and 0.04\,M$_{\odot}$ (ionised gas) for Hen2-436, respectively.
\citet{Dudziak:2000aa} argued that the outer boundary in Hen2-436 is
determined by radiation (ionisation bound). Indeed, Hen2-436 is very
compact; its outer radius is $\sim$0.03~pc. The small
O$^{2+}$/(O$^{+}$+O$^{2+}$) ratio in Hen2-436
\citep[0.85,][]{Otsuka:2011aa} seems to support this (c.f., 0.96 in
Wray16-423).

In Figure~\ref{sed}a, we present the observed SED plots (blue diamonds and
grey lines) and the modelled SED (red line). The dashed line is the
modelled incident SED of the CSPN, where the flux density in the F574M
band is consistent with the observed value.
In Figure~\ref{sed}b, we display a close-up of the SED in
the wavelength range covered by the \emph{Spitzer}/IRS spectrum.
The large mass fraction of the graphite grains is due to the rising mid-IR flux
density toward $\sim$25\,$\mu$m. Because we did not consider the
possibility of very large PAH molecules discussed in
Section~\ref{S:pah_size}, the 15-20\,$\mu$m PAH plateau is not reproduced by the
model. The complex of the 10-14\,$\mu$m PAHs and broad 11\,$\mu$m feature in
Wray16-423 could be reproduced with PAHs although the fitting is not
perfect. The observed band profile of the 10-14\,$\mu$m 
PAHs and broad 11\,$\mu$m feature is
nearly flat top around 12\,$\mu$m whereas the predicted band profile is
very similar to the observed one except for the small lack of
emission indicated by the arrow. The lack of emission 
could be filled if we add amorphous carbon
grains in the model, because the $Q_{abs}$($\lambda$) of amorphous
carbon calculated from the optical data of \citet{1991ApJ...377..526R} 
appears in a small bump around 12\,$\mu$m.

\section{Discussion}
\subsection{Progenitor mass estimate by comparison to AGB nucleosynthesis models}

    \begin{table}
     \centering
\caption{Comparison to AGB nucleosynthesis models of
   \citet{2010MNRAS.403.1413K}. The C abundance in Wray16-423 is the RL C
   value. See text regarding the estimate of the Mg abundance
     in Section~\ref{S-grains}. The core-mass of Wray16-423 was
     estimated from \citet{1994ApJS...92..125V}. \label{agb}}
\begin{tabular}{@{}lcccc@{}}
 \hline
 &\multicolumn{3}{c}{AGB models}\\
 \cline{2-4}
   &1.5\,M$_{\odot}$ &1.75\,M$_{\odot}$&1.90\,M$_{\odot}$&Present\\
            &$Z$=0.004          &$Z$=0.004&$Z$=0.004&work\\
 \hline 
He  & 10.97 & 10.98 & 10.99 & 11.02$\pm$0.04 \\ 
C       & ~~8.46 & ~~8.80 & ~~8.94 & ~~8.92$\pm$0.08 \\ 
N       & ~~7.65 & ~~7.69 & ~~7.71 & ~~7.69$\pm$0.04 \\ 
O       & ~~8.23 & ~~8.25 & ~~8.25 & ~~8.31$\pm$0.01 \\ 
Ne     & ~~7.42 & ~~7.51 & ~~7.61 & ~~7.61$\pm$0.03 \\ 
Mg     & ~~6.88 & ~~6.88 & ~~6.89 & ~~6.19$\pm$0.30 \\ 
S       & ~~6.69 & ~~6.70 & ~~6.70 & ~~6.40$\pm$0.03 \\ 
Fe    & ~~6.78 & ~~6.78 & ~~6.78 & ~~5.56$\pm$0.07 \\ 
\hline
Ejected mass& 0.04\,M$_{\odot}$& 0.15\,M$_{\odot}$& 0.44\,M$_{\odot}$& 0.59$\pm$0.10\,M$_{\odot}$\\ 
Core-mass & 0.62\,M$_{\odot}$& 0.63\,M$_{\odot}$& 0.64\,M$_{\odot}$& 0.62-0.67\,M$_{\odot}$\\ 
 \hline
\end{tabular}
    \end{table}

Because the HDS spectrum shows a signature of [WC] CSPN in the C\,{\sc
iv}\,$\lambda\lambda$\,5801/12\,{\AA} band, the CSPN could be He-rich.
Assuming that the CSPN is in the midst of He-burning, comparing the 
estimated $L_{\ast}$ and $T_{\rm eff}$ on He-burning post-AGB evolution
tracks with the initial $Z$=0.008 of \citet{1994ApJS...92..125V}
indicates a progenitor mass of 1.5-2.0\,M$_{\odot}$.

The presence of the C\,{\sc iv}\,$\lambda\lambda$\,5801/12\,{\AA} band
could be evidence that Wray16-423 experienced a very late thermal pulse (VLTP).
\citet{1994ApJS...92..125V} demonstrated the He-burning post-AGB track
for initially 1.5\,M$_{\odot}$ stars, with $Z$=0.008; these stars experienced a
VLTP at $\log_{10}$$T_{\rm eff}$$\sim$4.9 ($\sim$79\,430~K). The subsequent evolution was
dominated by He-shell burning. The difference from normal
He-burning evolution tracks is that there is no upturn luminosity, and
the evolution after VLTP is very similar to the H-burning tracks.
If Wray16-423 experienced a VLTP, then the progenitor mass is estimated to be
$>$1.5\,M$_{\odot}$.

Next, we verify whether AGB nucleosynthesis models can explain
observed elemental abundances in an initial $\sim$1.5-2.0\,M$_{\odot}$
star. In Table~\ref{agb}, we compare the AGB model results of
\citet{2010MNRAS.403.1413K} with our observation results. The $^{13}$C
pocket mass is not included. Besides
He/C/N/O/Ne/S/Fe abundances, we list the ejected mass at the
last thermal pulse and the core-mass of the CSPN in the last two lines of
Table~\ref{agb}. Here, we assume that the estimated gas
mass through our SED model is nearly equal to the ejected mass at the
last thermal pulse. The gas mass estimated in our SED models depends
on the outer nebula radius and the de-reddened {\hb} flux. So far, the gas
mass estimates in Wray16-423 range from 0.248\,M$_{\odot}$ \citep{Dudziak:2000aa} to
0.49$\pm$0.10\,M$_{\odot}$ (see section \ref{S-SEDr}). Among these
model predictions, the 1.9\,M$_{\odot}$ model provides an excellent fit
to the observed elemental abundances and the ejected mass. The
consistency between the observed and the predicted Ne abundances
indicates that the $^{13}$C pocket may not form in Wray16-423. Even if we
adopt the expected CEL C abundance of 8.70$\pm$0.14, as described in
section \ref{S-element}, our conclusion does not change considerably.

\subsection{Expected solid-phase Mg, S, and Fe\label{S-grains}}

The difference between the observed S and Fe abundances and the AGB model
predicted abundances could constrain how much mass of each element
could be locked by the dust grains, based on the assumption that
elemental abundances in the nebula are consistent with the predicted
values in the case of the 1.9\,M$_{\odot}$/$Z$=0.004 AGB model. 
The density fraction of an element X in the solid phase, $n_{\rm
solid}$(X)/$n$(X), can be roughly estimated using the following:

\begin{equation}
\frac{n_{\rm solid}({\rm X})}{n({\rm X})} = 1 -
  10^{\log_{10}\left(\frac{n({\rm X})}{n({\rm H})}\right)_{\rm obs} -
  \log_{10}\left(\frac{n({\rm X})}{n({\rm H})}\right)_{\rm model}},
\label{E-dust}
\end{equation}

\noindent where $\log_{10}$($n(X)$/$n$(H))$_{\rm obs}$ and
$\log_{10}$($n$(X)/$n$(H))$_{\rm model}$ are the observed and the
AGB star model predicted elemental abundances of the element X. Using the
Equation (\ref{E-dust}), $\sim$92-95\,$\%$ of the Fe atoms and
$\sim$46-53\,$\%$ of the S atoms in the nebula could be in the solid
phase.

The ionised Mg abundances in PNe have been calculated using the
forbidden lines at UV
wavelengths and recombination Mg\,{\sc ii}\,$\lambda$\,4481\,{\AA}. We were unable to
detect optical Mg\,{\sc ii} lines. Instead, we attempt to estimate the gas-phase Mg abundance using
Mg$^{0}$ and N$^{0}$ abundances based on the assumption that all N atoms exist as 
a gas. This assumption is possible, because the observed N
abundance (7.69$\pm$0.04) is consistent with AGB model predictions
(7.71). Under the assumption that Mg-containing grains did not form, 
the Mg/N ratio would be 0.151 from
the 1.9\,M$_{\odot}$ AGB model; thus, nearly
identical to Mg$^{0}$/N$^{0}$. In this case (i.e., Mg-atoms are not
depleted by dust grains), the Mg$^{0}$/H$^{+}$ should be 0.151$\times$N$^{0}$/H$^{+}$ =
8.98(--8)$\pm$4.92(--9). Because the actual Mg$^{0}$/H$^{+}$ is 
1.81(--8)$\pm$5.27(--9), about 14-26\,$\%$ of the Mg-atoms likely
exist as a gas and the remaining 74-86\,$\%$ are
in the solid phase. Using the modeled Mg/H of 6.89 dex
(7.76(--6) in linear scale), the gas-phase Mg/H would be 6.19$\pm$0.30
dex (1.53(--6)$\pm$4.63(--7) in linear scale).
Both the gas and solid phase Mg estimates are
optimistic and depend on the AGB model prediction.
UV Mg forbidden lines in \emph{HST}/STIS spectra
would provide a more exact value.

Using the estimated fractions of the solid phase Mg, S, and Fe atoms,
we estimate the solid-phase masses $m_{\rm solid}$({\rm X}) using the following equation:

\begin{equation}
m_{\rm solid}({\rm X}) = \mu({\rm X}) \cdot \frac{m_{g}}{1+4~n({\rm
 He})/n({\rm H})_{\rm obs}} \cdot \frac{n({\rm X})}{n({\rm H})}_{\rm
 model} \cdot \frac{n_{\rm solid}({\rm X})}{n({\rm X})},
\end{equation}

where $\mu$(X) is the atomic mass, and $n$({\rm He})/$n$({\rm H})$_{\rm obs}$
is the He number density (1.06(--1), See Table~\ref{abund}). 

In summary, $\sim$92--95\% of the Fe atoms, $\sim$74--86\% of the Mg and
$\sim$46--53\% of the S atoms in the nebula could be in the solid phase,
corresponding to masses of respectively 1.3$\pm$0.2, 0.7$\pm$0.2 and
0.33$\pm$0.01 $\times 10^{-4}$\,M$_{\odot}$.

\subsection{Can MgS or Fe50S grains explain the 30\,$\mu$m feature?}

   \begin{table}
    \centering
\caption{Required MgS and Fe50 grain masses for the 30\,$\mu$m
    feature.  The maximum Fe50S (Mg$_{0.5}$Fe$_{0.5}$S) and MgS mass could form
7.65(--5)\,M$_{\odot}$ and 5.95(--5)\,M$_{\odot}$, respectively. \label{T-mgs}}
  \begin{tabular}{@{}lcccc@{}}
\hline
Models& $T_{d}$(MgS) &$T_{d}$(Fe50S)&$m_{d}$(MgS) &  $m_{d}$(Fe50S) \\
     &(K)      &(K)& (M$_{\odot}$) &(M$_{\odot}$)\\
\hline
Case 1    &133.8$\pm$0.9&117.9$\pm$1.2       &5.5(--7)$\pm$1.0(--7) &1.8(--6)$\pm$3.2(--7) \\
Case 2    &  $''$       & $''$                    &6.0(--6)$\pm$1.1(--6)&1.9(--5)$\pm$3.6(--6) \\
   \hline   
  \end{tabular}
   \end{table}

In largely Fe-depleted C-rich PNe, such as Wray16-423, estimated carbon
dust masses may make up a portion of the dust grains in the nebula, not
including the broad 16-24\,$\mu$m and 30\,$\mu$m features. As discussed
in Section~\ref{S-30um}, Mg-Fe-S grains may form in the dusty
nebula of Wray16-423, and could be a carrier for the broad
30\,$\mu$m feature. Mg-Fe-S grain masses are controlled by the
solid-phase S-atom mass. Adopting the solid-phase Mg, S, and Fe masses
as we estimated above, the possible maximum Fe50S (Mg$_{0.5}$Fe$_{0.5}$S) and MgS masses could form
7.65(--5)\,M$_{\odot}$ and 5.95(--5)\,M$_{\odot}$, respectively.

As such, can MgS or Fe50S grains explain the 30\,$\mu$m feature? To answer
this question, we estimate their masses using the Plank function 
with a single dust temperature method described in \cite{Hony:2003aa}, who explained that the dust
mass $m_{d}$ in the optically thin nebula can be estimated using the
following equation: 

\begin{equation}
  m_{d} = \frac{4}{3} \, F_{\lambda_{0},\lambda_{1}} \, D^2 \, \rho \, \left(
  \int_{\lambda_0}^{\lambda_1} \frac{Q_{abs}(\lambda,a)}{a} \, B_{\lambda}(T_{d}) \, d\lambda
		\right)^{-1},
\end{equation}

\noindent where $F_{\lambda_{0},\lambda_{1}}$ is the integrated flux in the
wavelength range from $\lambda_{0}$ to $\lambda_{1}$. Here, we adopt the
central wavelength and FWHM listed in Table~\ref{band_measure} of Appendix~C.
$\lambda_{0}$ and $\lambda_{1}$ are 28\,$\mu$m and 36\,$\mu$m,
respectively, $D$ is the distance (24.8~kpc), and
$\rho$ is the grain density, 3\,g cm$^{-3}$ for MgS and 5\,g cm$^{-3}$
for Fe50S. When the grain radius
is small in comparison with the wavelength $\lambda$, the absorption
profile becomes independent of $a$ \citep{Pegourie:1988aa}. We calculated the
$Q_{abs}$($\lambda$)/$a$ values of the CDE MgS and Fe50S by following 
Equation (4) of \citet{2001A&A...378..228F}. The optical data of
MgS (Mg$_{0.9}$Fe$_{0.1}$S, actually) and Fe50S (Mg$_{0.5}$Fe$_{0.5}$S) were taken
from \citet{Begemann:1994aa}.

In the models, we use an average dust temperature for both
MgS and Fe50S derived by Plank function fitting with a single
temperature. For $F_{\lambda_{0},\lambda_{1}}$, we consider
two cases: (Case 1) the underlying continuum of the 30\,$\mu$m
feature is largely from graphite, as demonstrated in the {\sc Cloudy} model,
and (Case 2) instead of graphite, the contribution comes from amorphous carbon, 
whose absorption efficiency drops sharply for longer wavelengths,
compared with graphite; thus, MgS or Fe50S may be the main contributors
to this band radiation. For Case 1, we calculated
$F_{\lambda_{0},\lambda_{1}}$ of 4.88(--13) erg~s$^{-1}$~cm$^{-2}$
using the residual spectrum generated by subtracting the {\sc Cloudy}
synthesised spectrum from the observed \emph{Spitzer}/IRS spectrum. For
Case 2, we calculated $F_{\lambda_{0},\lambda_{1}}$ of 5.33(--12)
erg~s$^{-1}$~cm$^{-2}$ from the observed \emph{Spitzer}/IRS spectrum
(the {\sc Cloudy} spectrum was not subtracted).

The required masses to produce the band flux of the 30\,$\mu$m
feature are listed in the last column of Table~\ref{T-mgs}. If the
graphite grains exist in the nebula (Case 1), then the required MgS
or Fe50S mass is less than the possible maximum MgS and Fe50S masses.
Even if not graphite but amorphous carbon are the main dust grains
in the nebula, MgS and Fe50S grains can reproduce the
intensity peak of the observed broad 30\,$\mu$m feature (Case 2).

At present, we hold onto the possibility that MgS and Fe50S 
are carriers of the 30\,$\mu$m feature in the case
of Wray16-423. By adding the 1.8(--6)$\pm$3.2(--7)\,M$_{\odot}$ 
of Fe50S, we could estimate the maximum dust mass to be
1.6(--2)\,M$_{\odot}$. However, we understand that these estimates are optimistic,
given that we did not consider S-atom-based molecules, such as CS.
If any sulfide molecules form in the nebula, then it would be more
difficult to explain the 30\,$\mu$m feature with Fe50S and MgS. If
it is possible to estimate the mass of the S-atom existing as the molecule with
sub-mm spectra and the optical characteristics data of MgS, Fe50S and the new graphite
data by \citet{Jiang:2014aa} are available over a wider
wavelength range, then it may be possible to resolve which grains carry the
30\,$\mu$m feature.

  \section{Summary}

  We performed a detailed chemical abundance and dust analysis of Sgr dSph PN
  Wray16-423. Numerous gas emission lines were detected, from which the
  abundances of 11 elements were calculated. 
The [Kr/H] and [Fe/H] abundances were comparable to Sgr dSph PN Hen2-436, 
which has a similar metallicity. The [Kr/O] abundance was consistent with
the value expected by the [Kr/O]-C/O relation found amongst Galactic
  PNe. The extremely small [Fe/H] does not reflect the metallicity, but
it implies that most of the Fe atoms would be in the solid phase, given the
[Ar/H] abundance. 
The \emph{Spitzer}/IRS spectrum displays broad 16-24\,$\mu$m and 30\,$\mu$m features, as well
as PAH bands at 6-9\,$\mu$m (Class $\mathcal{B}$) and 10-14\,$\mu$m
  (Class $\alpha$). An unidentified broad feature in $\sim$16-24\,$\mu$m is
probably unrelated to FeS, amorphous silicate, and PAH. 
The progenitor was estimated to be a $\sim$1.5-2.0\,M$_{\odot}$ star.  
The observed elemental abundances and the derived gas mass were in good agreement
with the AGB nucleosynthesis model for an initial mass of
1.90\,M$_{\odot}$ and a $Z$=0.004 star. 
We found that about 80\,$\%$, 50\,$\%$, and 90\,$\%$ of the Mg, S, and Fe atoms
were likely in the solid phase, respectively. Using the derived
solid-phase Mg, S, and Fe masses, we estimated the possible maximum MgS
and Fe50S grain masses. We tested whether MgS and Fe50S grains can produce the
  intensity peak of the observed broad 30\,$\mu$m band with these
  masses. Depending on the total sulfide molecule masses, 
at present, MgS and Fe50S could be carriers of
the 30\,$\mu$m feature in Wray16-423.

   \section*{Acknowledgements}
I am grateful to the anonymous referee for carefully reading and 
the useful suggestions which greatly improved this article. I 
learned a lot of things from his/her comments. 
I sincerely thank Drs. Francisca Kemper, Mikako Matsuura, and Albert
   Zijlstra for critical reading and fruitful discussions. 
I thank Dr. Akito Tajitsu for supporting my Subaru HDS observations. 
I thank ESO La Silla Observatory staffs for supporting my FEROS
 observation. This work was partly based on archival data obtained with the
   \emph{Spitzer} Space Telescope, which is operated by the Jet
   Propulsion Laboratory, California Institute of Technology under a
   contract with NASA. Support for this work was provided by an award
   issued by JPL/Caltech. This work was, in part, based on observations made
   with the NASA/ESA Hubble Space Telescope, obtained from
   the data archive at the Space Telescope Science Institute. STScI
   is operated by the Association of Universities for Research in
   Astronomy, Inc. under NASA contract NAS 5-26555. A portion of this
   work was based on the use of the IAA clustering computing system.

\bibliographystyle{mnras}

\appendix
\section{The HDS spectrum between 3680 and 7100\,{\AA} and detected lines and identifications in the HDS, FEROS, \emph{Spitzer} spectra.}

\begin{figure*}
\centering
\raggedright
\includegraphics[width=\textwidth,angle=0]{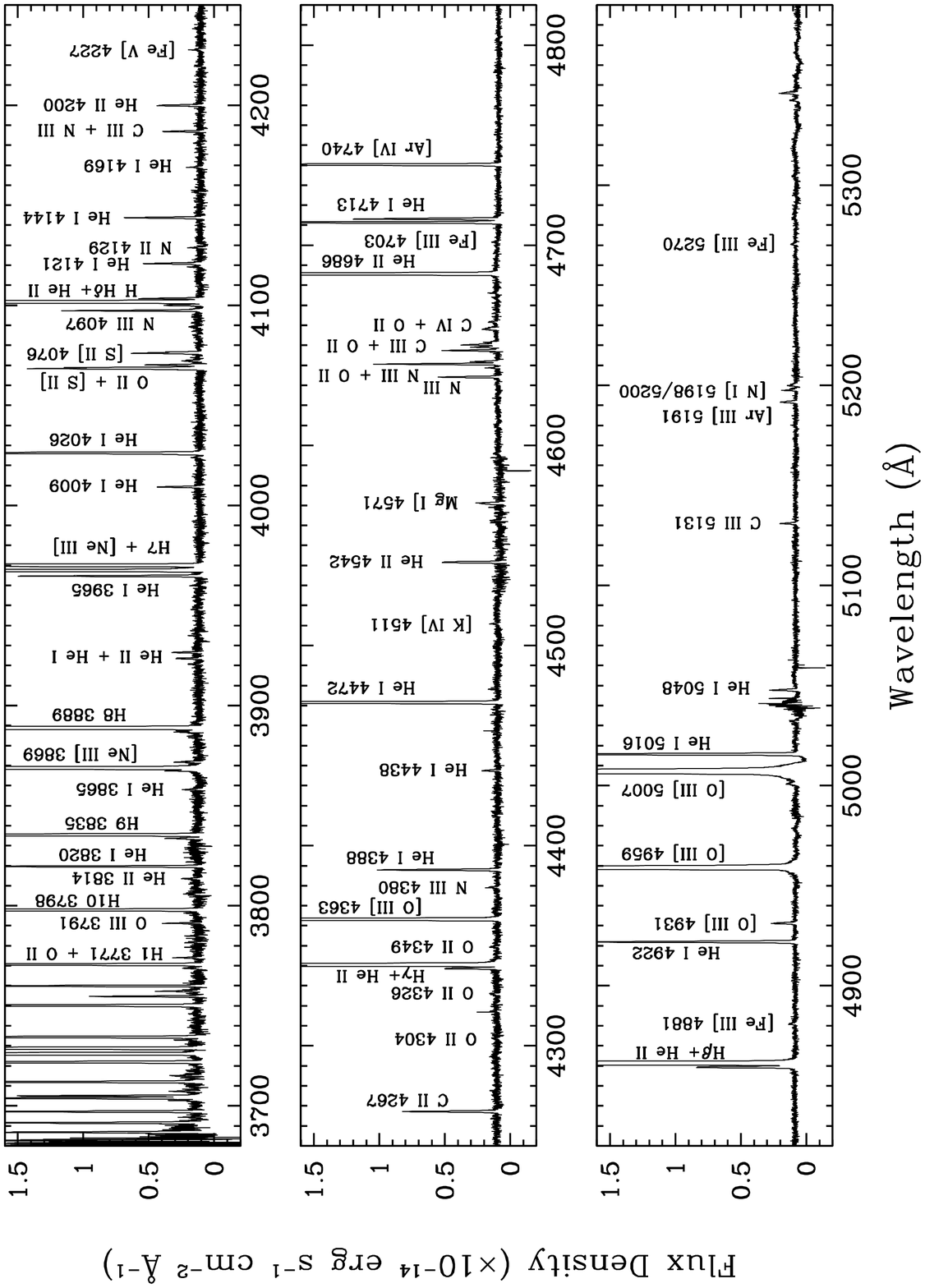}
\caption{Single 3680-5390\,{\AA} HDS spectrum of Wray16-423. The observed
 wavelength is corrected by the heliocentric radial velocity of +133.12
 {\kms}.}
\label{hdsspec1}
\end{figure*}

\begin{figure*}
\centering
\includegraphics[width=\textwidth,angle=0]{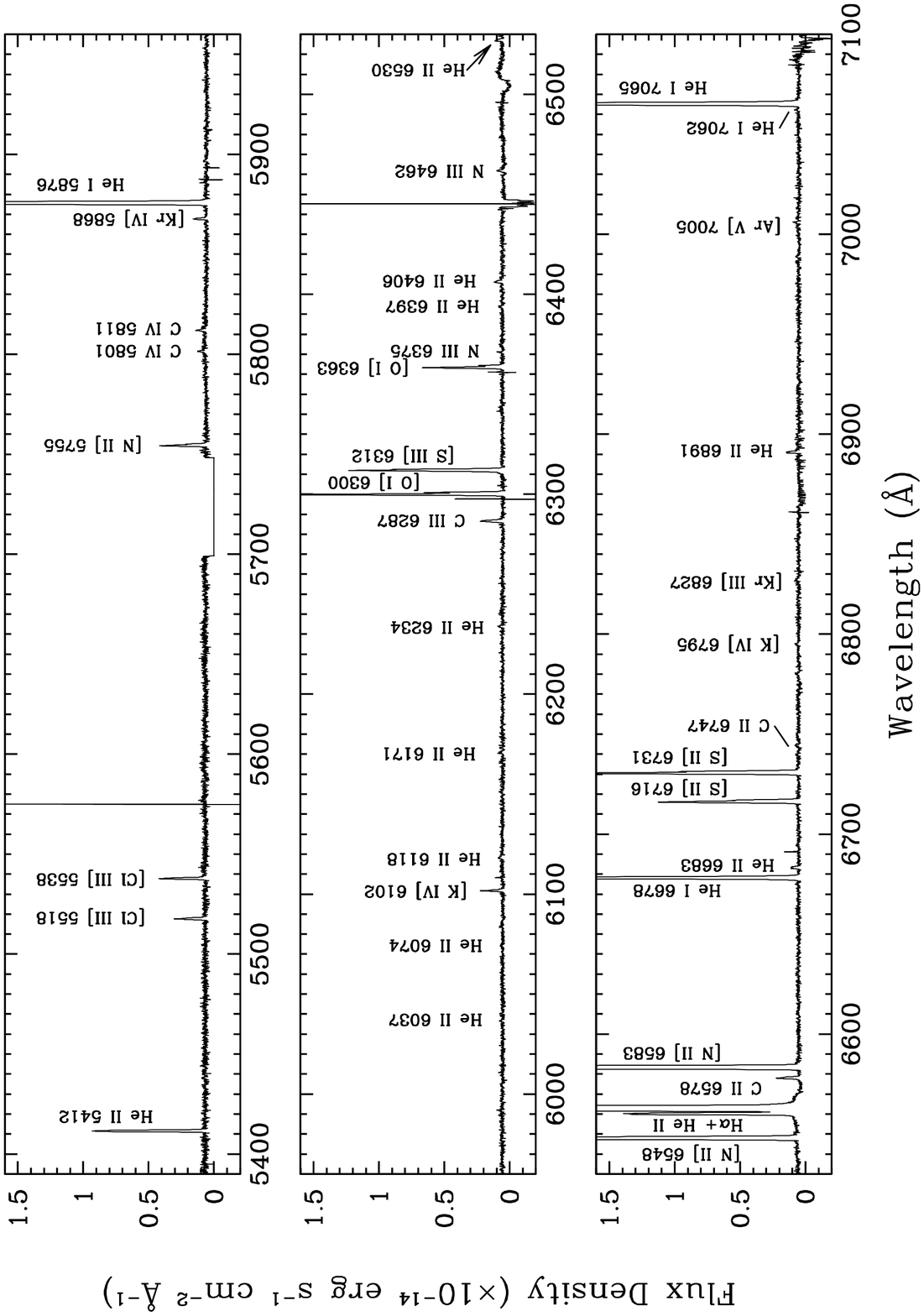}
\caption{Single 5390-7100\,{\AA} HDS spectrum of Wray16-423. The observed
 wavelength is corrected by the heliocentric radial velocity of +133.12
 {\kms}.}
\label{hdsspec2}
\end{figure*}

\renewcommand{\arraystretch}{0.80}
\begin{table*}
 \centering
\scriptsize
    \caption{Detected lines and identifications in the HDS/FEROS
 spectra. BC in component 2 of C\,{\sc iv}\,$\lambda$5801/11\,{\AA}
 means the broad component shown in Fig.~\ref{c4}. \label{hdstab}}
\begin{tabularx}{\textwidth}{@{}c@{\hspace{6.5pt} }l@{\hspace{6.5pt} }c@{\hspace{6.5pt} }c@{\hspace{6.5pt}}r@{\hspace{6.5pt} }r@{\hspace{6.5pt} }r|
c@{\hspace{6.5pt} }l@{\hspace{6.5pt} }c@{\hspace{6.5pt} }c@{\hspace{6.5pt} }r@{\hspace{6.5pt} }r@{\hspace{6.5pt}}r|
c@{\hspace{6.5pt} }l@{\hspace{6.5pt} }c@{\hspace{6.5pt} }c@{\hspace{6.5pt} }r@{\hspace{6.5pt} }r@{\hspace{6.5pt}}r@{}}
 \hline
 $\lambda_{\rm obs}$&Ion&$\lambda_{\rm lab}$&ID&$f$($\lambda$)&$I$($\lambda$)&$\delta$$I$($\lambda$)&
$\lambda_{\rm obs}$&Ion&$\lambda_{\rm lab}$&ID&$f$($\lambda$)&$I$($\lambda$)&$\delta$$I$($\lambda$)&
$\lambda_{\rm obs}$&Ion&$\lambda_{\rm lab}$&ID&$f$($\lambda$)&$I$($\lambda$)&$\delta$$I$($\lambda$)\\
 ({\AA})&&({\AA})&&&&&({\AA})&&({\AA})&&&&&({\AA})&&({\AA})\\
\hline
3688.46 & H19 & 3686.83 & 1 & 0.330 & 0.944 & 0.056 & 4543.65 & He\,{\sc ii}       & 4541.59 & 1 & 0.093 & 0.457 & 0.013 & 6208.10 & C  \,{\sc iii}      & 6205.55 & 1 & --0.250 & 0.014 & 0.002 \\ 
3693.17 & H18 & 3691.55 & 1 & 0.329 & 1.008 & 0.056 & 4573.11 & Mg\,{\sc i}$]$       & 4571.10 & 1 & 0.084 & 0.293 & 0.085 & 6236.64 & He\,{\sc ii}       & 6233.82 & 1 & --0.254 & 0.037 & 0.004 \\ 
3698.78 & H17 & 3697.15 & 1 & 0.328 & 1.125 & 0.060 & 4598.10 & O\,{\sc ii} & 4596.18 & 1 & 0.077 & 0.018 & 0.002 & 6275.14 & Ne\,{\sc ii} & 6272.43 & 1 & --0.259 & 0.021 & 0.005 \\ 
3704.33 & He\,{\sc i}        & 3702.62 & 1 & 0.327 & 0.074 & 0.009 & 4622.41 & C\,{\sc ii} & 4620.19 & 1 & 0.069 & 0.011 & 0.002 & 6289.39 & C\,{\sc iii} & 6286.80 & 1 & --0.261 & 0.219 & 0.007 \\ 
3705.49 & H16 & 3703.85 & 1 & 0.327 & 1.306 & 0.065 & 4636.21 & N  \,{\sc iii}      & 4634.12 & 1 & 0.065 & 0.278 & 0.007 & 6302.62 & {\oi} & 6300.30 & 1 & --0.263 & 1.496 & 0.040 \\ 
3706.65 & He\,{\sc i} & 3704.98 & 1 & 0.327 & 0.746 & 0.038 & 4640.84 & O\,{\sc ii} & 4638.86 & 1 & 0.064 & 0.042 & 0.005 & 6303.55 & {\oi} & 6300.30 & 2 & --0.263 & 0.560 & 0.027 \\ 
3708.92 & He\,{\sc i} & 3707.34 & 1 & 0.326 & 0.089 & 0.011 & 4642.69 & N  \,{\sc iii}      & 4640.64 & 1 & 0.063 & 0.561 & 0.010 &   &   &   & T &   & 2.056 & 0.055 \\ 
3713.61 & H15 & 3711.97 & 1 & 0.325 & 1.508 & 0.075 & 4643.89 & O\,{\sc ii} & 4641.81 & 1 & 0.063 & 0.124 & 0.007 & 6314.67 & [S  \,{\sc iii}]    & 6312.10 & 1 & --0.264 & 0.996 & 0.025 \\ 
3715.71 & O\,{\sc ii} & 3713.96 & 1 & 0.325 & 0.029 & 0.009 & 4648.51 & C\,{\sc iv} & 4646.64 & 1 & 0.061 & 0.026 & 0.004 & 6315.29 & [S  \,{\sc iii}]    & 6312.10 & 2 & --0.264 & 0.425 & 0.014 \\ 
3716.73 & He\,{\sc ii}       & 3715.16 & 1 & 0.325 & 0.085 & 0.013 & 4649.49 & C  \,{\sc iii}      & 4647.42 & 1 & 0.061 & 0.231 & 0.005 &   &   &   & T &   & 1.421 & 0.034 \\ 
3723.50 & H14 & 3721.94 & 1 & 0.323 & 2.686 & 0.131 & 4649.64 & C\,{\sc iv} & 4647.70 & 1 & 0.061 & 0.117 & 0.013 & 6366.13 & {\oi} & 6363.78 & 1 & --0.271 & 0.532 & 0.014 \\ 
3727.49 & {\oii} & 3726.03 & 1 & 0.322 & 13.353 & 0.653 & 4651.19 & O\,{\sc ii} & 4649.13 & 1 & 0.061 & 0.111 & 0.004 & 6367.08 & {\oi} & 6363.78 & 2 & --0.271 & 0.144 & 0.006 \\ 
3727.96 & {\oii} & 3726.03 & 2 & 0.322 & 6.167 & 0.307 & 4652.33 & C  \,{\sc iii}      & 4650.25 & 1 & 0.060 & 0.173 & 0.005 &   &   &   & T &   & 0.676 & 0.018 \\ 
  &   &  & T &   & 19.519 & 0.938 & 4653.06 & O\,{\sc ii} & 4650.84 & 1 & 0.060 & 0.009 & 0.001 & 6377.50 & N\,{\sc iii} & 6374.65 & 1 & --0.273 & 0.026 & 0.008 \\ 
3730.26 & {\oii} & 3728.81 & 1 & 0.322 & 7.282 & 0.373 & 4653.45 &
				 O\,{\sc ii} & 4651.33 & 1 & 0.060 &
						 0.067 & 0.006 & 6396.60
							 & [Mn\,{\sc v}]      & 6393.50 & 1 & --0.275 & 0.026 & 0.004 \\ 
3730.72 & {\oii} & 3728.81 & 2 & 0.322 & 3.441 & 0.185 & 4659.97 & C  \,{\sc iv}       & 4658.20 & 1 & 0.058 & 0.045 & 0.008 & 6399.53 & O\,{\sc ii} & 6396.79 & 1 & --0.275 & 0.010 & 0.003 \\ 
  &   &  & T &   & 10.724 & 0.531 & 4660.57 & C  \,{\sc iv}       & 4658.20 & 2 & 0.058 & 0.058 & 0.008 & 6409.12 & He\,{\sc ii}       & 6406.38 & 1 & --0.277 & 0.084 & 0.005 \\ 
3734.58 & He\,{\sc i} & 3732.86 & 1 & 0.321 & 0.084 & 0.011 &   &   &  & T &   & 0.103 & 0.011 & 6464.69 & N  \,{\sc iii}       & 6461.71 & 1 & --0.284 & 0.057 & 0.004 \\ 
3736.02 & H13 & 3734.37 & 1 & 0.321 & 2.269 & 0.111 & 4663.69 & O\,{\sc ii} & 4661.63 & 1 & 0.057 & 0.034 & 0.002 & 6529.92 & He\,{\sc ii}       & 6527.10 & 1 & --0.293 & 0.066 & 0.005 \\ 
3747.71 & He\,{\sc i} & 3745.94 & 1 & 0.318 & 0.019 & 0.005 & 4668.00 & C \,{\sc iv}        & 4665.26 & 1 & 0.056 & 0.015 & 0.002 & 6550.54 & {\nii} & 6548.04 & 1 & --0.296 & 3.899 & 0.100 \\ 
3751.81 & H12 & 3750.15 & 1 & 0.317 & 2.818 & 0.135 & 4675.70 & O\,{\sc ii} & 4673.73 & 1 & 0.053 & 0.017 & 0.003 & 6551.40 & {\nii} & 6548.04 & 2 & --0.296 & 1.605 & 0.049 \\ 
3756.39 & O \,{\sc iii}      & 3754.70 & 1 & 0.316 & 0.396 & 0.027 & 4678.32 & O\,{\sc ii} & 4676.23 & 1 & 0.053 & 0.037 & 0.004 &   &   &   & T &   & 5.504 & 0.140 \\ 
3758.94 & He\,{\sc i} & 3757.67 & 1 & 0.316 & 0.137 & 0.018 & 4687.64 & He\,{\sc ii}       & 4685.71 & 1 & 0.050 & 5.776 & 0.109 & 6563.07 & He\,{\sc ii}       & 6560.10 & 1 & --0.297 & 2.036 & 0.068 \\ 
3761.55 & O \,{\sc iii}      & 3759.88 & 1 & 0.315 & 1.194 & 0.060 & 4688.05 & He\,{\sc ii}       & 4685.71 & 2 & 0.050 & 5.361 & 0.098 & 6565.69 & H3 & 6562.80 & 1 & --0.298 & 285.000 & 5.321 \\ 
3770.55 & He\,{\sc i} & 3768.78 & 1 & 0.313 & 0.060 & 0.010 &   &   &  & T &   & 11.137 & 0.157 & 6580.97 & C\,{\sc ii} & 6578.05 & 1 & --0.300 & 0.211 & 0.008 \\ 
3772.30 & H11 & 3770.63 & 1 & 0.313 & 3.574 & 0.168 & 4703.69 & [Fe \,{\sc iii}]    & 4701.53 & 1 & 0.045 & 0.022 & 0.003 & 6585.94 & {\nii} & 6583.46 & 1 & --0.300 & 11.488 & 0.303 \\ 
3775.72 & O\,{\sc ii} & 3774.00 & 1 & 0.312 & 0.085 & 0.014 & 4713.37 & [Ar \,{\sc iv}]     & 4711.37 & 1 & 0.042 & 1.240 & 0.021 & 6586.81 & {\nii} & 6583.46 & 2 & --0.300 & 4.714 & 0.127 \\ 
3783.35 & He\,{\sc ii}       & 3781.68 & 1 & 0.311 & 0.054 & 0.009 & 4713.70 & [Ar \,{\sc iv}]     & 4711.37 & 2 & 0.042 & 0.618 & 0.014 &   &   &   & T &   & 16.202 & 0.413 \\ 
3792.94 & O\,{\sc iii}      & 3791.27 & 1 & 0.309 & 0.107 & 0.009 &   &   &  & T &   & 1.858 & 0.026 & 6680.98 & He\,{\sc i} & 6678.15 & 1 & --0.313 & 3.038 & 0.086 \\ 
3799.58 & H10 & 3797.90 & 1 & 0.307 & 4.821 & 0.225 & 4715.14 & He\,{\sc i} & 4713.14 & 1 & 0.042 & 0.504 & 0.017 & 6681.55 & He\,{\sc i} & 6678.15 & 2 & --0.313 & 0.937 & 0.039 \\ 
3815.15 & He\,{\sc ii}       & 3813.50 & 1 & 0.304 & 0.077 & 0.017 & 4715.52 & He\,{\sc i} & 4713.14 & 2 & 0.042 & 0.245 & 0.017 &   &   &   & T &   & 3.975 & 0.112 \\ 
3821.31 & He\,{\sc i} & 3819.60 & 1 & 0.302 & 1.114 & 0.052 &   &   &  & T &   & 0.749 & 0.024 & 6686.25 & He\,{\sc ii}       & 6683.20 & 1 & --0.313 & 0.092 & 0.007 \\ 
3835.34 & He\,{\sc ii}       & 3833.80 & 1 & 0.299 & 0.115 & 0.018 & 4726.26 & [Ne \,{\sc iv}]     & 4724.17 & 1 & 0.038 & 0.013 & 0.002 & 6718.96 & {\sii} & 6716.44 & 1 & --0.318 & 0.819 & 0.028 \\ 
3837.08 & H9 & 3835.38 & 1 & 0.299 & 6.471 & 0.292 & 4727.76 & [Ne \,{\sc iv}]     & 4725.64 & 1 & 0.038 & 0.011 & 0.003 & 6719.69 & {\sii} & 6716.44 & 2 & --0.318 & 0.759 & 0.030 \\ 
3859.77 & He\,{\sc ii}       & 3858.07 & 1 & 0.294 & 0.059 & 0.008 & 4742.20 & [Ar \,{\sc iv}]     & 4740.16 & 1 & 0.034 & 1.549 & 0.030 &   &   &   & T &   & 1.579 & 0.050 \\ 
3869.64 & He\,{\sc i} & 3867.47 & 1 & 0.291 & 0.547 & 0.403 & 4742.53 & [Ar \,{\sc iv}]     & 4740.16 & 2 & 0.034 & 1.168 & 0.022 & 6733.38 & {\sii} & 6730.81 & 1 & --0.320 & 1.744 & 0.049 \\ 
3870.40 & [Ne \,{\sc iii}]    & 3869.06 & 1 & 0.291 & 62.470 & 2.917 &   &   &  & T &   & 2.718 & 0.038 & 6734.25 & {\sii} & 6730.81 & 2 & --0.320 & 0.894 & 0.026 \\ 
3870.72 & [Ne \,{\sc iii}]    & 3869.06 & 2 & 0.291 & 23.243 & 1.367 & 4771.41 & O\,{\sc ii} & 4769.25 & 1 & 0.026 & 0.017 & 0.004 &   &   &   & T &   & 2.637 & 0.071 \\ 
  &   &  & T &   & 85.713 & 3.942 & 4804.70 & Ne\,{\sc ii} & 4802.58 & 1 & 0.016 & 0.018 & 0.003 & 6747.42 & C  \,{\sc iii}      & 6744.39 & 1 & --0.322 & 0.027 & 0.005 \\ 
3873.51 & He\,{\sc i} & 3871.79 & 1 & 0.290 & 0.101 & 0.012 & 4861.29 & He\,{\sc ii}       & 4859.32 & 1 & 0.001 & 0.269 & 0.095 & 6783.13 & C\,{\sc ii} & 6779.94 & 1 & --0.326 & 0.026 & 0.004 \\ 
3890.10 & H8 & 3889.05 & 1 & 0.287 & 1.400 & 0.123 & 4861.74 & He\,{\sc ii}       & 4859.32 & 2 & 0.000 & 0.430 & 0.096 & 6798.08 & [K \,{\sc iv}]      & 6795.10 & 1 & --0.328 & 0.030 & 0.004 \\ 
3890.63 & H8 & 3889.05 & 2 & 0.286 & 17.390 & 0.772 &   &   &  & T &   & 0.699 & 0.135 & 6830.07 & [Kr\,{\sc iii}] & 6826.85 & 1 & --0.333 & 0.030 & 0.007 \\ 
  &   &  & T &   & 18.790 & 0.834 & 4863.50 & H4 & 4861.32 & 1 & 0.000 & 100.000 & 0.882 & 6894.08 & He\,{\sc ii}       & 6890.90 & 1 & --0.341 & 0.117 & 0.011 \\ 
3922.37 & C\,{\sc ii} & 3920.68 & 1 & 0.279 & 0.034 & 0.006 & 4875.67 & N  \,{\sc iii}      & 4873.60 & 1 & --0.003 & 0.015 & 0.006 & 6992.04 & Ca\,{\sc i} & 6988.95 & 1 & --0.354 & 0.023 & 0.003 \\ 
3925.24 & He\,{\sc ii}       & 3923.48 & 1 & 0.278 & 0.097 & 0.009 & 4883.18 & [Fe \,{\sc iii}]    & 4881.00 & 1 & --0.005 & 0.032 & 0.002 & 7009.13 & $[$Ar\,{\sc v}$]$ & 7005.40 & 1 & --0.356 & 0.030 & 0.003 \\ 
3928.30 & He\,{\sc i} & 3926.54 & 1 & 0.277 & 0.116 & 0.008 & 4915.78 & O\,{\sc i} & 4913.60 & 1 & --0.014 & 0.040 & 0.007 & 7065.38 & He\,{\sc i} & 7062.28 & 1 & --0.364 & 0.017 & 0.003 \\ 
3966.48 & He\,{\sc i} & 3964.73 & 1 & 0.267 & 0.786 & 0.034 & 4924.11 & He\,{\sc i} & 4921.93 & 1 & --0.016 & 1.270 & 0.040 & 7068.11 & He\,{\sc i} & 7065.18 & 1 & --0.364 & 3.198 & 0.148 \\ 
3969.10 & [Ne \,{\sc iii}]    & 3967.79 & 1 & 0.267 & 13.079 & 0.884 & 4926.86 & O\,{\sc ii} & 4924.53 & 1 & --0.017 & 0.011 & 0.004 & 7068.59 & He\,{\sc i} & 7065.18 & 2 & --0.364 & 4.287 & 0.177 \\ 
3969.40 & [Ne \,{\sc iii}]    & 3967.79 & 2 & 0.267 & 13.068 & 0.880 & 4933.33 & [O  \,{\sc iii}]    & 4931.23 & 1 & --0.019 & 0.087 & 0.003 &   &   &   & T &   & 7.484 & 0.277 \\ 
  &   &  & T &   & 26.147 & 1.438 & 4933.76 & [O  \,{\sc iii}]    & 4931.23 & 2 & --0.019 & 0.031 & 0.003 & 7138.71 & [Ar \,{\sc iii}]    & 7135.80 & 1 & --0.374 & 7.116 & 0.159 \\ 
3971.83 & H7 & 3970.07 & 1 & 0.266 & 14.345 & 0.608 &   &   &  & T &   & 0.118 & 0.005 & 7139.36 & [Ar \,{\sc iii}]    & 7135.80 & 2 & --0.374 & 3.710 & 0.093 \\ 
4011.03 & He\,{\sc i} & 4009.26 & 1 & 0.256 & 0.192 & 0.010 & 4955.35 & N\,{\sc ii} & 4952.82 & 1 & --0.024 & 0.067 & 0.034 &   &   &   & T &   & 10.826 & 0.229 \\ 
4025.78 & He\,{\sc i} & 4023.98 & 1 & 0.252 & 0.033 & 0.008 & 4961.04 & [O  \,{\sc iii}]    & 4958.91 & 1 & --0.026 & 266.598 & 3.656 & 7180.68 & He\,{\sc ii}       & 7177.52 & 1 & --0.379 & 0.166 & 0.028 \\ 
4027.87 & He\,{\sc i} & 4026.18 & 1 & 0.251 & 1.049 & 0.055 & 4961.46 & [O  \,{\sc iii}]    & 4958.91 & 2 & --0.026 & 95.810 & 2.772 & 7240.00 & Ca\,{\sc i} & 7236.81 & 1 & --0.387 & 0.494 & 0.091 \\ 
4028.17 & He\,{\sc i} & 4026.18 & 2 & 0.251 & 1.035 & 0.042 &   &   &   & T &   & 362.407 & 4.612 & 7284.59 & He\,{\sc i} & 7281.35 & 1 & --0.393 & 1.011 & 0.032 \\ 
  &   &  & T &   & 2.084 & 0.087 & 5008.93 & [O  \,{\sc iii}]    & 5006.84 & 1 & --0.038 & 581.111 & 6.375 & 7321.86 & {\oii} & 7318.92 & 1 & --0.398 & 0.546 & 0.028 \\ 
4070.16 & {\sii} & 4068.60 & 1 & 0.239 & 0.732 & 0.028 & 5009.34 & [O  \,{\sc iii}]    & 5006.84 & 2 & --0.038 & 481.000 & 5.876 & 7322.85 & {\oii} & 7318.92 & 2 & --0.398 & 1.292 & 0.038 \\ 
4070.71 & {\sii} & 4068.60 & 2 & 0.239 & 0.375 & 0.017 &   &   &   & T &   & 1062.111 & 8.960 & 7323.67 & {\oii} & 7318.92 & 3 & --0.398 & 0.951 & 0.037 \\ 
  &   &  & T &   & 1.107 & 0.042 & 5017.90 & He\,{\sc i} & 5015.68 & 1 & --0.040 & 2.233 & 0.028 &   &   &   & T &   & 2.789 & 0.074 \\ 
4071.99 & O\,{\sc ii} & 4069.89 & 1 & 0.239 & 0.245 & 0.014 & 5049.96 & He\,{\sc i} & 5047.74 & 1 & --0.048 & 0.239 & 0.010 & 7332.46 & {\oii} & 7329.66 & 1 & --0.400 & 0.754 & 0.030 \\ 
4074.02 & O\,{\sc ii} & 4072.15 & 1 & 0.238 & 0.081 & 0.009 & 5133.34 & C \,{\sc iii}       & 5130.86 & 1 & --0.067 & 0.080 & 0.006 & 7333.44 & {\oii} & 7329.66 & 2 & --0.400 & 0.928 & 0.039 \\ 
4077.89 & {\sii} & 4076.35 & 1 & 0.237 & 0.284 & 0.013 & 5160.89 & $[$Fe\,{\sc ii}$]$ & 5158.78 & 1 & --0.073 & 0.010 & 0.002 & 7334.33 & {\oii} & 7329.66 & 3 & --0.400 & 0.467 & 0.031 \\ 
4078.46 & {\sii} & 4076.35 & 2 & 0.237 & 0.076 & 0.005 & 5182.38 & C  \,{\sc iii}      & 5179.90 & 1 & --0.078 & 0.021 & 0.005 &   &   &   & T &   & 2.150 & 0.068 \\ 
  &   &   & T &   & 0.361 & 0.015 & 5194.02 & [Ar \,{\sc iii}]    & 5191.82 & 1 & --0.081 & 0.126 & 0.006 & 7533.84 & C\,{\sc ii} & 7530.57 & 1 & --0.426 & 0.224 & 0.018 \\ 
4099.06 & N\,{\sc iii}      & 4097.35 & 1 & 0.231 & 0.355 & 0.016 & 5199.84 & {\Ni} & 5197.90 & 1 & --0.082 & 0.054 & 0.002 & 7750.93 & Fe\,{\sc i} & 7747.60 & 1 & --0.455 & 0.140 & 0.015 \\ 
4099.35 & N\,{\sc iii}      & 4097.35 & 2 & 0.231 & 0.157 & 0.010 & 5200.57 & {\Ni} & 5197.90 & 2 & --0.082 & 0.027 & 0.003 & 7754.24 & [Ar \,{\sc iii}]    & 7751.10 & 1 & --0.455 & 1.535 & 0.051 \\ 
  &   &  & T &   & 0.513 & 0.022 &   &   &  & T &   & 0.080 & 0.003 & 7754.94 & [Ar \,{\sc iii}]    & 7751.10 & 2 & --0.455 & 1.172 & 0.046 \\ 
4101.87 & He\,{\sc ii}       & 4100.04 & 1 & 0.230 & 0.117 & 0.043 & 5202.12 & {\Ni} & 5200.26 & 1 & --0.083 & 0.023 & 0.002 &   &   &   & T &   & 2.707 & 0.081 \\ 
4103.42 & H6 & 4101.73 & 1 & 0.230 & 12.501 & 0.454 & 5202.73 & {\Ni} & 5200.26 & 2 & --0.083 & 0.033 & 0.004 & 7806.18 & [V\,{\sc iii}] & 7802.76 & 1 & --0.462 & 0.077 & 0.023 \\ 
4103.72 & H6 & 4101.73 & 2 & 0.229 & 10.173 & 0.394 &   &   &  & T &   & 0.057 & 0.004 & 7819.78 & He\,{\sc i} & 7816.14 & 1 & --0.464 & 0.163 & 0.035 \\ 
  &   &  & T &   & 22.675 & 0.802 & 5243.20 & N\,{\sc ii} & 5240.86 & 1 & --0.091 & 0.018 & 0.004 & 8011.56 & Fe\,{\sc iii} & 8007.90 & 1 & --0.488 & 0.148 & 0.031 \\ 
4105.19 & N\,{\sc iii}      & 4103.39 & 1 & 0.229 & 0.183 & 0.016 & 5272.75 & [Fe \,{\sc iii}]    & 5270.40 & 1 & --0.098 & 0.040 & 0.005 & 8025.93 & He\,{\sc i} & 8021.30 & 1 & --0.490 & 0.141 & 0.018 \\ 
4121.13 & O\,{\sc ii} & 4119.22 & 1 & 0.224 & 0.036 & 0.007 & 5325.45 & [Cl \,{\sc iv}]     & 5322.99 & 1 & --0.108 & 0.021 & 0.004 & 8049.26 & [Cl \,{\sc iv}]     & 8046.30 & 1 & --0.492 & 0.557 & 0.026 \\ 
4122.66 & He\,{\sc i} & 4120.81 & 1 & 0.224 & 0.271 & 0.011 & 5344.85 & C\,{\sc ii} & 5342.43 & 1 & --0.112 & 0.042 & 0.003 & 8240.55 & He\,{\sc ii}       & 8236.79 & 1 & --0.515 & 0.460 & 0.038 \\ 
4130.60 & N\,{\sc ii} & 4128.66 & 1 & 0.222 & 0.034 & 0.006 & 5348.29 & [Kr \,{\sc iv}]     & 5346.02 & 1 & --0.113 & 0.093 & 0.008 & 8302.46 & P28 & 8298.83 & 1 & --0.522 & 0.160 & 0.020 \\ 
4145.59 & He\,{\sc i} & 4143.76 & 1 & 0.217 & 0.342 & 0.013 & 5377.63 & Ne\,{\sc ii} & 5375.14 & 1 & --0.119 & 0.014 & 0.002 & 8310.10 & P27 & 8306.11 & 1 & --0.523 & 0.204 & 0.026 \\ 
4158.28 & N\,{\sc ii} & 4156.36 & 1 & 0.213 & 0.029 & 0.006 & 5413.96 & He\,{\sc ii}       & 5411.52 & 1 & --0.126 & 1.036 & 0.023 & 8317.78 & P26 & 8314.26 & 1 & --0.524 & 0.235 & 0.028 \\ 
4164.82 & C  \,{\sc iii}      & 4162.88 & 1 & 0.211 & 0.020 & 0.006 & 5440.51 & Cu\,{\sc i} & 5438.08 & 1 & --0.131 & 0.022 & 0.005 & 8327.03 & P25 & 8323.42 & 1 & --0.525 & 0.207 & 0.020 \\ 
4170.84 & He\,{\sc i} & 4168.97 & 1 & 0.210 & 0.048 & 0.005 & 5520.13 & [Cl \,{\sc iii}]    & 5517.72 & 1 & --0.145 & 0.251 & 0.007 & 8337.53 & P24 & 8333.78 & 1 & --0.526 & 0.261 & 0.025 \\ 
4182.98 & Ar\,{\sc ii} & 4181.23 & 1 & 0.206 & 0.016 & 0.005 & 5540.28 & [Cl \,{\sc iii}]    & 5537.89 & 1 & --0.149 & 0.365 & 0.009 & 8349.30 & P23 & 8345.55 & 1 & --0.527 & 0.205 & 0.018 \\ 
4188.74 & C  \,{\sc iii}      & 4186.90 & 1 & 0.204 & 0.150 & 0.009 & 5579.42 & [I \,{\sc iii}] & 5576.94 & 1 & --0.155 & 0.015 & 0.004 & 8362.67 & P22 & 8359.00 & 1 & --0.529 & 0.297 & 0.025 \\ 
4188.90 & N \,{\sc iii} & 4187.06 & 1 & 0.204 & 0.314 & 0.059 & 5756.82 & {\nii} & 5754.64 & 1 & --0.185 & 0.293 & 0.010 & 8365.14 & He\,{\sc i} & 8361.73 & 1 & --0.529 & 0.136 & 0.020 \\ 
4197.62 & N  \,{\sc iii}      & 4195.74 & 1 & 0.202 & 0.015 & 0.007 & 5757.56 & {\nii} & 5754.64 & 2 & --0.185 & 0.106 & 0.007 & 8371.16 & P\,{\sc iv} & 8367.50 & 1 & --0.530 & 0.078 & 0.018 \\ 
4201.73 & He\,{\sc ii}       & 4199.83 & 1 & 0.200 & 0.236 & 0.010 &   &   &   & T &   & 0.399 & 0.013 & 8378.14 & P21 & 8374.48 & 1 & --0.531 & 0.294 & 0.018 \\ 
4229.54 & $[$Fe\,{\sc v}$]$ & 4227.19 & 1 & 0.192 & 0.040 & 0.005 & 5804.05 & C  \,{\sc iv}       & 5801.34 & 1 & --0.192 & 0.053 & 0.004 & 8396.10 & P20 & 8392.40 & 1 & --0.533 & 0.297 & 0.017 \\ 
4269.06 & C\,{\sc ii} & 4267.18 & 1 & 0.180 & 0.469 & 0.015 & 5808.55 & C  \,{\sc iv}       & BC & 2 & --0.193 & 0.660 & 0.056 & 8417.00 & P19 & 8413.32 & 1 & --0.535 & 0.302 & 0.019 \\ 
4305.64 & O\,{\sc ii} & 4303.82 & 1 & 0.168 & 0.018 & 0.006 & 5814.59 & C  \,{\sc iv}       & 5012.00 & 3 & --0.193 & 0.066 & 0.004 & 8441.71 & P18 & 8437.95 & 1 & --0.537 & 0.368 & 0.015 \\ 
4327.74 & O\,{\sc ii} & 4325.76 & 1 & 0.161 & 0.037 & 0.008 &  &  &  & T &  & 0.779 & 0.056 & 8470.98 & P17 & 8467.25 & 1 & --0.541 & 0.428 & 0.018 \\ 
4340.64 & He\,{\sc ii}       & 4338.67 & 1 & 0.157 & 0.341 & 0.234 & 5870.36 & [Kr \,{\sc iv}]     & 5867.74 & 1 & --0.202 & 0.111 & 0.005 & 8506.15 & P16 & 8502.48 & 1 & --0.544 & 0.535 & 0.033 \\ 
4342.39 & H5 & 4340.46 & 1 & 0.157 & 46.900 & 0.724 & 5878.10 & He\,{\sc i} & 5875.60 & 1 & --0.203 & 9.704 & 0.227 & 8549.06 & P15 & 8545.38 & 1 & --0.549 & 0.659 & 0.048 \\ 
4351.39 & O\,{\sc ii} & 4349.43 & 1 & 0.154 & 0.022 & 0.004 & 5878.59 & He\,{\sc i} & 5875.60 & 2 & --0.203 & 5.494 & 0.172 & 8602.21 & P14 & 8598.39 & 1 & --0.554 & 0.738 & 0.032 \\ 
4365.13 & [O  \,{\sc iii}]    & 4363.21 & 1 & 0.149 & 13.397 & 0.208 &   &   &   & T &   & 15.198 & 0.331 & 8668.84 & P13 & 8665.02 & 1 & --0.560 & 0.937 & 0.036 \\ 
4381.09 & N  \,{\sc iii}      & 4379.20 & 1 & 0.144 & 0.039 & 0.005 & 5899.35 & He\,{\sc ii}       & 5896.78 & 1 & --0.206 & 0.025 & 0.003 & 8754.33 & P12 & 8750.47 & 1 & --0.568 & 1.172 & 0.040 \\ 
4389.75 & He\,{\sc i} & 4387.93 & 1 & 0.142 & 0.302 & 0.011 & 5955.65 & He\,{\sc ii}       & 5952.93 & 1 & --0.214 & 0.024 & 0.004 & 9018.92 & P10 & 9014.91 & 1 & --0.590 & 1.874 & 0.061 \\ 
4390.06 & He\,{\sc i} & 4387.93 & 2 & 0.142 & 0.269 & 0.012 & 6007.32 & He\,{\sc ii}       & 6004.72 & 1 & --0.222 & 0.022 & 0.005 & 9071.93 & [S \,{\sc iii}]     & 9068.60 & 1 & --0.594 & 0.759 & 0.071 \\ 
  &   &  & T &   & 0.570 & 0.018 & 6039.29 & He\,{\sc ii}       & 6036.78 & 1 & --0.226 & 0.018 & 0.003 & 9072.70 & [S \,{\sc iii}]     & 9068.60 & 2 & --0.594 & 5.244 & 0.215 \\ 
4394.14 & Ne\,{\sc ii} & 4391.99 & 1 & 0.140 & 0.019 & 0.005 & 6076.85 & He\,{\sc ii}       & 6074.19 & 1 & --0.232 & 0.017 & 0.003 & 9073.07 & [S \,{\sc iii}]     & 9068.60 & 3 & --0.594 & 1.983 & 0.613 \\ 
4439.48 & He\,{\sc i} & 4437.55 & 1 & 0.126 & 0.073 & 0.005 & 6104.44 & [K  \,{\sc iv}]     & 6101.79 & 1 & --0.235 & 0.141 & 0.004 &   &   &   & T &   & 7.985 & 0.653 \\ 
4473.48 & He\,{\sc i} & 4471.47 & 1 & 0.115 & 5.167 & 0.126 & 6120.93 & He\,{\sc ii}       & 6118.26 & 1 & --0.238 & 0.038 & 0.005 &  &  &  &  &  &  &  \\ 
4512.86 & [K  \,{\sc iv}]     & 4510.92 & 1 & 0.103 & 0.024 & 0.003 & 6173.48 & He\,{\sc ii}       & 6170.69 & 1 & --0.245 & 0.045 & 0.007 &  &  &  &  &  &  &  \\

 \hline
\end{tabularx}
\end{table*}

\begin{table}
\caption{Line and band flux measurements in the \emph{Spitzer}
 spectrum.\label{spt_tbl}}
 \begin{tabularx}{\columnwidth}{l@{\hspace{8pt}}l@{\hspace{8pt}}c@{\hspace{8pt}}c@{\hspace{8pt}}c@{\hspace{8pt}}c}
\hline
$\lambda_{\rm obs}$&Ions/&$\lambda_{\rm
 lab}$&$F$($\lambda$)&$f$($\lambda$)&$I$($\lambda$)/$I$({\hb})\\
  ($\mu$m)&PAH&($\mu$m)&(erg s$^{-1}$ cm$^{-2}$)&&($I$({\hb})=100)\\
  \hline
10.50 & [S\,{\sc iv}] & 10.51 & 5.42(--13)$\pm$3.63(--14) & --0.959 & 27.796$\pm$2.340 \\ 
11.28 & PAH & 11.30 & 4.39(--14)$\pm$3.40(--15) & --0.970 & ~~2.245$\pm$0.208 \\ 
12.38 & H\,{\sc i} & 12.40 & 4.44(--14)$\pm$5.68(--15) & --0.980 & ~~2.266$\pm$0.312 \\ 
12.80 & [Ne\,{\sc ii}] & 12.80 & 2.10(--14)$\pm$2.45(--15) & --0.983 & ~~1.069$\pm$0.136 \\ 
15.56 & [Ne\,{\sc iii}] & 15.56 & 1.17(--12)$\pm$2.25(--14) & --0.985 & 59.760$\pm$3.249 \\ 
18.73 & [S\,{\sc iii}] & 18.71 & 1.94(--13)$\pm$5.97(--15) & --0.981 & ~~9.917$\pm$0.589 \\ 
25.92 & [O\,{\sc iv}] & 25.89 & 2.34(--13)$\pm$7.29(--15) & --0.989 & 11.921$\pm$0.711 \\ 
33.45 & [S\,{\sc iii}] & 33.50 & 8.49(--14)$\pm$2.31(--15) & --0.993 & ~~4.314$\pm$0.249 \\ 
36.02 & [Ne\,{\sc iii}] & 36.00 & 7.62(--14)$\pm$2.83(--15) & --0.993 & ~~3.874$\pm$0.244 \\
\hline
 \end{tabularx}
 \raggedright
Note -- $F$({\hb}) is (1.53$\pm$0.02)$\times$10$^{-12}$\,erg s$^{-1}$ cm$^{-2}$. \\ 
\end{table}

\section{Ionic Abundances derived from CELs and RLs, and the adopted ICFs.}

 \begin{table}
  \centering
  \caption{Ionic abundances from CELs.\label{celabund}}
\begin{tabularx}{\columnwidth}{@{}l@{\hspace{1.5pt}}c@{\hspace{1.5pt}}c@{\hspace{4pt}}c@{\hspace{1.5pt}}c@{\hspace{4pt}}c@{}}
\hline
X$^{\rm m+}$ &
$\lambda_{\rm lab}$&
$I$($\lambda_{\rm lab}$)/$I$({\hb})    &
$T_{\epsilon}$    &
$n_{\epsilon}$    &
X$^{\rm m+}$/H$^{+}$ \\
&
&
$[$$I$({\hb})=100$]$&
(K)&
		 (cm$^{-3}$)\\
 \hline    
N$^{0}$      & 5197.90\,{\AA}& 8.04(--2)$\pm$3.29(--3) & 10000 & 1010 & 6.11(--7)$\pm$2.50(--8) \\ 
             & 5200.26\,{\AA}& 5.65(--2)$\pm$4.29(--3) & 10000 & 1010 & 5.71(--7)$\pm$4.33(--8) \\ 
             &   &   &   &      & {\bf 5.94(--7)$\pm$3.26(--8)} \\ 	     
N$^{+}$ & 5754.64\,{\AA}& 3.99(--1)$\pm$1.25(--2) & 12090 & 3540 & 2.15(--6)$\pm$1.78(--7) \\ 
  & 6548.04\,{\AA}& 5.50(0)$\pm$1.40(--1) & 12090 & 3540 & 2.08(--6)$\pm$1.03(--7) \\ 
  & 6583.46\,{\AA}& 1.62(+1)$\pm$4.13(--1) & 12090 & 3540 & 2.07(--6)$\pm$1.02(--7) \\ 
  &   &   &   &      & {\bf 2.07(--6)$\pm$1.04(--7)} \\ 
O$^{+}$  & 3726.03\,{\AA}& 1.95(+1)$\pm$9.38(--1) & 12090 & 3540 & 8.33(--6)$\pm$6.83(--7) \\ 
  & 3728.81\,{\AA}& 1.07(+1)$\pm$5.31(--1) & 12090 & 3540 & 8.39(--6)$\pm$7.08(--7) \\ 
  & 7320/7330\,{\AA}& 4.08(0)$\pm$2.57(--1) & 12090 & 6870 &
 1.07(--5)$\pm$1.17(--6) \\ 
  &   &   &   &      & {\bf 8.63(--6)$\pm$7.49(--7)} \\ 
O$^{2+}$  & 4363.21\,{\AA}& 1.34(+1)$\pm$2.08(--1) & 12380 & 8820 &
 1.90(--4)$\pm$7.30(--6) \\ 
  & 4931.23\,{\AA}& 1.18(--1)$\pm$4.52(--3) & 12380 & 8820 & 1.57(--4)$\pm$6.67(--6) \\ 
  & 4958.91\,{\AA}& 3.62(+2)$\pm$4.61(0) & 12380 & 8820 & 1.88(--4)$\pm$4.22(--6) \\ 
  & 5006.84\,{\AA}& 1.06(+3)$\pm$8.96(0) & 12380 & 8820 & 1.91(--4)$\pm$3.91(--6) \\ 
  &   &   &   &      & {\bf 1.90(--4)$\pm$4.02(--6)} \\ 
O$^{3+}$  & 25.89\,$\mu$m& 1.19(+1)$\pm$7.11(--1) & 12800 & 11320 & {\bf 6.93(--6)$\pm$4.33(--7)} \\ 
Ne$^{+}$  & 12.81\,$\mu$m& 1.07(0)$\pm$1.36(--1) & 12070 & 7570 & {\bf 1.31(--6)$\pm$1.71(--7)} \\ 
Ne$^{2+}$     & 3869.06\,{\AA}& 8.57(+1)$\pm$3.94(0) & 12800 & 11320 & 3.75(--5)$\pm$3.26(--6) \\ 
  & 3967.79\,{\AA}& 2.61(+1)$\pm$1.44(0) & 12800 & 11320 & 3.80(--5)$\pm$3.49(--6) \\ 
  & 15.56\,$\mu$m& 5.98(+1)$\pm$3.25(0) & 12800 & 11320 & 3.76(--5)$\pm$2.10(--6) \\ 
  & 36.00\,$\mu$m& 3.87(0)$\pm$2.44(--1) & 12800 & 11320 & 3.26(--5)$\pm$2.11(--6) \\ 
  &   &   &   &      & {\bf 3.75(--5)$\pm$2.87(--6)} \\ 
Ne$^{3+}$      & 4724.17\,{\AA}& 1.28(--2)$\pm$2.36(--3) & 12800 & 11320 & 2.23(--6)$\pm$5.34(--7) \\ 
  & 4725.64\,{\AA}& 1.09(--2)$\pm$2.88(--3) & 12800 & 11320 & 1.98(--6)$\pm$6.07(--7) \\ 
  &   &   &   &      & {\bf 2.12(--6)$\pm$5.68(--7)} \\ 
Mg$^{0}$      & 4571.10\,{\AA}& 2.93(--1)$\pm$8.53(--2) & 10000 & 1010 &
		     {\bf 1.81(--8)$\pm$5.27(--9)} \\ 
S$^{+}$      & 4068.60\,{\AA}& 1.11(0)$\pm$4.17(--2) & 12090 & 3970 & 1.25(--7)$\pm$8.07(--9) \\ 
  & 6716.44\,{\AA}& 1.58(0)$\pm$5.01(--2) & 12090 & 3970 & 1.07(--7)$\pm$1.13(--8) \\ 
  & 6730.81\,{\AA}& 2.64(0)$\pm$7.14(--2) & 12090 & 3970 & 1.07(--7)$\pm$1.07(--8) \\ 
  &   &   &   &      & {\bf 1.11(--7)$\pm$1.03(--8)} \\ 
S$^{2+}$     & 6312.10\,{\AA}& 1.42(0)$\pm$3.44(--2) & 12070 & 6860 & 1.61(--6)$\pm$2.72(--7) \\ 
  & 9068.60\,{\AA}& 7.99(0)$\pm$6.53(--1) & 12070 & 6860 & 1.43(--6)$\pm$1.68(--7) \\ 
  & 18.71\,$\mu$m& 9.92(0)$\pm$5.89(--1) & 12070 & 6860 & 1.44(--6)$\pm$1.04(--7) \\ 
  & 33.50\,$\mu$m& 4.31(0)$\pm$2.49(--1) & 12070 & 6860 & 1.44(--6)$\pm$1.04(--7) \\ 
  &   &   &   &      & {\bf 1.44(--6)$\pm$1.36(--7)} \\ 
S$^{3+}$  & 10.51\,$\mu$m& 2.78(+1)$\pm$2.34(0) & 12380 & 8820 & {\bf 9.57(--7)$\pm$8.06(--8)} \\ 
Cl$^{2+}$     & 5517.72\,{\AA}& 2.51(--1)$\pm$6.81(--3) & 12070 & 8280 & 2.65(--8)$\pm$3.35(--9) \\ 
  & 5537.89\,{\AA}& 3.65(--1)$\pm$8.60(--3) & 12070 & 8280 & 2.64(--8)$\pm$3.26(--9) \\ 
  &   &   &   &      & {\bf 2.65(--8)$\pm$3.30(--9)} \\ 
Cl$^{3+}$      & 5322.99\,{\AA}& 2.08(--2)$\pm$4.28(--3) & 11800 & 8820 & 2.68(--8)$\pm$9.79(--9) \\ 
  & 8046.30\,{\AA}& 5.57(--1)$\pm$2.65(--2) & 11800 & 8820 & 2.78(--8)$\pm$4.41(--9) \\ 
  &   &   &   &      & {\bf 2.78(--8)$\pm$4.60(--9)} \\ 
Ar$^{2+}$     & 5191.82\,{\AA}& 1.26(--1)$\pm$5.62(--3) & 12000 & 7570 & 6.70(--7)$\pm$6.46(--8) \\ 
  & 7135.80\,{\AA}& 1.08(+1)$\pm$2.29(--1) & 12000 & 7570 & 6.66(--7)$\pm$3.09(--8) \\ 
  & 7751.10\,{\AA}& 2.71(0)$\pm$8.10(--2) & 12000 & 7570 & 6.94(--7)$\pm$3.55(--8) \\ 
  &   &   &   &      & {\bf 6.71(--7)$\pm$3.21(--8)} \\ 
Ar$^{3+}$      & 4711.37\,{\AA}& 1.86(0)$\pm$2.64(--2) & 12800 & 11320 & 3.25(--7)$\pm$2.18(--8) \\ 
  & 4740.16\,{\AA}& 2.72(0)$\pm$3.80(--2) & 12800 & 11320 & 3.28(--7)$\pm$2.10(--8) \\ 
  &   &   &   &      & {\bf 3.27(--7)$\pm$2.13(--8)} \\ 
Ar$^{4+}$       & 7005.40\,{\AA}& 2.99(--2)$\pm$3.40(--3) & 12800 & 11320
 & {\bf 3.11(--9)$\pm$3.87(--10)} \\ 
K$^{3+}$      & 6101.79\,{\AA}& 1.41(--1)$\pm$4.42(--3) & 12800 & 11320 & 6.62(--9)$\pm$4.84(--10) \\ 
  & 6795.10\,{\AA}& 3.02(--2)$\pm$3.78(--3) & 12800 & 11320 & 6.52(--9)$\pm$9.23(--10) \\ 
  &   &   &   &      & {\bf 6.60(--9)$\pm$5.61(--10)} \\ 
Fe$^{2+}$     & 4701.53\,{\AA}& 2.21(--2)$\pm$2.72(--3) & 12090 & 3540 & 1.49(--8)$\pm$1.98(--9) \\ 
 & 4881.00\,{\AA}& 3.17(--2)$\pm$2.07(--3) & 12090 & 3540 & 1.05(--8)$\pm$8.89(--10) \\ 
  & 5270.40\,{\AA}& 4.00(--2)$\pm$5.45(--3) & 12090 & 3540 & 1.91(--8)$\pm$2.74(--9) \\ 
  &   &   &   &      & {\bf 1.52(--8)$\pm$1.94(--9)} \\ 
Kr$^{2+}$  & 6826.85\,{\AA}& 2.99(--2)$\pm$7.37(--3) & 12070 & 7570 &
 {\bf 1.81(--9)$\pm$4.81(--10)} \\ 
Kr$^{3+}$      & 5867.74\,{\AA}& 1.11(--1)$\pm$5.07(--3) & 12380 & 8820 &
 {\bf 1.71(--9)$\pm$8.18(--11)}\\
\hline
\end{tabularx}
\raggedright
Note -- Corrected recombination contributions for the $[$O\,{\sc
 ii}$]$\,$\lambda\lambda$\,7320/30\,{\AA} and for the $[$O\,{\sc
 iii}$]$\,$\lambda$\,4363\,{\AA} lines.\\
\end{table}

 \begin{table}
  \centering
  \caption{Ionic abundances from RLs.\label{rlabund}}
\begin{tabularx}{\columnwidth}{@{}l@{\hspace{10pt}}c@{\hspace{10pt}}c@{\hspace{10pt}}c@{\hspace{10pt}}c@{}}
\hline
 X$^{\rm m+}$ &
$\lambda_{\rm lab}$&
Multi.&
$I$($\lambda_{\rm lab}$)    &
X$^{\rm m+}$/H$^{+}$ \\
&
&
&
$[$$I$({\hb})=100$]$\\
\hline     
He$^{+}$        & 5875.60\,{\AA} & V11 & 1.52(+1) $\pm$ 3.31(--1) & 9.56(--2) $\pm$ 6.75(--3) \\ 
        & 4471.47\,{\AA} & V14 & 5.17(0) $\pm$ 1.26(--1) & 9.68(--2) $\pm$ 2.12(--2) \\ 
        & 6678.15\,{\AA} & V46 & 3.97(0) $\pm$ 1.12(--1) & 9.89(--2) $\pm$ 6.35(--3) \\ 
        & 4921.93\,{\AA} & V48 & 1.27(0) $\pm$ 3.98(--2) & 9.55(--2) $\pm$ 5.47(--3) \\ 
        & 4387.93\,{\AA} & V51 & 5.70(--1) $\pm$ 1.81(--2) & 9.60(--2) $\pm$ 3.05(--3) \\ 
 &  &  &  &   {\bf 9.63(--2) $\pm$ 9.40(--3)} \\ 
He$^{2+}$       & 4685.71\,{\AA} & V3.4 & 1.11(+1) $\pm$ 1.57(--1) & {\bf 9.35(--3) $\pm$ 2.16(--3)} \\ 
C$^{2+}$       & 6578.05\,{\AA} & V2 & 2.11(--1) $\pm$ 7.76(--3) & 4.70(--4) $\pm$ 1.63(--4) \\ 
       & 4267.18\,{\AA} & V6 & 4.69(--1) $\pm$ 1.46(--2) & 4.73(--4) $\pm$ 1.29(--4) \\ 
 &  &  &    & {\bf 4.72(--4) $\pm$ 1.40(--4)} \\ 
C$^{3+}$      & 4647.42\,{\AA} & V1 & 2.31(--1) $\pm$ 4.84(--3) & 3.65(--4) $\pm$ 7.48(--5) \\ 
      & 4650.25\,{\AA} & V1 & 1.73(--1) $\pm$ 5.40(--3) &3.43(--4) $\pm$ 7.07(--5) \\ 
      & 4186.90\,{\AA} & V18 & 1.50(--1) $\pm$ 8.67(--3) & 2.69(--4) $\pm$ 6.86(--5) \\ 
 &  &  &    & {\bf 3.32(--4) $\pm$ 7.18(--5)} \\ 
C$^{4+}$       & 4658.20\,{\AA} & V8 & 1.03(--1) $\pm$ 1.13(--2) & {\bf 2.45(--5) $\pm$ 6.75(--6)} \\ 
O$^{2+}$       & 4638.86\,{\AA} & V1 & 4.15(--2) $\pm$ 4.79(--3) & 3.26(--4) $\pm$ 8.15(--5) \\ 
       & 4649.13\,{\AA} & V1 & 1.11(--1) $\pm$ 4.13(--3) & 2.44(--4) $\pm$ 5.43(--5) \\ 
       & 4661.63\,{\AA} & V1 & 3.44(--2) $\pm$ 1.69(--3) & 2.38(--4) $\pm$ 5.43(--5) \\ 
       & 4676.23\,{\AA} & V1 & 3.73(--2) $\pm$ 3.56(--3) & 3.90(--4) $\pm$ 9.56(--5) \\ 
       & 4349.43\,{\AA} & V2 & 2.15(--2) $\pm$ 4.01(--3) & 2.32(--4) $\pm$ 6.62(--5) \\ 
       & 4072.15\,{\AA} & V10 & 8.08(--2) $\pm$ 8.75(--3) & 3.44(--4) $\pm$ 8.50(--5) \\ 
       & 4119.22\,{\AA} & V20 & 3.61(--2) $\pm$ 7.37(--3) & 4.29(--4) $\pm$ 1.27(--4) \\ 
       & 4924.53\,{\AA} & V28 & 1.15(--2) $\pm$ 4.15(--3) & 2.66(--4) $\pm$ 1.13(--4) \\ 
       & 4303.82\,{\AA} & V53a & 1.76(--2) $\pm$ 5.83(--3) & 4.18(--4) $\pm$ 1.72(--4) \\ 
 &  &  &  &   {\bf 3.12(--4) $\pm$ 8.19(--5) }\\
 \hline
\end{tabularx}
\end{table}

  \begin{table}
   \centering
\caption{Adopted ionisation correction factors (ICFs).\label{icf}}
\begin{tabularx}{\columnwidth}{@{}l@{\hspace{35pt}}c@{\hspace{35pt}}c@{\hspace{35pt}}l@{}}
  \hline
X&
Line&
ICF(X)&
X/H\\
\hline
He &RL &1 & He$^{+}$+He$^{2+}$\\
\noalign{\smallskip}
C  &RL &1 & C$^{2+}$+C$^{3+}$+C$^{4}$\\
\noalign{\smallskip}
N &CEL &$\rm \left(\frac{O}{O^{+}}\right)_{CEL}$&ICF(N)N$^{+}$\\
\noalign{\smallskip}
O  &CEL & 1 &O$^{+}$+O$^{2+}$+O$^{3+}$\\
   &RL &$\rm \left(\frac{O}{O^{2+}}\right)_{CEL}$&ICF(O)O$^{2+}$\\
\noalign{\smallskip}
Ne &CEL&1&{$\rm Ne^{+}+Ne^{2+}+Ne^{3+}$}\\
\noalign{\smallskip}
S  &CEL &1 &$\rm S^{+}+S^{2+}+S^{3+}$\\
\noalign{\smallskip}
Cl &CEL &1  &Cl$^{2+}$+Cl$^{3+}$\\
\noalign{\smallskip}
Ar &CEL &1&$\rm Ar^{2+}+Ar^{3+}$\\
\noalign{\smallskip}
K &CEL &$\rm \left(\frac{Ar}{Ar^{3+}}\right)$&ICF(K)K$^{3+}$\\
\noalign{\smallskip}
Fe &CEL &$\rm \left(\frac{O}{O^{+}}\right)_{CEL}$ &ICF(Fe)Fe$^{2+}$\\
\noalign{\smallskip}
Kr &CEL &1 &Kr$^{2+}$ + Kr$^{3+}$\\
   \hline
  \end{tabularx}
  \end{table}

\section{The classifications of the PAHs and the measurements of the
 PAHs and dust features, the figure for a explanation of the
 15-18 micron plateau and 16.4 micron PAH}

  \begin{table*}
   \centering
   \caption{Classification of the 6.2/7.7/8.6\,$\mu$m PAH and
 the 10-14\,$\mu$m PAH bands and the measurements of the
 broad 11/16-24/30\,$\mu$m features. 
\label{band_measure}}
 \begin{tabularx}{\textwidth}{@{}l@{\hspace{15pt}}c@{\hspace{15pt}}
  c@{\hspace{15pt}}c@{\hspace{15pt}}c@{\hspace{15pt}}c@{\hspace{15pt}}
  c@{\hspace{15pt}}c@{\hspace{15pt}}c@{\hspace{15pt}}c@{\hspace{15pt}}c@{}}
\hline
PNe&
6.2\,$\mu$m&
7.7\,$\mu$m& 
8.6\,$\mu$m&
10-14\,$\mu$m& 
\multicolumn{2}{c}{11\,$\mu$m feature}& 
\multicolumn{2}{c}{16-24\,$\mu$m feature} &
\multicolumn{2}{c}{30\,$\mu$m feature}\\
& 
PAH&
PAH&
PAH&
PAH&
$\lambda_{c}$&FWHM&
$\lambda_{c}$&FWHM&
$\lambda_{c}$&FWHM
\\
& 
Class&
Class&
Class&
Class&
($\mu$m)&($\mu$m)&
($\mu$m)&($\mu$m)&
($\mu$m)&($\mu$m)\\
\hline
LMC19&$\mathcal{B}$&$\mathcal{A}$&$\mathcal{B}$&$\alpha$&11.98$\pm$0.06&1.85$\pm$0.17&19.32$\pm$0.06&6.09$\pm$0.25&31.83$\pm$0.05&5.58$\pm$0.22\\
LMC25&$\mathcal{B}$&$\mathcal{A}$&$\mathcal{B}$&$\delta$&11.46$\pm$0.01&1.91$\pm$0.03&18.47$\pm$0.03&4.34$\pm$0.14&29.82$\pm$0.02&6.74$\pm$0.09\\
LMC48&$\mathcal{B?}$&$\mathcal{A}$&$\mathcal{B}$&$\delta$&11.46$\pm$0.01&1.82$\pm$0.02&18.53$\pm$0.06&4.15$\pm$0.11&30.29$\pm$0.02&7.08$\pm$0.06\\
LMC78&$\mathcal{B}$&$\mathcal{A}$&$\mathcal{B}$&$\alpha$&11.87$\pm$0.02&1.62$\pm$0.04&18.81$\pm$0.05&4.39$\pm$0.11&31.43$\pm$0.05&5.43$\pm$0.12\\
Wray16-423&$\mathcal{B}$&$\mathcal{B}$&?&$\alpha$&11.90$\pm$0.06&1.76$\pm$0.11&19.50$\pm$0.03&6.47$\pm$0.08&31.87$\pm$0.06&5.57$\pm$0.13\\
 \hline
\end{tabularx}
  \end{table*}

\begin{figure}
\centering
\includegraphics[width=\columnwidth]{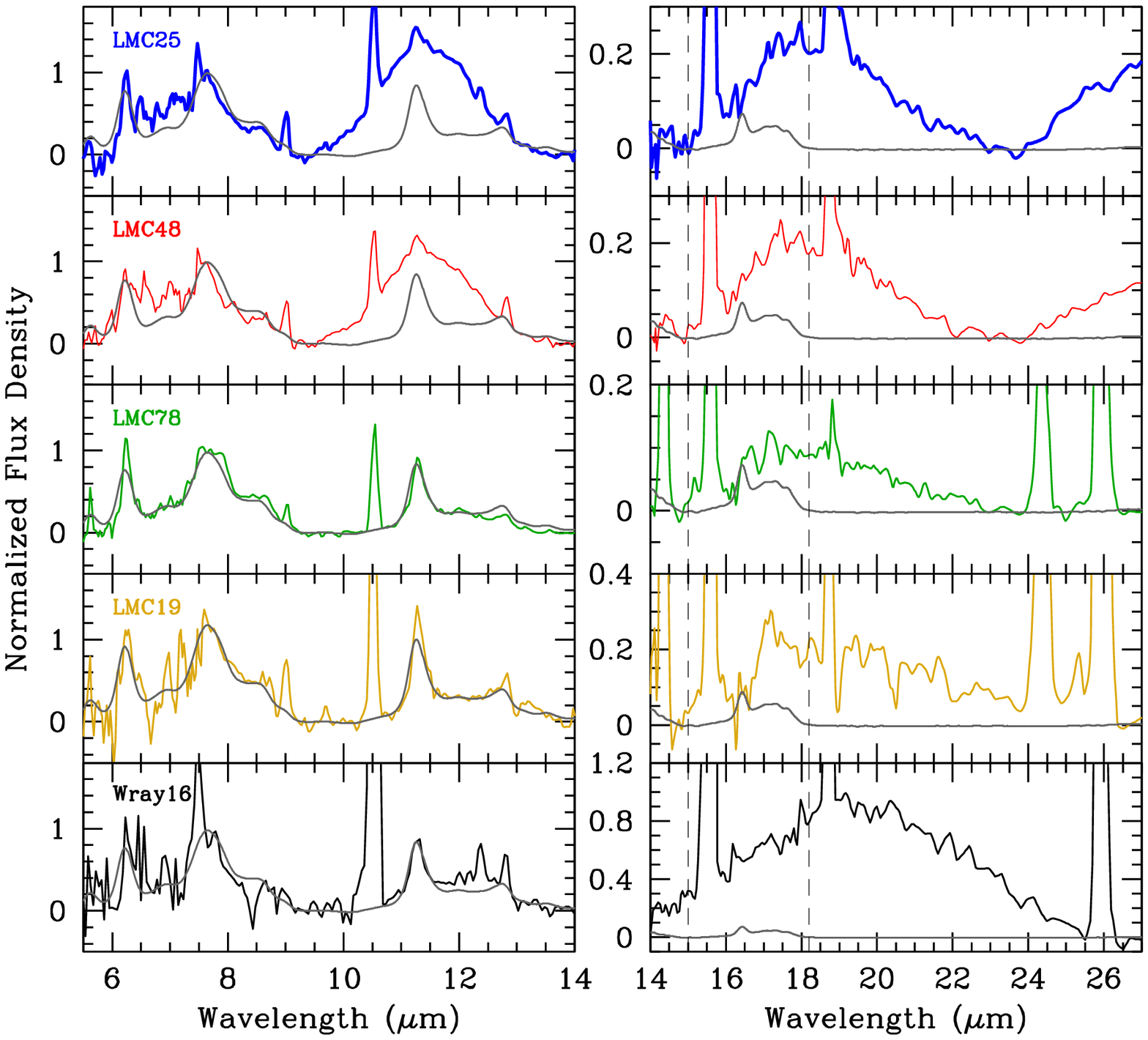}
\caption{
({\it left panels}) The 5.5-14\,$\mu$m continuum subtracted spectra, showing
 the 6-9\,$\mu$m and 10-14\,$\mu$m PAH bands and the broad 11\,$\mu$m feature. 
\emph{Spitzer}/IRS continuum subtracted spectrum of the Galactic 
PN K3-60, indicated by the grey line in each panel, 
 is presented as a template for the 16.4\,$\mu$m PAH and 15-18\,$\mu$m
 plateau. We convolved the SH and LH spectra with a Gaussian kernel 
for the SH and LH module spectra of K3-60 to smooth its spectral
resolving power down to those of the SL and LL spectra,
respectively. After that, we subtracted the local continuum and
gas-emission lines from this spectrum, and then normalised to the intensity peak of the 7.7\,$\mu$m
PAH band. ({\it right panels}) The 14-27\,$\mu$m continuum subtracted spectra, showing
 the broad 16-24\,$\mu$m feature. The spectrum of K3-60, indicated by
 the grey line, is superimposed on each spectrum of the sample PNe. The
 dashed lines show the wavelength range of the 15-18\,$\mu$m
 plateau + 16.4\,$\mu$m PAH. 
}
\label{spt_spec5}
\end{figure}

 \begin{table}
   \centering
   \caption{The ratios of the 6-9\,$\mu$m PAH band,
   the broad 16-24\,$\mu$m feature, the broad 30\,$\mu$m feature,
   and the 11\,$\mu$m SiC fluxes in the local continuum subtracted spectra 
to the integrated flux ($F$(IR)) between 5.75-36.8\,$\mu$m.
 \label{band_measure2}}
   \begin{tabularx}{\columnwidth}{@{}l@{\hspace{4pt}}c@{\hspace{4pt}}c@{\hspace{4pt}}c@{\hspace{4pt}}c@{}}
\hline
    PNe&$F$(6-9\,$\mu$m
    PAH)/&$F$(16-24\,$\mu$m)/&$F$(30\,$\mu$m)/&$F$(11\,$\mu$m SiC)/\\
       &$F$(IR)&$F$(IR)&$F$(IR)&$F$(IR)\\
    \hline
LMC8 & 1.96(--2) & 5.43(--2) & 2.45(--2) & 9.01(--2) \\ 
LMC19 & 7.28(--2) & 4.75(--2) & 3.13(--2) & $\cdots$  \\ 
LMC25 & 4.14(--2) & 2.61(--2) & 5.52(--2) & 7.05(--2) \\ 
LMC48 & 5.19(--2) & 2.64(--2) & 4.85(--2) & 6.88(--2)  \\ 
LMC78 & 8.40(--2) & 2.55(--2) & 2.92(--2) & $\cdots$ \\ 
LMC85 & 5.58(--2) & 3.39(--2) & 2.41(--2) & 1.13(--1) \\ 
LMC99 & 1.07(--1) & 4.17(--2) & 1.29(--2) & $\cdots$  \\ 
Wray16-423 & 1.77(--2) & 8.23(--2) & 3.33(--2) &$\cdots$   \\ 
\hline
\end{tabularx}
\end{table}

\label{lastpage}

\end{document}